\documentclass[12pt]{article}
\usepackage{amsmath,amsthm,amsfonts,amssymb,amscd,cite,color}
\usepackage[colorlinks=true,citecolor=blue,linkcolor=blue,urlcolor=blue]{hyperref}
\usepackage{tikz-cd,pstricks-add}
\usetikzlibrary{positioning}

\newlength{\xtrawidth}
\newlength{\xtraheight}
\allowdisplaybreaks

\makeatletter
\renewcommand\section{\@startsection {section}{1}{\z@}%
                                   {-5.0ex \@plus -1ex \@minus -.2ex}%
                                   {1.5ex \@plus .2ex}%
                                   {\normalfont\Large\bfseries}}
\renewcommand\subsection{\@startsection {subsection}{2}{\z@}%
                                   {-4.0ex \@plus -1ex \@minus -.2ex}%
                                   {1.2ex \@plus .2ex}%
                                   {\normalfont\normalsize\bfseries}}                                   
\makeatother

\numberwithin{equation}{section}
\numberwithin{table}{section}

\newcommand*{\bigtimes}{\mathop{\raisebox{-.5ex}{\hbox{\Huge{$\times$}}}}}
\newcommand{\sslash}{\mathbin{/\mkern-6mu/}}
 
\newcommand*{\bC}{\ensuremath\mathbb{C}}

\newcommand*{\bP}{\ensuremath\mathbb{P}}

\newcommand*{\bZ}{\ensuremath\mathbb{Z}}
\newcommand*{\cD}{\ensuremath\mathcal{D}}
\newcommand*{\cH}{\ensuremath\mathcal{H}}
\newcommand*{\cL}{\ensuremath\mathcal{L}}
\newcommand*{\cM}{\ensuremath\mathcal{M}}
\newcommand*{\cN}{\ensuremath\mathcal{N}}
\newcommand*{\cO}{\ensuremath\mathcal{O}}

\newcommand*{\cW}{\ensuremath\mathcal{W}}

\newcommand*{\cC}{\ensuremath\mathcal{C}}
\newcommand*{\cQ}{\ensuremath\mathcal{Q}}
\newcommand*{\cE}{\ensuremath\mathcal{E}}
\newcommand*{\cV}{\ensuremath\mathcal{V}}
\newcommand*{\cT}{\ensuremath\mathcal{T}}

\newcommand*{\tz}{\Tilde{\zeta}}

\newcommand{\qt}{q_{\tau}}
\newcommand{\kG}{\kappa_\text{G}}
\newcommand{\kR}{\kappa_\text{R}}
\newcommand{\Lq}{\mathfrak{L}_q}
\newcommand{\Uone}{\operatorname{U}(1)}
\newcommand{\spec}{\operatorname{Spec}}
\newcommand{\Ecurve}{E_{\qt}}

\newcommand*{\cEL}{\ensuremath\mathcal{E}ll}
\newcommand{\Hom}{\operatorname{Hom}}

\newcommand{\dd}{\mathrm{d}}
\newcommand{\ii}{\mathrm{i}}
\newcommand{\ee}{\mathrm{e}}

\begin{document}
\begin{titlepage}
\begin{center}
\hfill LMU-ASC 04/26\\
\hfill MITP/26-041\\
\vskip 0.4in
{\hskip -0\textwidth \hbox to 1\textwidth%
{\centerline{\Large\bf{\noindent%
\parbox[t]{1.4\textwidth}{\begin{center}
Quantum Elliptic Cohomology From\\[1ex]
Four-Dimensional Minimal Supersymmetric Gauge Theories
\end{center}}}}%
}}%
\vskip 0.3in
{\large{Ilka Brunner${}^{1,a}$, Peng Cheng${}^{2,b}$, Hans Jockers${}^{2,c}$}}
\vskip 0.4in
{\it ${}^{1\,}$Arnold Sommerfeld Center, Ludwig--Maximilians--Universit\"at M\"unchen\\Theresienstraße 37, 80333 M\"unchen, Germany}
\vskip 0.15in
{\it ${}^{2\,}$Mainz Institute for Theoretical Physics, Johannes Gutenberg-Universit\"at\\ Staudinger Weg 7, 55128 Mainz, Germany}
\vskip 0.15in
{\tt%
${}^{a\,}$\href{mail:ilka.brunner@lmu.de}{ilka.brunner@lmu.de}, 
${}^{b\,}$\href{mail:pcheng@uni-mainz.de}{pcheng@uni-mainz.de},
${}^{c\,}$\href{mail:jockers@uni-mainz.de}{jockers@uni-mainz.de}}
\end{center}

\vskip 0.4in

{\begin{center} {\bf Abstract} \end{center}}
In this paper, we study non-perturbative vortex partition functions of four-dimensional $\cN=1$ supersymmetric gauge theories on the space-time geometry $T^2\times D^2$, and we propose that these partition functions offer a notion of quantum elliptic cohomology. For particular $\Uone$-gauge theories we calculate these partition functions explicitly by applying equivariant localization methods to Handsaw quiver varieties that realize for this particular class of $\Uone$-gauge theories the moduli spaces of the non-perturbative vortex sectors. The determined vortex partition functions are annihilated by difference operators, which are interpreted as Ward identities among $\mathcal{N}=(0,2)$ BPS surface defects. Compared to lower dimensional gauge theories with four supercharges, anomalies play an essential role for a consistent formulation of the four-dimensional partition functions. In our proposal towards a mathematical formulation of the vortex partition functions in terms of equivariant elliptic cohomology, the gauge theory anomalies relate to geometric properties of the Thom sheaves corresponding to the relevant quasimap moduli spaces. Motivated by the explicit computations we reflect on the existence of a `virtual structure sheaf' on the moduli space of quasimaps for a general mathematical theory of quantum elliptic cohomology.
\vfill
\noindent \hfill September 2026
\end{titlepage}
\newpage

{
\hypersetup{linkcolor=black}
\tableofcontents
}

\newpage
\section{Introduction} \label{sec:intro}
Supersymmetry is a powerful symmetry of gauge theories that often allows us to determine suitable correlation functions or the geometry of their moduli spaces exactly by including both perturbative and non-perturbative quantum corrections \cite{Seiberg:1994bz,Intriligator:1994jr,Seiberg:1994rs,Kaplunovsky:1994fg}. Such quantum exact correlation functions or quantum-corrected branches of moduli spaces define a notion of quantum geometry.

We explore the quantum geometries for gauge theories with four supercharges, which --- due to the standard anti-commutation relations of the supercharges transforming in spinorial representations of the Lorentz group --- implies together with Poincar\'e invariance that the space-time dimension must be smaller than or equal to four. Starting from four-dimensional $\cN=1$ supersymmetric gauge theories dimensionally reduced on $S^1$ in two steps, we successively obtain $\cN=2$ supersymmetric gauge theories in three space-time dimensions and $\cN=(2,2)$ supersymmetric gauge theories in two space-time dimensions. 

It is established that in two-dimensional $\mathcal{N}=(2,2)$ gauge theories --- known as two-dimensional $\mathcal{N}=(2,2)$ gauged linear sigma models --- correlation functions of point-like chiral ring elements can be computed exactly \cite{Lerche:1989uy,Morrison:1994fr,Closset:2015rna,Gerhardus:2018zwb}. In the context of three-dimensional $\mathcal{N}=2$ gauge theories correlators among BPS line operators admit a quantum-exact description \cite{Kapustin:2009kz,Beem:2012mb,Yoshida:2014ssa,Koroteev:2017nab,Jockers:2018sfl,Jockers:2019lwe,Jockers:2021omw,Ueda:2019qhg,Gu:2020zpg,Gu:2022yvj,Gu:2023tcv,Closset:2023bdr}. In four-dimensional $\mathcal{N}=1$ gauge theories the natural objects to study are $\mathcal{N}=(0,2)$ BPS~surface defects. The quantum geometry attributed to such surface defects is less explored and is the topic of this work.  

The point-like, the line-like, and the surface-like BPS objects in two-dimensional, three-dimensional and four-dimensional gauge theories enjoy a mathematical formulation in terms of cohomology classes, K~theory classes and elliptic cohomology classes, respectively \cite{Nekrasov4:2009rc,Nekrasov:2009uh}. In particular, the quantum exact fusion products of point-like BPS~observables in two-dimensional gauge theories realize the notion of quantum cohomology \cite{Lerche:1989uy}, which is a deformation of the classical cup product of the cohomology ring to a quantum product that captures non-perturbative instanton corrections \cite{Witten:1988xj,Witten:1990hr,MR1492534,MR1442525}. Similarly, the quantum exact fusion product of BPS~line operators in three-dimensional gauge theories is governed by quantum K~theory, which is a quantum deformation of the tensor product of K~theory classes that includes non-perturbative vortex corrections \cite{Koroteev:2017nab,Jockers:2018sfl,Jockers:2019lwe,Jockers:2021omw,Ueda:2019qhg,Gu:2020zpg,Gu:2022yvj,Gu:2023tcv,Closset:2023bdr}. Hence, a notion of quantum elliptic cohomology is expected to govern fusions of BPS~surface defects in the context of four-dimensional $\mathcal{N}=1$ gauge theories.

Modern gauge theoretic localization techniques offer a powerful tool to investigate the quantum nature of such generalized cohomology theories. We study quantum corrections to the classical Higgs branch~$X_H$ by analyzing the Euclidean gauge theory partition function on the hemispheres $D^2$, which for $\mathcal{N}=2$ three-dimensional and $\mathcal{N}=1$ four-dimensional gauge theories is fibered over the circle $S^1$ and the torus $T^2$. These gauge theory hemisphere partition functions can be calculated with gauge theory localization techniques on curved backgrounds \cite{Festuccia:2011ws,Dumitrescu:2012ha,Closset:2013vra}, as pioneered in refs.~\cite{Pestun:2007rz}. Furthermore, for three- and four-dimensional gauge theories with four supercharges these partition functions correspond to indices of suitable bulk-boundary operators \cite{Beem:2012mb,Gadde:2013wq,Dimofte:2017tpi}, because they can be viewed as Euclidean path integrals with a thermal circle appearing in the base of the above described hemisphere fibrations. The appearance of an index indicates the relevance of enumerative geometry in the context of quantum-deformed generalized cohomology theories. 

The correspondence between gauge theoretic hemisphere partition functions, which include the afore-mentioned non-perturbative quantum corrections, and the quantum geometry of the Higgs branch target spaces~$X_H$ is well-established for two- and three-dimensional gauge theories with four supercharges \cite{Hori:2013ika,Honda:2013uca,Sugishita:2013jca,Jockers:2018sfl}. 

The correspondences between two-dimensional $\mathcal{N}=(2,2)$ gauge theories and quantum cohomology and between three-dimensional $\mathcal{N}=2$ gauge theories and quantum K~theory suggest a similar correspondence between four-dimensional $\mathcal{N}=1$ gauge theories and a suitable definition of quantum elliptic cohomology. However, the precise formulation of such a correspondence is more intricate and subtle for both mathematical and physical reasons.

From a physics perspective, four-dimensional $\mathcal{N}=1$ gauge theories are chiral and as a consequence their spectra are constrained by anomalies, which for instance imposes strong conditions on the possible Higgs branch geometries. In this paper we focus on perturbative chiral anomalies. 

From a mathematical perspective, elliptic cohomology does not exhibit a Thom isomorphism. Furthermore, roughly speaking, K~theory (and ordinary cohomology) exhibit a ring structure, and one may associate to such rings an affine space. In elliptic cohomology, these affine spaces are replaced by projective schemes. The Thom sheaf is a specific line bundle living over this projective scheme -- while in K~theory or cohomology it is a trivial line bundle over affine space. This has many important consequences. First of all, elliptic cohomology does not exhibit a explicit ring structure in the ordinary sense, and hence, quantum elliptic cohomology is not a quantum deformation of a ring, in contrast to cohomology and K theory. Furthermore, the non-trivial Thom sheaf has to be taken into account when formulating a push-forward in elliptic cohomology.

In this paper, we propose the elliptic vortex partition function as basic ingredient for a formulation of a quantum version of elliptic cohomology. From a physics perspective, vortex strings are specific types of surface operators that preserve $\cN=(0,2)$ supersymmetry and couple to a four-dimensional gauge theory containing a $\Uone$ factor. The canonical quantity protected by supersymmetry is the index \cite{Kinney:2005ej,Romelsberger:2005eg}
\begin{equation} \label{eq:IndexIntro}
  Z_\text{4d}(y,q) = \operatorname{Tr}_{\cH} (-1)^F y^P q^{J_3 + R/2} \ ,
\end{equation}
where $y$ and $q$ are the fugacities, referring to global flavor and $\Uone_R$-symmetries. The non-perturbative contributions to this index are captured by the weighted sum over different topological sectors
\begin{equation} \label{eq:IndexIntroSum}
 Z_\text{4d}^\text{np}(Q,q,y) = \sum_{d \ge 0} \, Q^d Z_{\text{vor},d}(q,y) \ .
\end{equation}
The weight $Q$ is determined in terms of the $\theta$-angle and Fayet--Iliopoulos parameter with respect to the $\Uone$-gauge group factor of the four-dimensional $\cN=1$ supersymmetric gauge theory, and $d$ is vortex degree of the topological sector. At the heart of our arguments are the degree-$d$ vortex partition functions $Z_{\text{vor},d}$ that we analyze from various perspectives. 

In Section~\ref{sec:4d N=1 localization} we calculate the index~\eqref{eq:IndexIntro} for a certain class of four-dimensional $\cN=1$ supersymmetric $\Uone$-gauge theories. To this end, using the Hirzebruch--Riemann--Roch index theorem we first formulate the non-perturbative contributions $Z_{\text{vor},d}$ as integrals over the degree $d$ vortex sector moduli spaces. By realizing these moduli spaces in terms of suitable Handsaw quiver varieties, we explicitly evaluate the vortex sector contributions~$Z_{\text{vor},d}$ with the help of equivariant localization techniques. Analyzing the behavior of $Z_{\text{vor},d}$ under large gauge transformations, we observe that $Z_{\text{vor},d}$ is a meromorphic section of a certain  line bundle $\cL_{d}$
over an abelian variety that is associated to the symmetries of the considered gauge theory. For anomaly-free --- and hence quantum-consistent --- gauge theories with a non-anomalous $\Uone_R$-symmetry, we find that the line bundles~$\cL_{d}$ are trivial and the vortex sectors~$Z_{\text{vor},d}$ become meromorphic sections thereof. As the weight $Q$ is determined from a constant non-dynamical background, which is in particular invariant under large gauge transformations, the entire non-perturbative partition function~\eqref{eq:IndexIntroSum} becomes a meromorphic function on the afore-mentioned abelian variety. Furthermore, based on indicative examples, we show recursion relations among vortex sectors $Z_{\text{vor},d}$ at different degrees $d$. As a consequence, the partition functions~\eqref{eq:IndexIntroSum} satisfies a difference equation, which in turn has a physical interpretation in terms of a Ward identity for surface operators. 

By introducing and reviewing aspects of equivariant elliptic cohomology in Section~\ref{sec: equi ell coho}, we set the stage for the proposed mathematical formulation of the vortex partition function in next the section. In particular, we illustrate with simple geometric examples the differences between equivariant K~theory and equivariant elliptic cohomology, which become relevant in the context of the vortex partition function.

In Section~\ref{subsect: vortex partition math} we suggest a mathematical interpretation for the vortex partition function. To the physical setting, we associate the moduli space of quasimaps from $\bP^1$ to the Higgs branch target space $X_H$ with a single marked point $p$. Quasimaps appear naturally in a gauged linear sigma model context \cite{Witten:1993yc,Morrison:1994fr}. As opposed to proper maps, quasimaps are allowed to take values outside the Higgs branch~$X_H$ for a finite number of points, which correspond to special `small vortex' configurations. As the image of the marked point $p$ is required to be a proper point on the Higgs branch $X_H$, we can relate the moduli space of quasimaps to the Higgs branch via the evaluation map
\begin{equation}
  ev: \left\{ \operatorname{Maps}(\phi:\ \bP^1 \dashrightarrow X_H;\, p \in \bP^1) \right\}\to X_H \ , 
  \quad (\phi;p)  \mapsto \phi(p) \in X_H \ .
\end{equation}
We propose that the non-perturbative vortex contributions~$Z_{\text{vor},d}$ are computed using the push-forward of the evaluation map $ev$ in elliptic cohomology. The strategy and arguments are much like in K~theory and ordinary cohomology, namely one uses the equivariant Atiyah--Bott localization formula with respect to the symmetries of the quasimap moduli space. The elliptic version of these arguments are collected in Section~\ref{subsec: Elliptic vortex partition function and anomalies}, where we in particular explain how the Thom sheaf enters the stage. Motivated by the discussed examples and by concrete constructions for quantum cohomology and quantum K~theory, we contemplate on the existence of a `virtual elliptic structure sheaf' for a general theory of quantum elliptic cohomology.  Under this assumption, the calculation in terms of quasimaps yields the same result as the physical vortex partition function. Our physical observations obtain new mathematical interpretations: The anomaly relates to the Thom sheaf. The difference equation is a consequence of the structure of the quasimap moduli space together with the factorization property of the Euler class that are involved in the localization arguments.

Finally, let us remark that for the presented formulation of the vortex partition function and for the resulting proposal of quantum elliptic cohomology, an essential ingredient is the elliptic curve $T_\tau^2$ with complex structure $\tau$, which arises as the compact component of the space-time geometry of the four-dimensional $\cN=1$ supersymmetric gauge theory. As a result, we study equivariant elliptic cohomology rather than its universal version based on topological modular forms. Instead of considering a fixed elliptic curve $T_\tau^2$, the cohomology theory of topological modular forms are formulated over the universal family of elliptic curves, see ref.~\cite{Hopkins:2002} and references there in and ref.~\cite{Gepner:2023} for the equivariant version. A field theoretic realization of topological modular forms is the topic of ref.~\cite{stolz-teichner}. It is certainly interesting to generalize the discussion in this paper to such a universal version of quantum elliptic cohomology.

\section{Vortex Partition Function}  \label{sec:4d N=1 localization}
The key ingredient in this work is the partition function of four-dimensional $\cN=1$ supersymmetric gauge theories on the space-time manifold $T^2\times D^2$ with a Riemannian metric. Following refs.~\cite{Festuccia:2011ws,Dumitrescu:2012ha,Closset:2013vra}, we first review the geometry of curved Riemannian space-time manifolds, which preserve two of the four supercharges of the four-dimensional $\cN=1$ supersymmetric quantum field theory. The partition function on the space-time geometry $T^2\times D^2$ can explicitly be computed with supersymmetric localization techniques \cite{Pestun:2007rz}.  As it receives non-trivial contributions from $\cN=(0,2)$ BPS vortices along the space-time torus $T^2$, we also refer to the partition function on $T^2\times D^2$ as the vortex partition function of the four-dimensional $\cN=1$ gauge theory. 

There is yet another perspective on the vortex partition function. Namely, viewing a circle~$S^1$ of the space-time torus $T^2$ as a thermal circle, we can alternatively interpret the vortex partition function as a supersymmetric index. This index is graded by the degree of the vortex sectors, such that the individual graded pieces realize a collection of four-dimensional indices coupled to two-dimensional $\cN=(0,2)$ BPS vortex strings. In this section we develop this perspective systematically, by studying a particular class of four-dimensional $\cN=1$ supersymmetric $\Uone$-gauge theories. After defining the two-dimensional $\cN=(0,2)$ BPS index of vortex strings, we compute for this gauge theory the contributions of the individual vortex sectors to the vortex partition function. In this process we find that an anomaly-free supersymmetric gauge theory is essential to define consistently the vortex partition function at all vortex degrees. Finally, we connect the obtained vortex partition functions to the twisted chiral ring relations of $\cN=(0,2)$ BPS surface defects. We further show that these chiral ring relations are closely tied to the difference equation that annihilates the vortex partition function. We see that this difference equation in turn enjoys the physical interpretation as a Ward identity among $\cN=(0,2)$ BPS surface defects.

\subsection{Four-Dimensional Riemannian Manifolds Preserving two Supercharges} \label{sec:M4SUSY2}
The conditions for two preserved supercharges of four-dimensional $\mathcal{N}=1$ supersymmetric gauge theories with a $\Uone_R$~symmetry on a curved Riemannian manifold $(M_4,g)$ with a spin structure are systematically analyzed and determined in ref.~\cite{Dumitrescu:2012ha}. The authors couple the gauge theory to the off-shell formulation of $\mathcal{N}=1$ supergravity theory of refs.~\cite{Stelle:1978ye,Sohnius:1981tp} that is suitable for $\mathcal{N}=1$ gauge theories with $\Uone_R$~symmetries. Then --- in addition to the bosonic graviton field and its superpartner the gravitino field $\Psi$ --- the off-shell gravity multiplet consists of the auxiliary two-form gauge field $B_{\mu\nu}$ and the abelian gauge field $A_\mu$. The former auxiliary field possesses the dual one-form field strength $V_\mu$ with $V = \star \dd B$ in terms of the Hodge star $\star$ of the Riemannian manifold $(M_4,g)$. The latter auxiliary field couples to the $\Uone_R$ symmetry current of the supersymmetric gauge theory. In this setup the BPS equations read \cite{Dumitrescu:2012ha}
\begin{equation} \label{vanish of gravitino variation 2nd}
 0 = \left(\nabla_\mu-i A_\mu\right) \zeta + i V_\mu \zeta +  i V^\nu \sigma_{\mu \nu} \zeta \ , \qquad
 0 = \left(\nabla_\mu+i A_\mu\right) \widetilde{\zeta}+ i V_\mu \widetilde{\zeta}+ i V^\nu \widetilde{\sigma}_{\mu \nu} \widetilde{\zeta} \ .
\end{equation}
Here $\nabla_\mu$ is the covariant derivative of the Riemannian manifold $(M_4,g)$. $\zeta$ and $\widetilde{\zeta}$ are the two-component fermionic parameters of the supersymmetry transformation $\delta_{(\zeta, \widetilde{\zeta})}$ in the Weyl representations $\mathbf{2}_+$ and $\mathbf{2}_-$ of the group $\operatorname{Spin}(4)\simeq \operatorname{SU}(2)_+ \times \operatorname{SU}(2)_-$,\footnote{Unlike in four-dimensional Lorentzian signature, the two Weyl representations in Euclidean signature are not related by complex conjugation.} and $\sigma_{\mu\nu}$ and $\widetilde{\sigma}_{\mu\nu}$ are constructed from the generators of Lie algebra $\mathfrak{spin}(4)\simeq \mathfrak{so}(4)$ in these two Weyl representations in the orthonormal frame bundle. As the parameters $\zeta$ and $\tz$ are not complex conjugates in Euclidean signature, the auxiliary connection~$A_\mu$ and field strength~$V_\mu$ can be complex. 

The simultaneous existence of globally non-vanishing solutions~$\zeta$ and $\tz$ to both equations~\eqref{vanish of gravitino variation 2nd} yields strong constraints on the Riemannian manifolds $(M_4,g)$ and implies additional geometric structures. First of all, the spinors $\zeta$ and $\tz$ give rise to non-degenerate self-dual and anti-self-dual two-forms
\begin{equation} \label{eq:twoforms}
  \omega_{\mu \nu}=-\frac{2 i}{|\zeta|^2} \zeta^{\dagger} \sigma_{\mu \nu} \zeta \ ,  \quad  \star\omega = +\omega \ , \qquad
  \widetilde{\omega}_{\mu \nu}=-\frac{2 i}{|\tz|^2} \tz^{\dagger} \Tilde{\sigma}_{\mu \nu} \tz \ ,
   \quad  \star\widetilde{\omega} = - \widetilde{\omega} \ .
\end{equation}
The duality properties of these forms imply --- with the chosen normalization in eq.~\eqref{eq:twoforms} --- that the relations $g(J(X), Y) = \omega(X ,Y)$ and $g(\widetilde{J}(X), Y) = \widetilde{\omega}(X,Y)$ for arbitrary vector fields $X$ and $Y$ define two distinct commuting almost complex structure $J: TM_4 \to TM_4$ and $\widetilde{J}: TM_4 \to TM_4$ \cite{Dumitrescu:2012ha,MR2108214}
\begin{equation}\label{two complex structures}
  J^{\mu}_{\phantom{\mu}\nu} = g^{\mu\kappa} \omega_{\kappa\nu} \ , \qquad
  \widetilde{J}^{\mu}_{\phantom{\mu}\nu} = g^{\mu\kappa} \widetilde{\omega}_{\kappa\nu} \ , \qquad
  J^2 = \widetilde{J}^2 = - \mathbf{1} \ , \qquad [J,\widetilde{J}] = 0 \ .
\end{equation}
Moreover, the two-forms $\omega$ and $\widetilde{\omega}$ are closed and --- due to their duality property --- also co-closed, and hence harmonic, i.e.,  $\Delta \omega = \Delta\widetilde{\omega}=0$ with the Laplace operator $\Delta$ of $(M_4,g)$. As a consequence the associated almost complex structures are integrable \cite{MR2108214}, and hence furnish two commuting complex structures \cite{Dumitrescu:2012ha}. Note that by construction the Riemannian metric $g$ is compatible and hence Hermitian with respect to both complex structures $J$ and $\widetilde{J}$. 

In addition, contracting the spinors $\zeta$ and $\tz$ with the matrices $\sigma^\mu$ --- constructed from the Pauli matrices $(\vec \sigma,- i \mathbf{1} )$ in the orthonormal frame bundle --- yields a (complex) nowhere vanishing Killing vector field $K=K^\mu\partial_\mu$ that is anti-holo\-mor\-phic with respect to both complex structures $J$ and $\widetilde{J}$ \cite{Dumitrescu:2012ha}
\begin{equation} \label{eq:KillingVectorField}
    K^\mu = \zeta \sigma^\mu \widetilde{\zeta}  \ , \qquad 
    \mathcal{L}_K g = 0 \ , \qquad
    J(K) = \widetilde{J}(K) = - i K \ .
\end{equation}    
The real and imaginary part of the anti-holomorphic Killing vector field $K$ generate two isometries of the Riemannian manifold $(M_4,g)$.

Let us further assume that $[K, \overline{K}]=0$.\footnote{A non-vanishing commutator $[K, \overline{K}]$ implies a non-trivial Lie algebra of Killing vector field, which then further constrains the geometry $(M_4,g)$. See ref.~\cite{Dumitrescu:2012ha} for the discussion of these exceptional cases.} Then the Hermitian metric $g$ of the manifold $M_4$ is locally described in terms of the holomorphic coordinates $(z,w)$
\begin{equation} \label{eq:4dMet}
  \dd s_4^2=\Omega(z, \bar{z})^2\left((\dd w+h(z, \bar{z}) \dd z)(\dd \bar{w}+\bar{h}(z, \bar{z}) \dd \bar{z})+c(z, \bar{z})^2 \dd z d \bar{z}\right) \ .
\end{equation}
Here, $\Omega(z,\bar z)$ and $c(z,\bar z)$ are non-vanishing real analytic functions, and $h(z,\bar z)$ is a complex analytic function. In these local coordinates the Killing vector field $K$ is proportional to $\partial_{\bar w}$. Note that the holomorphic transition functions between the holomorphic coordinates $(z',w')$ and $(z,w)$ of two open stets are given by $(z',w')=(G(z),w+F(z))$ with $G(z)$ a biholomorphic function and $F(z)$ a holomorphic function on the overlap of these open sets.

For a $\mathcal{N}=1$ supersymmetric gauge theory on the Riemannian manifold $(M_4,g)$ with two non-vanishing solutions $\zeta$ and $\widetilde{\zeta}$ to eqs.~\eqref{vanish of gravitino variation 2nd} the supersymmetry variations $\delta_\zeta = \zeta \cQ$ and $\delta_{\widetilde{\zeta}}= \widetilde{\zeta}\widetilde{\cQ}$ encode the preserved supercharges obeying the supersymmetry algebra
\begin{equation} \label{eq:preservedSUSY}
  \big[\delta_\zeta,\delta_{\widetilde{\zeta}}\big] = 2 \ii  \delta_{K} \ , \qquad
 \big[\delta_{{K}}, \delta_\zeta\big]=\big[\delta_{{K}}, \delta_{\widetilde{\zeta}}\big]=0  \ , \qquad
\delta^2_\zeta = \delta^2_{\widetilde{\zeta}}=0 \ ,
\end{equation}
where $\delta_K = K^\mu \mathcal{P}_\mu$ denotes the variation with respect to the momentum operator $\mathcal{P}_\mu$ that generates translations along the $w$-direction.

In this work, we consider the four-dimensional manifold $T_\tau^2\times_{(\varepsilon_1,\varepsilon_2)} D^2$ with the Riemannian metric
\begin{equation}  \label{eq:4dcigar}
   \dd s_4^2 = f(\theta)^2\, \dd \theta^2 +\ell^2 \sin^2\theta\,(\dd\phi+\varepsilon_1 \, \dd x 
   + \varepsilon_2 \, \dd y)^2+\beta^2 (\dd x + \tau_1\, \dd y)^2 + \beta^2 \tau_2^2 \, \dd y^2 \ .
\end{equation}
Here $0 \le \theta \le \frac\pi2$ and $0 \le \phi \le 2\pi$ are polar coordinates of the disk $D^2$, which is fibered over the torus $T^2_\tau$ parametrized by the two circular coordinates $0\le x \le 2 \pi$, $0\le y \le 2 \pi$, $f(\theta)$ is a positive monotone increasing function with $f(0)=\ell$, and the constants $\tau_1$, $\beta$, $\ell$ are real and positive, whereas the constants $\varepsilon_1, \varepsilon_2$, $\tau_1$ are real. The complex variable $\tau =\tau_1 +\tau_2$ resides in the upper-half plane $\mathcal{H} = \{\tau \in \mathbb{C} \,|\, \operatorname{Im}(\tau) > 0 \}$ and becomes the complex structure modulus of the torus $T^2_\tau$. 

This geometry possesses the abelian isometry group $\Uone^3$, which is generated by the mutually commuting Killing vectors $\partial_\phi$, $\partial_x$, and $\partial_y$. The isometry group $\Uone^3$ acts transitively on $T^2_\tau\times_{(\varepsilon_1,\varepsilon_2)} {D^{*2}}$ in terms of the punctured disk ${D^{*2}} \equiv D^2 \setminus \{\theta=0\}$ fibered over the torus~$T^2_\tau$. The two-torus $T^2_{\tau,0} \subset T^2_\tau\times_{(\varepsilon_1,\varepsilon_2)} D^2$ at the zero section $\theta=0$ of the fibration is the locus stabilized by the $\Uone$ isometry subgroup generated by the Killing vector $\partial_\phi$. In the equivariant setting discussed in the following, the fixed locus $T^2_{\tau,0}$ becomes relevant for the insertion of $\cN=(0,2)$ BPS surface defects.

As in refs.~\cite{Closset:2013sxa,Longhi:2019hdh}, we introduce local holomorphic coordinates $(w,z)$ upon setting
\begin{equation}
\begin{aligned}
    &w = \frac1{2\pi} \left(  x + \tau y  \right) 
      &&\text{with}\quad  \tau = \tau_1 + \ii \,\tau_2  \ ,  \\
    &z = \ee^{F(\theta) + \ii \left( \phi + \varepsilon_1 x + \varepsilon_2 y\right)} 
      && \text{with} \quad  F'(\theta) = -\frac{f(\theta)}{\ell \sin\theta} \ , 
\end{aligned}  
\end{equation}
where the function~$F(\theta)$ is analytic for $\theta \in (0,\frac\pi2)$ with a logarithmic singularity at $\theta=0$. The disk $|z| \le \ee^{F(\pi/2)}$ parametrized in terms of the holomorphic coordinate $z$ is fibered over the torus $T^2_\tau$ of the holomorphic coordinate $w$. The holomorphic coordinates $(w,z)$ enjoy the identification
\begin{equation}
  \left(w,z\right) \sim \left(w+1,\ee^{2\pi \ii \varepsilon_1} z  \right) \  , \qquad
  \left(w,z\right) \sim \left(w+\tau, \ee^{2\pi \ii \varepsilon_2} z \right) \  ,
\end{equation} 
The parameters $\varepsilon_1$ and $\varepsilon_2$ determine the fibration and the Riemannian metric becomes
\begin{equation}
  \dd s_4^2 = (2 \pi \beta)^2 \left(  \dd w \dd \bar w + c(z,\bar z)^2 \dd z \dd\bar z \right) \quad
  \text{with}\quad c(z,\bar z) = \frac{\ell \sin \theta}{2 \pi \beta  |z|} \ . 
\end{equation}
With respect to the orthonormal frame of one-forms given by the vier-bein $e^a$, $a=1,2,3,4$, defined by
\begin{equation}
  e^1 + \ii e^2 =  2\pi \beta\, \dd w \ , \qquad e^3 +  \ii e^4 = 2\pi \beta \, c(z,\bar z) \,\dd z   \ ,
\end{equation}  
the non-vanishing spinorial solutions $\zeta$ and $\tz$ that solve the BPS~equations~\eqref{vanish of gravitino variation 2nd} read\cite{Dumitrescu:2012ha}
\begin{equation} 
  \zeta = \frac12 \begin{pmatrix} 0 \\ 1 \end{pmatrix} \ , \qquad 
  \tz = 2 \pi \beta \begin{pmatrix} 0 \\ 1 \end{pmatrix}\ .
\end{equation}
The stated spinors give rise to the self-dual and anti-self-dual two forms~\eqref{eq:twoforms}
\begin{equation}
\begin{aligned}
    \omega &= \ii \,(2\pi\beta)^2  \left( \dd w \wedge \dd \bar w + c(z,\bar z)^2 \dd z \wedge \dd \bar z \right) \ , \\
    \widetilde{\omega} &=\ii \, (2\pi\beta)^2 \left( \dd w \wedge \dd \bar w - c(z,\bar z)^2 \dd z \wedge \dd \bar z \right) \ ,
\end{aligned}    
\end{equation}
and the resulting commuting complex structures given by $J(\partial_w) = \ii \partial_w$, $J(\partial_z) = \ii \partial_z$ and $\widetilde{J}(\partial_w) = \ii \partial_w$, $\widetilde{J}(\partial_z) = -\ii \partial_z$. For more details on this space-time geometry and the discussion of the required background values of the auxiliary fields in the off-shell gravity multiplet we refer to refs.~\cite{Dumitrescu:2012ha,Closset:2013sxa,Longhi:2019hdh}.

For the four-dimensional space-time geometry $T^2_\tau\times_{(\varepsilon_1,\varepsilon_2)} D^2$ the supersymmetry algebra~\eqref{eq:preservedSUSY} becomes explicitly in terms of the anti-holomorphic Killing vector $K=\partial_{\bar w}$ 
\begin{equation} \label{eq:SUSYexplicit}
   \big[\delta_\zeta,\delta_{\widetilde{\zeta}}\big] = 2 \ii  \delta_{\partial_{\bar w}} \ .
\end{equation}
At the central torus $T^2_{\tau,0}$ at $\theta=0$ this algebra restricts to a two-dimensional $\cN=(0,2)$ supersymmetry algebra. As a consequence supersymmetric observables in the space-time background $T^2_\tau\times_{(\varepsilon_1,\varepsilon_2)} D^2$ are realized by coupling two-dimensional $\cN=(0,2)$ supersymmetric quantum field theories to the four-dimensional bulk theory at the central torus $T^2_{\tau,0}$, at least at computational level. Contrary to Wilson line observables in three-dimensional $\cN=2$ supersymmetric gauge theories, such two-dimensional $\cN=(0,2)$ supersymmetric quantum field theories cannot be described in terms of the dynamics of the four-dimensional bulk fields. 

At the space-time boundary $T^3 = \partial\left(T^2_\tau\times_{(\varepsilon_1,\varepsilon_2)} D^2\right)$ at $\theta=\frac\pi2$, the algebra~\eqref{eq:SUSYexplicit} is complex and hence cannot realize the real three-dimensional $\cN=1$ supersymmetry algebra. As a consequence the theory restricted to the space-time boundary~$T^3$ cannot be viewed as a three-dimensional $\cN=1$ supersymmetric quantum field theory. This again differs form the three-dimensional $\cN=2$ quantum field theories on the space-time geometry~$S^1\times D^2$ with boundary $T^2 = \partial(S^1\times D^2)$ discussed in ref.~\cite{Yoshida:2014ssa,Bullimore:2020jdq}, where the three-dimensional bulk theory restricted to the boundary~$T^2$ enjoys a description in terms of two-dimensional $\cN=(0,2)$ quantum field theories.  

Let us present another point of view on the preserved supersymmetry and boundary conditions, starting from flat space and appealing to the holomorphic twist. 
The $\cN=1$ supersymmetry algebra in four dimensions reads
\[
\{ Q_\alpha, \bar{Q}_{\dot\alpha} \}= 2\sigma_{\alpha \dot{\alpha}}^\mu P_\mu \ ,
\]
with all other brackets vanishing. 
Imposing boundary conditions implies to choose a subalgebra of the supersymmetry algebra that does not involve translations perpendicular to the boundary. If one wants all $3$ momenta to be part of the chosen supersymmetry algebra, or, in other words, full three-dimensional Lorentz symmetry, then one is automatically led to preserving three-dimensional $\cN=1$ supersymmetry at the boundary.

This is not the kind of boundary condition we want to consider in this paper, as seen above in terms of the preserved supersymmetries of the background and surface defects.

From the flat space point of view, one may want to discuss this by twisting holomorphically. This is the only possible twist in four dimensional minimal supersymmetry, see ref.~\cite{saberietal} for a systematic discussion of twists. Twisting  implies to pick a supercharge $Q_h$ that squares to $0$, and to take cohomology with respect to that supercharge. Choosing an adapted  complex structure with coordinates $w, z$, the anti-holomorphic translations become $Q_h$ exact, i.e., 
\[
Q_h^2=0\ , \quad \{Q_h, G_{\bar{w}} \} = 2\ii P_{\bar{w}} \ , \quad \{Q_h, G_{\bar{z}} \}=2\ii P_{\bar{z}} \ .
\]
Passing to cohomology makes anti-holomorphic derivatives exact. The choice of a three-dimensional $\cN=1$ subalgebra, and with it a boundary condition preserving that symmetry is not compatible with the holomorphic twist, as three-dimensional $\cN=1$ does not allow for a holomorphic twist.

To formulate a subalgebra that does not contain the momentum transverse to the boundary and in addition is compatible with a holomorphic twist, we are thus led to consider
\[
Q_h^2=0\ , \quad \{Q_h, G_{\bar{w}} \} = 2\ii P_{\bar{w}} \ .
\]
Note that this boundary condition no longer exhibits the full three-dimensional Poincare symmetry. However, matching $Q_h$ with $\delta_\zeta$ and $G_{\bar{w}}$ with $\delta_{\tilde\zeta}$, this is exactly the form of the supersymmetry algebra \eqref{eq:SUSYexplicit}. From this point of view, it is an algebra singled out by compatibility of the supersymmetric space-time with boundary with a holomorphic twist. The algebra \eqref{eq:SUSYexplicit} arising from the curved background thus expresses  the compatibility of the holomorphic twist with the chosen background geometry. 

\subsection{\texorpdfstring{$\cN=(0,2)$}{N=(0,2)} BPS Surface Defects} \label{sec:N02BPSDefects}
In this section we probe the four-dimensional supersymmetric $\mathcal{N}=1$ gauge theory on the compact four-dimensional geometry $T_{\tau}^2 \times_{\varepsilon_1,\varepsilon_2}D^2$ with two-dimensional $\cN=(0,2)$ BPS surface defects extending over the two-dimensional torus $T_{\tau}^2$ along the center of the disk $D^2$. Hence, such defects are invariant with respect to the $\Uone$ isometry subgroup generated by the Killing vector $\partial_\phi$ of the four-dimensional metric~\eqref{eq:4dcigar}. This allows us to define equivariant indices of the degrees of freedom of such two-dimensional $\cN=(0,2)$ BPS surface that are localized at the center of the disk~$D^2$ with respect to this unbroken isometry group. 

The $\cN=(0,2)$ BPS surface defects are extended BPS observables of the four-di\-men\-sion\-al $\mathcal{N}=1$ gauge theory that preserve $\cN=(0,2)$ supersymmetry, and their worldvolume degrees of freedom couple to the four-dimensional gauge theory \cite{Gukov:2006jk,Gadde:2013dda}. For surface defects that respect the flavor symmetry group $G_F$ of the four-dimensional gauge theory, the flavor symmetry group $G_F$ is inherited by the defect as well. Furthermore, the gauge group $G$ occurs as an additional flavor symmetry of defect degrees of freedom along the defect.  

We engineer a large class of $\mathcal{N}=(0,2)$ BPS surface defects by assigning degrees of freedom to the worldvolume theory, which come in two-dimensional $\mathcal{N}=(0,2)$ multiplets transforming in representations with respect to the preserved symmetry group $T\times T_F$, the maximal torus of $G\times G_F$. The corresponding flavor symmetry backgrounds along surface defects are:
\begin{equation}
  \mu_i = \oint_\beta a_i - \tau \oint_\alpha a_i \ , \ i=1,\ldots,r \ , \qquad
  \nu_j = \oint_\beta A_j - \tau \oint_\alpha A_j \ ,  \ j=1,\ldots,r_F \ .
\end{equation}
Here the holonomies are computed over the one cycles $(\alpha,\beta)$ generating $H_1(T_\tau^2,\mathbb{Z})$. Moreover, $r = \operatorname{rank}(T)$ and $r_F = \operatorname{rank}(T_F)$, and $a=(a_1,\ldots,a_r)$ and $A = (A_1,\ldots,A_{r_F})$ are the flat background connection one-forms of the respective torus bundles $T$ and $T_F$.

By construction the holonomies take values in the Jacobian $\mathcal{J}(T_\tau^2) = H^{0,1}_{\bar\partial}(T_\tau^2)/H^1(T_\tau^2,\mathbb{Z})$ of the torus $T_\tau^2$. The Jacobian $\mathcal{J}(T_\tau^2)$ is isomorphic to the elliptic curve $\Ecurve$ \cite{Griffiths:1994prl}
\begin{equation}
    \mathcal{J}(T_\tau^2) \simeq  \Ecurve = \bC^{*}/\qt^{\bZ}  \ , \qquad
    \qt = \ee^{2\pi \ii \tau} \ .
\end{equation}    
Altogether, the moduli space of the flat $T \times T_F$-bundles over the torus $T_\tau^2$ is the $(r+r_F)$-dimensional abelian variety $\mathcal{M}_{\text{flat},T_\tau^2}$ given by the $(r+r_F)$-fold Cartesian product of elliptic curves
\begin{equation}  \label{eq:ModFlat}
  \cM_{\text{flat}, T_{\tau}^2} = \bigtimes_{k=1}^{r+r_F} \Ecurve \ ,
\end{equation}
which is parametrized by the $(r+r_F)$-tuple of holonomies
\begin{equation} \label{eq:Defmunu}
  (\mu_1,\ldots,\mu_r,\nu_1,\ldots,\nu_{r_F}) \in  \cM_{\text{flat}, T_{\tau}^2} \ .
\end{equation}
The elliptic genus of $\mathcal{N}=(0,2)$ BPS surface defects is given by the index \cite{Witten:1986bf,Witten:1987cg,Gadde:2013dda,Benini:2013nda}
\begin{equation}
   I_\text{2d}^{(0,2)}(y; \qt) = \operatorname{Tr}_{\cH} (-1)^{F} \qt^{H_L}{{\bar{q}}_\tau}^{H_R} y^P \ , 
\end{equation}
with the fermion number operator~$F$, the left- and right-moving Hamiltonians~$H_L$ and $H_R$, and with the trace over the Hilbert space~$\cH$ of the surface defect states. The index is refined by the fugacities $y$ of the global symmetries that are generated by the operators $P$ commuting with the $\mathcal{N}=(0,2)$ supercharges. Note that the elliptic genus depends only holomorphically on the complex structure modulus $\qt$ of the worldvolume torus $T_\tau^2$ because  the right-moving modes cancel due to the two right-moving supercharges and the insertion of $(-)^F$. 

For the discussed surface defects preserving the flavor symmetries $T\times T_F$, the elliptic genus depends on the fugacities
\begin{equation} \label{eq:Defsf}
   s_i = \ee^{2 \pi \ii \mu_i} \ , \ i=1,\ldots,r \ , \qquad  f_j = \ee^{2 \pi \ii \nu_j} \ , \  j=1,\ldots,r_F \ .
\end{equation}
Geometrically, the index~$I_\text{2d}^{(0,2)}(s,f;\qt)$ is a global meromorphic section of a line bundle~$\cL$ over the moduli space~$\cM_{\text{flat}, T_{\tau}^2}$. As the moduli space $\cM_{\text{flat}, T_{\tau}^2}$ is an abelian variety, the meromorphic section becomes a rational function of theta functions of the abelian variety, which --- due to expression~\eqref{eq:ModFlat} --- factorizes further into a rational term of theta functions of the elliptic curve~$\Ecurve$\footnote{Another definition of Theta function widely used in the literature is $\theta(y;\qt) = (y^{1/2}-y^{-1/2})\prod_{i=1}^{\infty}(1-y\qt^i)(1-y^{-1}\qt^i)$. In this paper we will stick to \eqref{eq:deftheta}. Both can be regarded as the unique section of $\cO(0_{E})$ on $\Ecurve$ and the discussions in this paper apply to both conventions.}
\begin{equation} \label{eq:deftheta}
  \theta(y; \qt) = \big(y;\qt\big)_\infty \, \big(y^{-1}\qt;\qt\big)_\infty 
  = \prod_{n=0}^{+\infty}  \left( 1  - y \qt^n \right)  \left( 1  - y^{-1} \qt^{n+1} \right)\ ,
\end{equation}
with the transformation behavior
\begin{equation} \label{eq:ThetaTrans}
  \theta(\qt^k y; \qt) = (-y)^{-k} \qt^{-\frac12 k(k-1)} \theta(y; \qt) \ ,
\end{equation}
where the argument $y = \prod_{i,j} s_i^{\alpha_i} f_j^{\beta_j}$ is given in terms of the fugacities $s_i$ and $f_j$ and the flavor charges $\alpha_i,\beta_j\in\mathbb{Z}$, which are determined by the $G\times G_F$ representations of the $\mathcal{N}=(0,2)$ worldvolume multiplets. 

While the index~$I_\text{2d}^{(0,2)}(s,f;\qt)$ depends holomorphically on the moduli $\mu_i$ and $\nu_j$ of the flat $T\times T_F$-bundle, the topological data of the line bundle $\cL$ is characterized by its first Chern class~$c_1(\cL)$, which in turn determines the 't~Hooft anomaly of the flavor symmetries of the two-dimensional worldvolume theory of the BPS surface defect.

\bigskip

Let us illustrate the Witten index of BPS $\mathcal{N}=(0,2)$ surface defects explicitly with a few representative examples:
\paragraph{Free $\mathcal{N}=(0,2)$ Chiral Multiplet:}
A two-dimensional $\mathcal{N}=(0,2)$ chiral multiplet consists of a free complex scalar field $\phi$ and a right-moving free chiral fermion $\psi_R$. Let us assume that the chiral multiplet carries a charge $u$ with respect to a global $\Uone$ flavor symmetry. For a flavor background connection $a$ with holonomy $\mu$, the elliptic genus of such a chiral multiplet reads \cite{Benini:2013nda}\footnote{Ref.~\cite{Benini:2013nda} uses another quantization scheme, in which the two left-moving vacua have flavor charges~$\pm\frac{u}{2}$.}
\begin{equation}
  I_{\text{2d},u}^{\text{chiral}}(s;\qt) = \frac{s^{u/2}}{\qt^{1/12}\theta(s^u;\qt)} \ ,
\end{equation}
in terms of the fugacity~$s=\ee^{2\pi\ii \mu}$. Under large gauge transformations of the flavor background fields
\begin{equation} \label{eq:LargeFlavorGauge}
   s \mapsto \qt^k \,s \quad \text{with} \quad k\in\mathbb{Z} \ ,
\end{equation} 
the Witten index transforms as
\begin{equation}
  I_{\text{2d},u}^{\text{chiral}}(s;\qt) \mapsto  (-s)^{u^2k} \qt^{\frac{(uk)^2}2} I_{\text{2d},u}^\text{chiral}(s;\qt) \ . 
\end{equation}
This transformation behavior reflects that the elliptic genus of this free $\mathcal{N}=(0,2)$ chiral field is a meromorphic section of a line bundle $\cL_\text{chiral}$ over $\mathcal{M}_{\text{flat},T_\tau^2} = \Ecurve$  with first Chern class
\begin{equation}
   c_1(\cL_\text{chiral}) = -u^2\, [F] \ , 
\end{equation}
where $[F]$ is the (positive) two-form cohomology class generating $H^2(\Ecurve,\mathbb{Z})$. The coefficient of the first Chern class of the bundle $\cL_\text{chiral}$ is determined 't~Hooft anomaly of the $\Uone$ global symmetry, which is encoded in the four-form anomaly polynomial of the two-dimensional chiral field theory in terms of the second Chern class of the background field strength $F_{\Uone}$ of the global $\Uone$ symmetry, i.e., 
\begin{equation}
  \mathcal{I}^{(4)}_\text{chiral} = - u^2 \operatorname{ch}_2( F_{\Uone} ) \ .
\end{equation}
\paragraph{Free $\mathcal{N}=(0,2)$ Fermi Multiplet:}
The on-shell degrees of freedom of a two-dimensional free $\mathcal{N}=(0,2)$ fermi multiplet arise from a left-moving free chiral fermion~$\psi_L$. Let us assume that the fermion $\psi_L$ has charge $v$ with respect to a global $\Uone$ flavor symmetry. With the flavor background connection $a$ and the associated fugacity $s$, the elliptic genus of this fermi multiplet with periodic left moving spin structure is given by \cite{Benini:2013nda} 
\begin{equation}
   I_{\text{2d},v}^{\text{fermi}}(s;\qt) = \frac{\qt^{1/12}\theta(s^v;\qt)}{s^{v/2}} \ .
\end{equation}
 
For large gauge transformations~\eqref{eq:LargeFlavorGauge} of the flavor background the elliptic genus transforms as
\begin{equation}
  I_{\text{2d},v}^{\text{fermi}}(s;\qt) \mapsto  (-s)^{-v^2k} \qt^{-\frac{(vk)^2}2} I_{\text{2d},v}^\text{fermi}(s;\qt) \ ,
\end{equation}
such that $I_{\text{2d},v}^{\text{fermi}}(s;\qt)$ is a meromorphic section of a line bundle $\cL_\text{fermi}$ over $\mathcal{M}_{\text{flat},T_\tau^2} = \Ecurve$ with first Chern class
\begin{equation}
   c_1(\cL_\text{fermi}) = v^2\, [F] \ ,
\end{equation}
with the (positive) generator $[F]$ of $H^2(\Ecurve,\mathbb{Z})$. The first Chern class is again determined by the 't~Hooft anomaly of the global $\Uone$ symmetry in terms of the four-form anomaly polynomial
\begin{equation}
  \mathcal{I}^{(4)}_\text{fermi} = v^2 \operatorname{ch}_2( F_{\Uone} ) \ ,
\end{equation}
with the background field strength $F_{\Uone}$. 
\paragraph{Free $\mathcal{N}=(0,2)$ Chiral and Fermi multiplets:}
Generalizing the first two examples, we finally consider a $\mathcal{N}=(0,2)$ surface defect with an unbroken $\Uone^K$~flavor symmetry that is described by $n_c$ free $\mathcal{N}=(0,2)$ chiral multiplets and $n_f$ free $\mathcal{N}=(0,2)$ fermi multiplets (with periodic left moving spin structure). The chiral and fermi multiplets carry the $\Uone^K$ flavor charges $u_{\alpha,i}$ with the index $\alpha=1,\ldots,n_c$, and $v_{\beta,i}$ with the index $\beta=1,\ldots,n_f$ and with the common flavor charge index $i=1,\ldots,K$. The elliptic genus of this surface defect reads 
\begin{equation}
  I_{\text{2d}}^{n_c,n_f}(s; \qt)  = 
  \qt^\frac{n_f-n_c}{12} \cdot
  \prod_{i=1}^K \left(\frac{\prod_{\alpha=1}^{n_c}  s_i^{u_{\alpha,i}/2}}
  {\prod_{\beta=1}^{n_f}  s_i^{v_{\beta,i}/2}}\right)
  \cdot
  \frac{\prod_{\beta=1}^{n_f} \theta(s_1^{v_{\beta,1}} \cdot\ldots\cdot s_K^{v_{\beta,K}};\qt)}{\prod_{\alpha=1}^{n_c} \theta(s_1^{u_{\alpha,1}} \cdot\ldots\cdot s_K^{u_{\alpha,K}};\qt)} \ ,
\end{equation}  
where $s_i = \ee^{2\pi \ii \mu_i}$, $i=1,\ldots K$, are the flavor fugacities. For a large gauge transformation of the $\Uone^K$ flavor background of the form 
\begin{equation}
   s_i \mapsto \qt^{k_i} s_i \quad \text{for} \quad k_i \in \mathbb{Z}, \ i=1,\ldots,K \ ,
\end{equation}
the elliptic genus transforms as 
\begin{multline}
 I_{\text{2d}}^{n_c,n_f}(s; \qt) \mapsto 
 (-1)^{\sum_{i,\alpha} u_{\alpha,i} + \sum_{i,\beta} v_{\beta,i}}
 \qt^{\frac12 \sum_\alpha \left(\sum_i u_{\alpha,i}k_i\right)^2 - \frac12 \sum_\beta \left(\sum_i v_{\beta,i}k_i \right)^2 } \\
 \cdot \prod_{j=1}^K s_j^{ - \sum_{\alpha,i} u_{\alpha,i}u_{\alpha,j} k_i + \sum_{\beta,i} v_{\beta,i}v_{\beta,j} k_i } \,
 I_{\text{2d}}^{n_c,n_f}(s; \qt) \ .
\end{multline}
Thus, the elliptic genus~$I_{\text{2d}}^{n_c,n_f}$ is a meromorphic section of a line bundle $\cL_{n_c,n_f}$ over $\mathcal{M}_{\text{flat},T_\tau^2} = \Ecurve \times \ldots \times \Ecurve$ with the first Chern class 
\begin{equation}
  c_1(\cL_{n_c,n_f}) = - \sum_{i,j=1}^K \left( \sum_{\alpha=1}^{n_c} u_{\alpha,i} u_{\alpha,j} 
  - \sum_{\beta=1}^{n_f} v_{\beta,i} v_{\beta,j}\right)  [F_{ij}] \ ,
\end{equation}
with the cohomology classes $[F_{ij}] \in H^2(\mathcal{M}_{\text{flat},T_\tau^2},\mathbb{Z})$ represented by the two-forms $F_{ij} = \frac1{2\operatorname{Im} \tau} ( \mu_i \wedge \bar\mu_j + \mu_j \wedge \bar\mu_i )$ in terms of the holomorphic coordinates $(\mu_1,\ldots,\mu_K) \in \mathcal{M}_{\text{flat},T_\tau^2}$. 

Note that the first Chern class of line bundle $\mathcal{L}_{n_c,n_f}$ of the elliptic genus again directly relates to  the 't~Hooft anomaly, which for the $\Uone^K$ global symmetries is encoded in the four-form anomaly polynomial 
\begin{equation}
  \mathcal{I}^{(4)}_{n_c,n_f} = - \frac12 \sum_{i,j=1}^K \left( \sum_{\alpha=1}^{n_c} u_{\alpha,i} u_{\alpha,j} 
  - \sum_{\beta=1}^{n_f} v_{\beta,i} v_{\beta,j}\right)  \operatorname{c}_1(F_{\Uone_i}) \operatorname{c}_1(F_{\Uone_j}) \ ,
\end{equation}
in terms of the first Chern classes of the background field strengths $F_{\Uone_i}$, $i=1,\ldots,K$, of the global $\Uone$~symmetry factors. The coefficients of $\frac12 \operatorname{c}_1(F_{\Uone_i}) \operatorname{c}_1(F_{\Uone_i}) \equiv \operatorname{ch}_2( F_{\Uone_i})$ determine the $\Uone_i$--$\Uone_i$ 't~Hooft anomaly, while the coefficients $\operatorname{c}_1(F_{\Uone_i}) \operatorname{c}_1(F_{\Uone_j})$ for $i\ne j$ realize the mixed $\Uone_i$--$\Uone_j$ 't~Hooft anomalies. 

These examples illustrate that the 't~Hooft anomaly of the global symmetries of the $\mathcal{N}=(0,2)$ surface defects determines the topological data of a line bundle $\mathcal{L}$ that is associated to the refined elliptic genus. However, the line bundle $\mathcal{L}$ and hence the elliptic genus itself contains more information than just its topological data, as the line bundle $\cL$ depends continuously on the moduli of flat background connections~\eqref{eq:ModFlat} without altering the Chern class of the line bundle $\cL$. As detailed in ref.~\cite{Cheng:2026abk}, the relationship between the first Chern class~$c_1(\cL)$ of the line bundle $\cL$ and the degree two anomaly polynomial of the corresponding triangular one-loop Feynman diagrams is established with the Atiyah--Singer family index theorem.

\subsection{Vortex Strings in Four-Dimensional \texorpdfstring{$\cN=1$}{N=1} Gauge Theory} \label{sec:vstrings}
$\mathcal{N}=(0,2)$ vortex strings furnish a specific type of $\mathcal{N}=(0,2)$ BPS surface defects. Unlike general $\mathcal{N}=(0,2)$ BPS surface defects, the field content of the vortex string relates to the dynamics of the bulk gauge theory, which in our case is a $\mathcal{N}=1$ gauge theory with gauge group $G$. Vortex strings are topological non-trivial solutions in the Higgs branch that are characterized by the non-trivial holonomy of the four-dimensional gauge group $G$ upon encircling the vortex string \cite{Hanany:2004ea,Shifman:2015kla}. As a consequence the existence of vortex strings requires that the compact gauge group $G$ is not semi-simple but contains abelian $\Uone$ factors.

For the four-dimensional $\mathcal{N}=1$ gauge theory on the geometry $T_{\tau}^2 \times_{\varepsilon_1,\varepsilon_2}D^2$, a $\mathcal{N}=(0,2)$ BPS vortex string extends along the two-dimensional torus $T_\tau^2$ at the center of the disk $D^2$. The partition function $Z_\text{4d}$ of the vortex string coupled to the bulk gauge theory is given in terms of the index \cite{Kinney:2005ej,Romelsberger:2005eg}
\begin{equation} \label{eq:4d2dIndex}
   Z_\text{4d}(y,q) = \operatorname{Tr}_{\cH} (-1)^F y^P q^{J_3 + R/2} \ ,
\end{equation}
where $F$ is the fermion number operator, $P$ are the symmetry generators commuting with the preserved supercharges together with their associated fugacities $y$. For the background geometry $T_{\tau}^2 \times_{\varepsilon_1,\varepsilon_2}D^2$ the operator $J_3 + \frac{R}2$ with the fugacity $q=\ee^{2\pi\ii (\varepsilon_2 - \tau \varepsilon_1) }$ commutes with the unbroken supercharges as well \cite{Kinney:2005ej,Romelsberger:2005eg,Assel:2015nca}, where $R$ is the $\Uone_R$ symmetry generator and $J_3$ the generator for rotations of the disk $D$. From the perspective of the vortex string, the operator $J_3 + \frac{R}2$ with the fugacity $q$ is a flavor symmetry, as the worldvolume of the vortex string preserves the $\Uone_R$ symmetry and is the fixed-point locus with respect to rotations of the disk $D$. 

The trace is taken over the Hilbert space $\cH$, which describes the states of the four-dimensional $\mathcal{N}=1$ gauge theory in the presence of vortex strings. The holonomies characterizing the vortex strings are measured by the dual topological symmetries with the fugacities
\begin{equation} \label{eq:defQ}
  Q_a = \ee^{-\zeta_a + 2\pi \ii \vartheta_a} \ ,
\end{equation}   
in terms of the $\vartheta_a(= \int_{T^2} B)$ -angles and the Fayet--Iliopoulos parameters $\zeta_a$ for each $\Uone$-factor of the gauge group $G$. As a consequence the partition function $Z_\text{4d}$ decomposes as
\begin{equation} \label{eq:Znp}
  Z_\text{4d}(Q,q,y) = Z_\text{4d}^\text{per}(q,y) Z_\text{4d}^\text{np}(Q,q,y) \ ,
\end{equation} 
where $Z_\text{4d}^\text{per}(q,y)$ denotes the perturbative contributions to the index, which is independent of the multi-index fugacities~$Q$ of the dual topological symmetries. Instead, it depends only on the fugacities $y$ of the remaining symmetries, which also includes the fugacity $\qt$ attributed to the translations along isometries of the torus $T_{\tau}^2$. The contribution $Z_\text{4d}^\text{np}(Q,q,y)$ captures the non-perturbative field configurations of the $\mathcal{N}=(0,2)$ BPS vortex strings and decomposes into topological sectors according to
\begin{equation}
  Z_\text{4d}^\text{np}(Q,q,y) = \sum_{d \ge 0} \, Q^d Z_{\text{vor},d}(q,y) \ ,
\end{equation}  
where the individual topological sectors are graded by the fugacities $Q$ and where the sum runs over the multi-index $d$ with non-negative entries. In the remainder of this work, we focus on the computation and the analysis of the non-perturbative contributions~$Z_{\text{vor},d}(q,y)$ for certain classes of four-dimensional $\mathcal{N}=1$ abelian gauge theories.

\subsection{The \texorpdfstring{$\cN=(0,2)$}{N=(0,2)} BPS Vortex Moduli Space} \label{sec:BPSVorMSpace}
We consider the four-dimensional $\mathcal{N}=1$ $\Uone$~gauge theories $\cT$ given by the spectrum listed in Table~\ref{tab:TheoryT}. These gauge theories are consistent for
\begin{equation} \label{eq:Anom}
   \kG = 0 \quad\text{with}\quad
   \kG = n+1 - \sum_{i=1}^a l_i^3  \ ,
\end{equation}
as then there is no four-dimensional triangular $\Uone$-gauge anomaly.
\begin{table}
\def\vr{\vrule height2.5ex depth1.25ex width 0.6pt}
\def\vrth{\vrule height2.5ex depth1.25ex width 1pt}
\hfil\vbox{
\offinterlineskip
\tabskip=0pt
\halign{\vrth~#~\hfil&\vr~#~\hfil&\vr\hfil~#~\hfil&\vr\hfil~#~\hfil&\vr\hfil~#~\hfil\vrth\cr
\noalign{\hrule height 1pt}
Multiplet&Multiplicity&$\Uone$-Charge&$\operatorname{U}_R(1)$-Charge&$\Uone^{n+1+a}$-Flavor Charges\cr
\noalign{\hrule height 1pt}
$\Uone$-vector&\hfil$1$&$0$&$0$&$0$\cr
chiral $\phi^I$&$I=1,\ldots,n+1$&$+1$&$\mathfrak{q}_\phi$&$f_I$\cr
chiral $\varphi^i$&$i=1,\ldots,a$&$-l_i < 0$&$\mathfrak{q}_i$&$f_{n+i}$\cr
\noalign{\hrule height 1pt}
}}\hfil
\caption{The spectrum of the four-dimensional $\mathcal{N}=1$ abelian gauge theories of type~$\cT$. The chiral fields have flavor charges $f_\alpha=(f_I,f_{i+n+1})$ with respect to the classical $\Uone^{n+1+a}$-flavor symmetry, which take values in the lattice $f_{\alpha} \in \mathbb{Z}^{n+1+a}$, $\alpha=1,\ldots,n+a+1$.} \label{tab:TheoryT}
\end{table}
In addition, we require that the global $\Uone_R$-symmetry of the gauge theory $\cT$ is free of anomalies as well. The triangular diagram of the $\Uone_R$-current is determined by the gauge-charged Weyl fermions in the chiral multiplets, and hence the vanishing of the mixed $\Uone_R$--$\Uone$-gauge anomaly yields the condition
\begin{equation} \label{eq:AnomMixed}
   \kR = 0 \quad \text{with} \quad
  \kR = (n+1)(\mathfrak{q}_\phi -1) + \sum_{i=0}^a l_i^2 (\mathfrak{q}_i -1)\ ,
\end{equation}
where the shift of the R-charges reflects the difference between the R-charge of the chiral multiplet and the R-charge of the Weyl fermion within this multiplet.

Part of the classical $\Uone^{n+1+a}$-flavor symmetry is not entirely preserved at the quantum level, but instead suffers an Adler--Bell--Jackiw anomaly determined by the triangular diagram with the $\Uone$-flavor currents coupling to two $\Uone$-gauge bosons. Hence, the anomaly free subgroups of the $\Uone^{n+1+a}$-flavor symmetry are determined by the kernel of the $1\times (n+a+1)$-matrix $\left(f_{1,s}+ \ldots f_{n+1,s}+l_1^2 f_{n+2,s}+\ldots +l_a^2f_{n+a+1,s}\right)_{s=1,\ldots,n+a+1}$ of rank one, which yields a non-anomalous subgroup $\Uone^{n+a}$ of the classical $\Uone^{n+1+a}$-flavor symmetry group.

For unbroken flavor symmetries and hence in the absence of a superpotential, the scalar potential is minimized semi-classically by the D-term equation
\begin{equation}
  \sum_{I=1}^{n+1} |\phi^I|^2 -  \sum_{i=1}^{a} l_i  |\varphi^i|^2 = \zeta \ .
\end{equation}
For the Fayet--Iliopoulos parameter $\zeta > 0$ the theory possesses the semi-classical Higgs branch $X_H$ given by the total space $\operatorname{Tot}(\cE \to \mathbb{P}^n)$ of the vector bundle $\cE = \cO(-l_1) \oplus \ldots \oplus \cO(-l_a)$ over the complex projective space $\mathbb{P}^n$, which arises form the symplectic quotient
\begin{equation}\label{Higgs branch XH}
 X_H = \operatorname{Tot}(\cE \to \mathbb{P}^n)   \ \simeq\  
  \left\{ \sum_{I=1}^{n+1} |\phi^I|^2 -  \sum_{i=1}^{a} l_i  |\varphi^i|^2 - \zeta = 0 \right\} 
  \Big/ \Uone \quad \text{for} \quad \zeta>0 \  ,
\end{equation}
where the projective and fiber coordinates are the complex scalar fields $\phi^I$ and $\varphi^i$ of the chiral multiplets, respectively. 

Note that the symplectic quotient~\eqref{Higgs branch XH} of the Higgs branch $X_H$ is independent of the R-charge assignments of the chiral multiplets of the gauge theory $\cT$ as given in Table~\ref{tab:TheoryT}. Nevertheless, the index~\eqref{eq:4d2dIndex} of the gauge theory $\cT$ does explicitly depend on these R-charges, which is consistent with an appropriate enumerative interpretation. Namely, including the effect of the R-charges, there are at least two kinds of (sub-)bundles of $\cE$ that occur in the total space of the Higgs branch $X_H = \operatorname{Tot}(\cE \to \mathbb{P}^n)$. Assuming vanishing R-charges $\mathfrak{q}_\phi=0$ of the chiral multiplets $\phi^I$, the chiral multiplets $\varphi^i$ with vanishing R-charge $\mathfrak{q}_i$ parametrize ordinary line bundles $\cO(-l_i)$ over the base $\mathbb{P}^n$, while chiral multiplets $\varphi^j$ with R-charge $\mathfrak{q}_j=2$ realize super-line bundles $\Pi\cO(-l_j)$ over the base $\mathbb{P}^n$ \cite{Givental:2015Kvi}. The super-line bundles $\Pi\cO(-l_j)$ indicate that the gauge theory $\cT$ comes with a non-vanishing $\mathcal{N}=1$ superpotential \cite{Witten:1993yc}
\begin{equation}
   W = \sum_{\{j| \mathfrak{q}_j =2 \} } \varphi^j \, G_j^{(l_j)} \ .
\end{equation}
Here $G_j^{(l_j)}$ are generic holomorphic functions with $\Uone$-gauge charge $l_j$ of the chiral fields $\phi^I$ and $\varphi^i$ that have vanishing R-charges. As a consequence, the low energy effective Higgs branch $X_H^\text{eff}$ is in general not given by the symplectic quotient $X_H$ alone, but rather by the intersection of the symplectic quotient $X_H$ with the critical locus of the superpotential~$W$,\footnote{Depending on the spectrum of the gauge theory $\cT$, even for generic choices of the functions $G_j^{(l_j)}$ the effective Higgs branch~$X_H^\text{eff}$ may not be a smooth manifold or orbifold. Then the described semi-classical analysis for the effective Higgs branch $X_H^\text{eff}$ becomes insufficient.} i.e.,
\begin{equation}
  X_H^\text{eff} = X_H \cap \left\{ dW = 0 \right\} \ .
\end{equation}
It is really the geometry of the low-energy effective Higgs branch~$X_H^\text{eff}$ --- arising for generic functions $G_J^{(l_j)}$ ---- for which the index~\eqref{eq:4d2dIndex} --- sensitive to the R-charges of the theory $\cT$ --- encodes enumerative information. Note that for a non-vanishing superpotential~$W$ it is necessary to choose the boundary conditions of the four-dimensional $\cN=1$ chiral multiplets along the three-dimensional space-time boundary $T^3=\partial(T^2_\tau \times_{\epsilon_1,\epsilon_2} D^2)$ such that the superpotential terms at the boundary respect the supersymmetries~\eqref{eq:SUSYexplicit}. For the gauge theory $\cT$ this can for instance be achieved by imposing Dirichlet boundary conditions on those chiral fields $\varphi^j$ with R-charge $\mathfrak{q}_j=2$. 

For the four-dimensional $\mathcal{N}=1$ gauge theories $\cT$ we want to determine the vortex partition function~$Z_\text{4d}^\text{np}(Q,q,\qt,y)$, where $Q$, $q$, $\qt$, and $y$ are the fugacities of the dual topological symmetry of the $\Uone$-gauge symmetry, of the generator $J_3+\frac{R}2$, of the isometries of the torus $T_\tau^2$, and of the flavor symmetry $\Uone^{n+1+a}$, respectively. We explicitly illustrate the computation of $Z_\text{4d}^\text{np}(Q,q,\qt,y)$ for the canonical gauge theory~$\cT_\text{can}$ that is defined by the representations of the classical theory 
\begin{equation} \label{eq:ChargeCan}
\begin{aligned}
    &\mathfrak{q}_\phi = 0 \ , \
    &&\mathfrak{q}_i = 2 \ , \ l_i = 1 \quad \text{for} \quad i=1,\ldots,n+1 \quad \text{with} \quad a=n+1 \ , \\
    &f_{\alpha,s} = \delta_{\alpha,s} \ &&\text{with}\quad \alpha,s \in\{1, \ldots 2n+2\} \ , 
\end{aligned}    
\end{equation}  
and --- for the Fayet--Iliopoulos parameter $\zeta > 0$ --- gives rise to the Higgs branch
\begin{equation} \label{eq:DefXH}
   X_H = \operatorname{Tot}(\cO(-1)^{\oplus (n+1)} \to \mathbb{P}^n)  \ .
\end{equation}
Note that the charge assignments~\eqref{eq:ChargeCan} fulfill the anomaly cancellation conditions \eqref{eq:Anom} and \eqref{eq:AnomMixed}.

Vortex solutions arise in the Higgs branch $X_H$ only if the expectation values of the negatively charged scalar field $\varphi^i$ vanish \cite{Hanany:2003hp}.  Hence, we can consider the reduced Higgs branch $X_{H,\text{red}}\simeq \mathbb{P}^{n}$ defined as the zero section of the bundle $\cO(-1)^{\oplus (n+1)}$ over $\mathbb{P}^n$. For a particular point on the reduced Higgs branch $X_{H,\text{red}}$, the moduli space $\cM_{\text{vor},d}$ of $\mathcal{N}=(0,2)$ BPS vortex configurations of degree $d$ is realized by a complex $d(n+1)$-dimensional K\"ahler manifold \cite{nakajima2011handsaw}. 

Let us parametrize the two-dimensional transverse directions of the $\mathcal{N}=(0,2)$ BPS vortex configurations in terms of the holomorphic coordinate $z$. A degree $d$ vortex is then described by a field configuration of the scalar fields $\phi^I$ in the transverse direction that is given by the degree $d$ holomorphic map $\phi: \mathbb{C} \to \mathbb{C}^{n+1}$ of the form \cite{Witten:1993yc}
\begin{equation} \label{eq:SheafHomPhi}
  \phi(z) = \phi_0 + \phi_1 z + \ldots + \phi_{d} z^d 
   \quad \text{with} \quad \phi_{d} \ne 0 \ .
\end{equation}
For large values of $z$ the vortex configuration approaches the vacuum of the Higgs branch. Hence, $\phi_d$ describes a projective point in the Higgs branch, i.e., 
\begin{equation} \label{eq:HiggsVac}
  \phi_d \in X_{H,\text{red}} \simeq \mathbb{P}^n \ .
\end{equation}
Upon compactifiying the complex plane $\mathbb{C}$ parametrized by $z$ to the one-dimensional complex projective space $\mathbb{P}^1$, the moduli space $\cM_{\text{vor},d}$ of BPS vortex solutions enjoys the algebraic description in terms of the moduli space of flags of holomorphic bundles over $\mathbb{P}^1$ of the type
\begin{equation} \label{eq:flag}
\begin{CD}
  0 \subset \mathcal{O}(-d) \subset 
  \mathcal{O}^{\oplus(n+1)} \\
  @VVV\\
  \mathbb{P}^1 
\end{CD}  \ .
\end{equation}
In this setup $\phi$ becomes an injective sheaf homomorphism $\phi \in \operatorname{Hom}(\mathcal{O}(-d), \mathcal{O}^{\oplus(n+1)})$, which gives rise to an inclusion of holomorphic vector bundles $\phi:\mathcal{O}(-d) \hookrightarrow  \mathcal{O}^{\oplus(n+1)}$ and hence a flag of the from~\eqref{eq:flag}, if the quotient sheaf $\mathcal{O}^{\oplus(n+1)} / \mathcal{O}(-d)$ is a locally free, which means that the quotient sheaf is a holomorphic vector bundle as well. This condition is fulfilled if the injective sheaf homomorphism is an injective vector space homomorphism $\phi(z) \in \operatorname{Hom}(\mathbb{C}, \mathbb{C}^{n+1})$ for any point $z \in \mathbb{P}^1$. If, however, there are solutions to the equation $\phi(z)=0$ --- which we refer to small vortex solutions in analogy to small instantons in the context of two-dimensional gauged linear sigma models \cite{Witten:1993yc,Morrison:1994fr,Losev:1999nt,Losev:1999tu} --- then there are finitely many points in $\mathbb{P}^1$, where the vector space homomorphism~$\phi(z) \in \operatorname{Hom}(\mathbb{C}, \mathbb{C}^{n+1})$ fails to be injective. Even in this case, a flag of holomorphic vector bundles can be assigned to $\phi$ by restricting to the locally free part of the quotient $\mathcal{O}^{\oplus(n+1)} / \mathcal{O}(-d)$. This construction is discussed in detail in ref.~\cite{nakajima2011handsaw}. We denote the (possibly empty) subspace of vortex configurations of degree $d$ without small vortex solutions as the regular moduli space $\cM_{\text{vor},d}^\text{reg}$ of $\mathcal{N}=(0,2)$ BPS vortex solutions.  

In this algebraic formulation the Higgs vacuum~\eqref{eq:HiggsVac} is determined by the flag over the point $z=\infty \in \mathbb{P}^1$, where the injective map $\phi(\infty) \equiv \phi_d: \mathbb{C} \hookrightarrow \mathbb{C}^{n+1}$ determines a Higgs vacuum in $X_{H,\text{red}}$. As we describe the BPS vortex moduli space of degree $d$ for a fixed Higgs vacuum, the points in moduli space $\cM_{\text{vor},d}$ are given by the space of framed flags of holomorphic vector bundles. These flags determine the Higgs vacuum in terms of the fixed fiber at $z=\infty$. Thus the complex dimension of the moduli space $\cM_{\text{vor},d}$ of framed flags is readily seen to be $d(n+1)$, which is parametrized by the coefficient vectors $\phi_0, \ldots, \phi_{d-1}$ of complex dimension $n+1$, while the coefficient vector $\phi_d$ determines the framing. This description of the moduli space $\cM_{\text{vor},d}$ directly relates to the quasimap moduli space to be discussed in Section~\ref{sec:QuasiMapModuliSpace}.

There is yet another formulation of the moduli space $\cM_{\text{vor},d}$ of $\mathcal{N}=(0,2)$ BPS vortices in terms of a Handsaw quiver variety~\cite{nakajima2011handsaw}, which is based on the ADHM construction~\cite{Atiyah:1978ri}. For the specific four-dimensional $\mathcal{N}=1$ gauge theory $\mathcal{T}_\text{can}$ the Handsaw  quiver simplifies to
\begin{equation} \label{eq:Quiver}
\begin{tikzpicture}
[roundnode/.style={circle, draw=black!60, fill=green!5, thick, minimum size=7mm},
squarednode/.style={rectangle, draw=black!60, fill=red!5, thick, minimum size=5mm},]
\node[roundnode](maintopic){$d$};
\node[squarednode](lowercircle)[below=of maintopic] {$1$};
\node[squarednode](rightsquare)[right=of lowercircle] {$n$};
\draw[->] (lowercircle.north) --  node[pos=0.5, left] {$a$}(maintopic.south);
\draw[->] (maintopic.south east) -- node[midway,above] {$b$} (rightsquare.north west);
\draw[black,thick,->] (maintopic.90) arc (0:264:4mm) node[pos=0.5, left] {$B$};;
\end{tikzpicture} \ .
\end{equation}
Here, the round node represents the $d$-dimensional complex vector space $V \simeq \mathbb{C}^d$ transforming in the defining representation of the Lie group $\operatorname{GL}(V)$. The square nodes represent the one-dimensional and the $n$-dimensional complex vector spaces $W_0\simeq \mathbb{C}$ and $W_1\simeq \mathbb{C}^n$, which do not transform under $\operatorname{GL}(V)$. The arrows  of the quiver represent vector space endomorphisms $B \in \operatorname{End}(V)$, and vector space homomorphisms $a\in \operatorname{Hom}(W_0,V)$ and $b \in \operatorname{Hom}(V,W_1)$, respectively. The induced action of the Lie group $\operatorname{GL}(V)$ on the triple $(B,a,b)$ reads
\begin{equation} \label{eq:GLaction}
  g \in \operatorname{GL}(V): \ (B,a,b) \mapsto (g^{-1} B g, g^{-1} a, b g) \ .
\end{equation}
The Handsaw quiver variety $\cD_d$ associated to the quiver~\eqref{eq:Quiver} is given by orbits of stable triples~$(B,a,b)$, namely 
\begin{equation}
  \cD_d =  \left\{ (B,a,b)\ \text{stable}  \right\} \sslash \operatorname{GL}(V)  \ ,
\end{equation} 
with a stable triple $(B,a,b)$ defined as \cite{nakajima2011handsaw}
\begin{equation} \label{eq:BabStable}
    (B,a,b)\ \text{stable}  \quad :\Longleftrightarrow \quad \mathbb{C}[B] \circ a(W_0) = V \ ,
\end{equation} 
where $\mathbb{C}[B]$ denotes the subring of the endomorphism ring $\operatorname{End}(V)$ generated by the endomorphism $B$. This means that there is no non-trivial subvector space of $V$ that is invariant with respect to the action of $B$ containing the image of $a$. Moreover, we define the regular subvariety $\cD_d^\text{reg}$ of $\cD_d$ as
\begin{equation}
  \cD_d^\text{reg} =  \left\{ (B,a,b)\ \text{stable and co-stable}  \right\} \sslash \operatorname{GL}(V)  \ ,
\end{equation} 
where a co-stable triple $(B,a,b)$ is given by
\begin{equation}
  (B,a,b)\ \text{co-stable} \quad :\Longleftrightarrow \quad 
  \left\{v \in V \ \middle| \ b \circ \mathbb{C}[B](v)  = 0 \right\} = 0  \ .
\end{equation}
This says that there is no non-zero subvector space invariant under the action of $B$ in the kernel of $b$. Note that the definitions of stable and co-stable triples $(B,a,b)$ respect the group action~\eqref{eq:GLaction}.

The moduli space $\cM_{\text{vor},d}$ of degree $d$ BPS vortices is isomorphic to the Handsaw quiver variety $\cD_d$ \cite{nakajima2011handsaw}, i.e.,
\begin{equation} \label{eq:IsoD}
  \cM_{\text{vor},d} \simeq \cD_d \ .
\end{equation}  
Moreover, the regular subvariety $\mathcal{D}^\text{reg}$ maps to  the regular submoduli space $\cM_{\text{vor},d}^\text{reg}$ of degree $d$ BPS vortex solutions. This isomorphism can be seen as follows. For a stable triple $(B,a,b)$ the sheaf homomorphism
\begin{equation}
  \alpha: (W_0 \oplus V) \otimes \mathcal{O} \to V \otimes \mathcal{O}, \quad \alpha := \left( a, z - B\right) \ ,
\end{equation} 
is surjective for all points $z \in \mathbb{P}^1$ \cite{nakajima2011handsaw}. Furthermore, given the sheaf homomorphism
\begin{equation}
  \beta: (W_0 \oplus V) \otimes \mathcal{O} \to (W_0 \oplus W_1) \otimes \mathcal{O}, \quad \beta := \operatorname{id} \oplus b \ ,
\end{equation} 
for a co-stable triple $(B,a,b)$ the intersection $\operatorname{ker} \alpha |_z \cap \operatorname{ker} \beta |_z$ is empty for all points $z \in \mathbb{P}^1$ \cite{nakajima2011handsaw}. As a consequence, for stable triples $(B,a,b)$ the kernel $\operatorname{ker} \alpha$ becomes the line bundle $\mathcal{O}(-d)$ and for both stable and costable triples $(B,a,b)$ the map $\beta|_{\operatorname{ker} \alpha}: \mathcal{O}(-d) \hookrightarrow \mathcal{O}^{\oplus(n+1)}$ describes an inclusion of the line bundle $\mathcal{O}(-d)$ into the rank $(n+1)$ vector bundle $\mathcal{O}^{\oplus(n+1)}$. Identifying the map $\beta|_{\operatorname{ker} \alpha}$ with the sheaf homomorphism $\phi(z)$ in eq.~\eqref{eq:SheafHomPhi}, the $\operatorname{GL}(V)$-orbit of a stable and co-stable triple $(B,a,b)$ is map to a flag of holomorphic bundles of the type~\eqref{eq:flag} that furnishes a point in the regular submoduli space $\cM_{\text{vor},d}^\text{reg}$. Upon picking a representative in the $\operatorname{GL}(V)$-orbit a framing of the flag of holomorphic bundles is chosen in terms of the Higgs vacuum
\begin{equation}
  \beta|_{\operatorname{ker} \alpha,z=\infty}(W_0\oplus V)\equiv b(V) \in \mathbb{P}^n \simeq X_{H,\text{red}}  \ .
\end{equation}
Note that for any orbit of a co-stable triple $(B,a,b)$ any point in the Higgs branch $X_{H,\text{red}}$ can be reached because the group $\operatorname{GL}(V)$ acts transitively on the set of non-vanishing endomorphisms~$b$. Conversely, given a framed flag of holomorphic bundles of the form~\eqref{eq:flag} an orbit of a co-stable triple $(B,a,b)$ can unambiguously be identified, which shows $\cM_{\text{vor},d}^\text{reg} \simeq \mathcal{D}^\text{reg}$. This identification extends to the isomorphism $\cM_{\text{vor},d} \simeq \cD_d$. For further details and proofs of these statements, we refer the reader to the thorough exposition in ref.~\cite{nakajima2011handsaw}.

As opposed to the framed flags realizing points in $\cM_{\text{vor},d}$, the Handsaw quiver variety $\cD_d$ does not depend on a particular choice of Higgs vacuum, e.g. a fixed point of some symmetry actions. Instead, as the points of $\cD_d$ are given by orbits of stable triples $(B,a,b)$ --- the framing runs over the entire base of Higgs branch --- the Handsaw quiver variety $\cD_d$ describes the Higgs branch in a manifestly gauge-invariant manner.  

\subsection{Vortex Partition Function of Four-Dimensional
\texorpdfstring{$\Uone$}{U(1)} Gauge Theory} \label{sec:VorParTheoryT}
The non-perturbative index~\eqref{eq:4d2dIndex} in the vortex sector of degree $d$ arises from the fluctuations along the space-time torus $T^2_\tau$ about the semi-classical field configurations of the degree $d$ $\mathcal{N}=(0,2)$ BPS vortex solutions. For the positively and negatively charged chiral multiplets these fluctuations come from two-dimensional $\mathcal{N}=(0,2)$ chiral multiplets and two-dimensional $\mathcal{N}=(0,2)$ fermi multiplets, respectively. 

The on-shell degrees of freedom of the two-dimensional $\mathcal{N}=(0,2)$ chiral multiplets are given by complex bosonic scalar fields. As the right-moving modes of these fields do not contribute to the supersymmetric $\mathcal{N}=(0,2)$ index, we only need to consider the left-moving fluctuations about the $\mathcal{N}=(0,2)$ BPS vortex solution, which take values in the holomorphic tangent bundle $T\cM_{\text{vor},d}$ and the anti-holomorphic tangent bundle $\overline{ T\cM}_{\text{vor},d}$. As the moduli space $\cM_{\text{vor},d}$ is K\"ahler, the anti-holomorphic tangent bundle $\overline{ T\cM}_{\text{vor},d}$ is identified with the holomorphic co-tangent bundle $T^*\cM_{\text{vor},d}$. Altogether the modes of the $\mathcal{N}=(0,2)$ chiral multiplets contributing to the index  are described by the bundle
\begin{equation} \label{eq:Vbd}
  V_{\text{bos},d} = \bigotimes_{k=0}^{+\infty} 
  \left( S_{\qt^k} T\cM_{\text{vor},d} \otimes S_{\qt^k} T^*\cM_{\text{vor},d} \right) \ .
\end{equation}  
Here the bundle $S_t \mathcal{V}$ comprises the $t$-graded infinite direct sum of symmetric tensor products~$S^k\mathcal{V}$ of the holomorphic vector bundle~$\mathcal{V}$, i.e., 
\begin{equation}
  S_{t} \mathcal{V} = \bigoplus_{k=0}^{+\infty} \left( t^k \, S^k \mathcal{V}  \right)
  = \mathcal{O} \oplus t\, \mathcal{V} \oplus t^2\, S^2\mathcal{V} \oplus \ldots \ .
\end{equation}
In our case the modes of the complex scalar fields are the symmetric products of the tangent and cotangent bundles of $\cM_{\text{vor},d}$, graded by powers of the fugacity $q_\tau$ associated to isometries of the torus $T_\tau^2$.

The relevant on-shell degrees of freedom of the two-dimensional $\mathcal{N}=(0,2)$ fermi multiplets are the left-moving modes of a complex fermion, which takes values in both the holomorphic and anti-holomorphic bundles $\cE_d$ and $\overline{\cE_d}$. The bundle $\cE_d$ over the moduli space $\cM_{\text{vor},d}$ is induced from the bundle $\cE$ in the total space of the Higgs branch $X_H$. Similarly as the anti-holomorphic tangent bundle $\overline{ T\cM}_{\text{vor},d}$, the anti-holomorphic bundle $\overline{\cE_d}$ gets identifed with $\cE_d^*$ as a consequence of the K\"ahler property of the total space~\eqref{Higgs branch XH} with the identification of $\overline{\cE}$ and $\cE^*$ on $X_H$. As a result the left-moving fermionic modes relevant for the index are captured by the bundle
\begin{equation}\label{eq:Vfd}
  V^{\varepsilon}_{\text{ferm},d} = \bigotimes_{n=0}^{+\infty}
  \left( \Lambda_{\qt^{n+1-\varepsilon}} \cE_d  \otimes \Lambda_{\qt^{n+\varepsilon}}\cE_d^* \right)  
  \quad \text{for} \quad
  \varepsilon \in \{ 0, \tfrac12 \} \ ,
\end{equation}
in terms of the infinite direct sum of the anti-symmetric tensor products~$\Lambda^k\cV$
\begin{equation}
  \Lambda_t \cV = \bigoplus_{k=0}^{+\infty} \left( t^k \, \Lambda^k\cV  \right)
  = \cO \oplus t \, \cV \oplus t^2\,  \Lambda^2\cV \oplus \ldots \ .
\end{equation}
The assignment of the fugacity~$\qt$ in eq.~\eqref{eq:Vfd} depends on the spin structure $\varepsilon \in \{0,\frac12\}$ of the left-movers, where these two choices relate to the boundary conditions of the left-moving fermionic modes as
\begin{equation}
  \text{Ramond sector:} \quad \varepsilon=0 \ , \qquad
  \text{Neveu--Schwarz sector:} \quad \varepsilon=\frac12 \ . 
\end{equation}
In the Ramond sector $\varepsilon=0$ the contribution~$\Lambda_{\qt^0}\cE_d^*$ describes the Ramond ground states that realize an irreducible representation of the Clifford algebra of the left-moving fermionic zero modes.

The non-perturbative contribution $Z_{\text{vor},d}$ of the degree $d$ vortices to the index~\eqref{eq:Znp} can be determined with the Hirzebruch--Riemann--Roch index theorem according to
\begin{equation}
  Z_{\text{vor},d}(\qt) =  \operatorname{Tr}_{\cH} (-1)^F \qt^{H_L} 
  = \int_{\cM_{\text{vor},d}} \operatorname{Td}\left(\cM_{\text{vor},d}\right) \, \operatorname{ch}\left( V_{\text{bos},d}\otimes V_{\text{ferm},d}^\varepsilon \right) \ .
\end{equation}
Here $\operatorname{Td}\left(\cM_{\text{vor},d}\right)$ is the Todd class of the tangent bundle~$T\cM_{\text{vor},d}$. Since the Chern character map is a ring homomorphism, the Chern character of the tensor product of the bundles~\eqref{eq:Vbd} and \eqref{eq:Vfd} reads 
\begin{equation}
\operatorname{ch}\left( V_{\text{bos},d} \otimes V_{\text{ferm},d}^\varepsilon \right)=\frac{\theta(\qt^{\varepsilon}\cE^*_d; \qt)}{\theta(T\cM_{\text{vor},d}^*; \qt)}  \ ,
\end{equation}
with the bundle-valued theta functions~$\theta(\cV;\qt)$ of a holomorphic vector bundle $\cV$ are defined in terms of the Chern character $\operatorname{ch}(\cV)$ as 
\begin{equation} \label{eq:deftheta2}
  \theta(\cV; \qt) = \prod_{n=0}^{+\infty}  \left(\strut 1  - \operatorname{ch}(\cV) \qt^n \right) 
   \left(\strut 1  - \operatorname{ch}(\cV^*) \qt^{n+1} \right)\ .
\end{equation}
Using the splitting principle these bundle-valued theta functions become products of ordinary theta functions~\eqref{eq:deftheta} in terms of the Chern roots of the holomorphic bundle~$\cV$. Finally, inserting for the Todd class the formula
\begin{equation}
  \operatorname{Td}\left(\cM_{\text{vor},d}\right) = \frac{e\left(\cM_{\text{vor},d}\right)}{1-\operatorname{ch}(T\cM_{\text{vor},d}^*) } \ ,
\end{equation}
where $e\left(\cM_{\text{vor},d}\right)$ is the Euler class of the tangent bundle $T\cM_{\text{vor},d}$, we arrive for $Z_{\text{vor},d}$ at
\begin{equation} \label{eq:ZnpIntegral}
   Z_{\text{vor},d}(\qt) =  \int_{\cM_{\text{vor},d}} \frac{\theta(\qt^\varepsilon\cE^*_d; \qt)}{\theta(T\cM_{\text{vor},d}^*; \qt)}  \, 
   e\left(\cM_{\text{vor},d}\right) \ .
\end{equation}

In order to evaluate the integral~\eqref{eq:ZnpIntegral} in the context of the four-dimensional $\cN=1$ gauge theory $\cT_\text{can}$, we apply equivariant localization techniques using the Handsaw quiver formulation $\cD_d$ for the degree $d$ vortex sector moduli space~$\cM_{\text{vor},d}$. The quiver variety possesses the abelian $T_q \times T_F$ symmetry with with $T_q = \mathbb{C}_q^*$ and $T_F = (\mathbb{C}^*)^{n+1}$.\footnote{We will not distinguish the real and complexified version of symmetry groups in this paper.} The generator of $T_q$ is the operator $J_3 + \frac{R}2$, which --- tracing the details of the identification~\eqref{eq:IsoD} --- acts on the triple $(B,a,b)$ as
\begin{equation} \label{eq:qAction}
   q \in T_q: \ (B,a,b) \mapsto (q B, a, q b) \ .
\end{equation} 
The symmetry $T_F$ realizes the abelian flavor symmetry transforming the chiral fields $\phi^I$ in Table~\ref{tab:TheoryT}. It acts via a representation $\rho_F: T_F \to \operatorname{GL}(W_0\oplus W_1)$ on the $(n+1)$-dimensional vector space $W_0\oplus W_1$ and induces an action on the triple $(B,a,b)$ according to the quiver~\eqref{eq:Quiver}. We denote the weights of the abelian symmetry $T_q \times T_F$ by $q$ and $y_I$, $I=1,\ldots, n+1$, which in the physical context of the gauge theory $\cT_\text{can}$ become the fugacities of the vortex partition functions $Z_{\text{vor},d}$. 

The fixed point loci of the quiver variety $\cD_d$ with respect to $T_q\times T_F \simeq ( \mathbb{C}^*)^{n+2}$ are the $\operatorname{GL}(V)$-orbits of the triple $(B,a,b)$ that are invariant with respect to the $(n+2)$ individual $\mathbb{C}^*$-symmetries. This implies that the vector spaces $W_0$, $W_1$, and $V$ decompose into direct sums of irreducible $\mathbb{C}^*$ representations such that these decompositions are invariant with respect to the action of the triple $(B,a,b)$ \cite{nakajima2011handsaw}. The vector space $W_0$ determines a Higgs vacuum in $X_{H,\text{red}}$ upon which the flavor symmetry $T_F$ acts according to Table~\ref{tab:TheoryT}. Hence, the $(n+1)$-fixed points of $\cD_d$ arise from a choice of a Higgs vacuum given by a one-dimensional equivariant vector space
\begin{equation} \label{eq:W0equiv}
  W_0 = \mathbb{C}_{y_I} \ , \quad I=1,\ldots,n+1 \ ,
\end{equation}
together with a decomposition of $W_1$ with respect to the remaining $\mathbb{C}^*$-symmetries
\begin{equation} \label{eq:W1equiv}
  W_1 = \bigoplus_{\substack{J=1 \\ J \ne I}}^{n+1} \mathbb{C}_{y_J} \ .
\end{equation}
Here the index of $\mathbb{C}_{w}$ determines the weight $w$ of the equivariant $\mathbb{C}^*$-action on the one-dimensional vector space $\mathbb{C}$ at the fixed point labeled by $I$. Due to the stability condition~\eqref{eq:BabStable} of a triple $(B,a,b)$, we decompose the vector space $V$ into the direct sum of one-dimensional vector spaces as
\begin{equation} 
   V = \bigoplus_{\alpha=0}^{d-1} B^\alpha \circ a(W_0) \ .
\end{equation}  
For a basis respecting this decomposition of $V$, a representative $(B,a,b)$ of the fixed point with respect to the symmetry $T_q \times T_F$ reads
\begin{equation} \label{eq:reppoint}
   B = \begin{pmatrix} 0 & \cdots & 0 & 0 \\ 
   \lambda_1 & \cdots & 0 & 0 \\
   \vdots & \ddots & \vdots & \vdots \\
   0 & \cdots & \lambda_{d-1} & 0 
    \end{pmatrix}  \ , \qquad
    a = \begin{pmatrix} \lambda_0 \\ 0 \\ \vdots \\ 0 \end{pmatrix} \ ,  \qquad
    b = 0 \ ,
\end{equation} 
with the non-vanishing entries $\lambda_0,\ldots,\lambda_{d-1} \in \mathbb{C}^*$. The $T_q \times T_F$-transformation maps the triple $(B,a,b)$ to $(q B, a\,y_I^{-1}, q \operatorname{Diag}(y_1,\ldots,\widehat{y_I},\ldots,y_{n+1}) b)$, which is compensated by a $\operatorname{GL}(V)$-transformation~\eqref{eq:GLaction} with the group element
\begin{equation} \label{eq:gTrans}
   g = \operatorname{Diag}\left(y_I^{-1}, y_I^{-1} q, \ldots, y_I^{-1} q^{d-1}\right)  \in \operatorname{GL}(V) \ .
\end{equation}
This demonstrates that the triple~\eqref{eq:reppoint} indeed represents a fixed $\operatorname{GL}(V)$-orbit. As a result, the deformations $(\delta B,\delta a,\delta b)$ modulo $\operatorname{GL}(V)$-transformations can be realized by the pair $(\delta\widetilde B,\delta b)$, where $\delta\widetilde{B}$ deforms the last column of $B$ in the triple~\eqref{eq:reppoint} while $\delta b$ deforms all entries of $b$.\footnote{Infinitesimal deformations to the entries of $a$ and to the first $(d-1)$ columns of $B$ realize infinitesimal transformations along the $\operatorname{GL}(V)$-orbits.}  Geometrically, these deformations realize equivariant sections of the conormal bundle  $\cN^*_{d,I}$ to the $I$-th fixed point, and their equivariant weights are determined from the combined $T_q\times T_F$- and $\operatorname{GL}(V)$-transformation
\begin{equation}
  (\delta B, \delta b) \mapsto (q \,g^{-1} \,\delta B \, g, 
  q  \operatorname{Diag}\left(y_1,\ldots,\widehat{y_I},\ldots,y_{n+1}\right) \delta b  \,g ) \ .
\end{equation}
Therefore, the conormal bundle enjoys the equivariant decomposition
\begin{equation} \label{eq:wNd}
    \cN^*_{d,I} = \bigoplus_{J=1}^{n+1} \bigoplus_{k=1}^d \mathbb{C}_{y_J y_I^{-1}q^{-k}} \ .
\end{equation}

Moreover, the bundle $\cE_d^*$ of the unreduced Higgs branch~$X_H$  is identified in the quiver variety $\cD_d$ with $\operatorname{Hom}(\bigoplus_{j=1}^{n+1} \mathbb{C}_{y_{j+n+1}},V)$, where  $y_{n+2}, \ldots, y_{2n+2}$ are the additional weights associated to the flavor symmetries of the chiral fields $\varphi_j$, $j=1,\ldots,n+1$, of the canonical gauge theory $\cT_\text{can}$. Taking into account the $\operatorname{GL}(V)$-transformation~\eqref{eq:gTrans}, the bundle $\cE_d^*$ at the $I$-th fixed point decomposes as
\begin{equation} \label{eq:wEd}
  \cE_{d,I} = \bigoplus_{j=1}^{n+1} \bigoplus_{k=1}^d \mathbb{C}_{y_{j+n+1} y_Iq^{k-1}} \ .
\end{equation}

Now, all the ingredients are assembled to calculate the vortex partition function~\eqref{eq:ZnpIntegral} using equivariant localization. Namely, with the Atiyah--Bott localization formula, upon dividing by the equivariant Euler class $e(\cD_d)$ the integral over the vortex moduli space $\cM_{\text{vor},d} \simeq \cD_d$ simplifies to a sum over the described equivariant fixed points, which applied to our situation results in
\begin{equation}
 Z_{\text{vor},d}(\qt,q,y_1,\ldots,y_{2n+2})
   = \sum_{I=1}^{n+1} \frac{\theta(\qt^\varepsilon\cE^*_{d,I})}{\theta(\cN^*_{d,I})} \ .
\end{equation}
Inserting the weights~\eqref{eq:wNd} and \eqref{eq:wEd}, we arrive at the expression
\begin{equation} \label{eq:Zdvor}
   Z_{\text{vor},d} = 
   \sum_{I=1}^{n+1} \left. 
    \frac{\prod_{j=1}^{n+1}
    \prod_{k=1}^d\theta(y_{j+n+1}s^{-1} q^{k}\qt^\varepsilon;\qt)}
    {\prod_{J=1}^{n+1} \prod_{k=1}^d\theta(y_J sq^{-k};\qt)} 
   \right|_{s = y_I^{-1}} \ .
\end{equation}
Here,  the $I$-th fixed point of $T_F$ action on $X_H$ si specified by relation the relation $s = y_i^{-1}$. Finally, adding the contributions of all degree $d$ vortices yields for the canonical gauge theory $\cT_\text{can}$ the non-perturbative vortex partition function
\begin{equation}
  Z_\text{vor} = \sum_{d=0}^{+\infty}  \sum_{I=1}^{n+1} \left. 
    \frac{\prod_{j=1}^{n+1}
    \prod_{k=1}^d\theta(y_{j+n+1}s^{-1} q^{k}\qt^\varepsilon;\qt)}
    {\prod_{J=1}^{n+1} \prod_{k=1}^d\theta(y_J sq^{-k};\qt)} 
   \right|_{s = y_I^{-1}} Q^d  \ .
\end{equation}
Using the more powerful machinery of quasimaps, we rederive this result in Section~\ref{subsect: vortex partition math}, which allows us to generalize our findings beyond the canonical gauge theory $\cT_\text{can}$. In particular, the generalization of the vortex partition function for the more general gauge theory $\cT$ reads\footnote{Note the peculiar R charge shift $q^{\mathfrak{q}_{j,\phi}/2}$ comes from additional $q^{R/2}$ twists in the index counting \eqref{eq:4d2dIndex}}
\begin{equation} \label{eq:ZnpT}
  Z_\text{vor} = \sum_{d=0}^{+\infty}  \sum_{I=1}^{n+1} \left. 
    \frac{\prod_{j=1}^{a}
\prod_{k=1}^{d\,l_j}\theta(y_{j+n+1}s^{-l_j} q^{k-1+\mathfrak{q}_j/2}\qt^\varepsilon;\qt)}
    {\prod_{J=1}^{n+1} \prod_{k=1}^d\theta(y_J sq^{-k+\mathfrak{q}_J/2};\qt)} 
   \right|_{s = y_I^{-1}q^{-\mathfrak{q}_I/2}} Q^d  \ .
\end{equation}
In the following, we focus on the Ramond sector $\varepsilon=0$. The following discussions work analogously for the Neveu--Schwarz sector $\varepsilon=\frac{1}{2}$ as well.

\subsection{Integral Representation of the Vortex Partition Function}\label{sec:4dlocformula}
In this section we interpret the non-perturbative partition function $Z_\text{4d}^\text{np}$ of eq.~\eqref{eq:Znp} as arising from four-dimensional supersymmetric $\cN=1$ localization on the space-time geometry $T_{\tau}^2\times_{\varepsilon_1,\varepsilon_2} D^2$. In this formulation the gauge theory path integral localizes on $\mathcal{N}=(0,2)$ vortex configurations and simplifies to a finite dimensional integral, which can be evaluated by standard residue calculus techniques. 

In order to arrive at the integral representation of the non-perturbative partition function $Z_\text{4d}^\text{np}$, we consider as a toy model the classical $\Uone$ gauge theory with a single $\cN=1$ chiral multiplet of gauge charge one and R-charge zero with Neumann boundary conditions at the space-time boundary~$\partial(T_{\tau}^2\times_{\varepsilon_1,\varepsilon_2} D^2) \simeq T^3$. This chiral theory is anomalous and hence inconsistent at the quantum level. Nevertheless, the integral representation of the non-perturbative partition~$Z_\text{4d}^\text{np}$ can be deduced from the classical field theory perspective. From eq.~\eqref{eq:Zdvor} we see that such a chiral multiplet yields the non-perturbative partition function
\begin{equation} \label{eq:ZnpChiral}
     Z_\text{4d,chiral}^\text{np}(Q,\qt,q,y) = \sum_{d=0}^{+\infty} \left. \frac{1}{\prod_{k=1}^{d}\theta(ysq^{-k};\qt)}\right|_{s=y^{-1}} Q^d  \ ,
 \end{equation}
where $y$ is the fugacity of the classical $\Uone$-flavor symmetry.\footnote{For a single chiral multiplet the $\Uone$-gauge symmetry coincides with the $\Uone$-flavor symmetry. However, in order to generalize to theories with multiple chiral multiplets, it is convenient here to introduce the flavor fugacity $y$ as well.} From the localization techniques of supersymmetric $\Uone$-gauge theories \cite{Pestun:2007rz}, we expect that this partition functions together with its perturbative decorations enjoys an integral representation of the form
\begin{equation} \label{eq:ZnpChiralInt}
   Z_\text{4d,chiral} (Q,\qt,q,y) = \frac1{2\pi\ii} \int_\cC \frac{\dd s}{s} F(Q,\qt,q,sy) \ ,
 \end{equation}   
in terms of a holomorphic integrand $F(Q,\qt,q,sy)$ and along a suitable contour $\cC$. From the expression~\eqref{eq:ZnpChiral} we deduce that the non-perturbative vortex contributions up to degree~$d$ from the integrand $F(Q,\qt,q,sy)$ arise from a tower of poles in the $s$-plane --- encircled by the contour $\cC$ --- at $s=y^{-1} q^k$, $k=1,\ldots,d$. The form of the partition function~\eqref{eq:ZnpChiral} implies that the integrand satisfies the difference equation with respect to the variable $q$
 \begin{equation} \label{eq:DiffEqChiral}
     \theta(z;\qt) \, F(Q,\qt,q,z) = Q\, F(Q,\qt,q,z q) \ .
  \end{equation}
It has the solution
\begin{equation} \label{eq:Fsol}
    F(Q,\qt,q,z) = \ee^{-\frac1{\log q} \log Q \,\cdot \, \log z} \, \Gamma(z;q,\qt) \ ,
\end{equation}
where $\Gamma(z;q,\qt)$ is the double elliptic Gamma function \cite{Narukawa:2004} 
\begin{equation} \label{doubel elliptic gamma function 1st}
      \Gamma(z;q,\qt) = \prod_{k,l\geq 0}\frac{1-q^{k+1}\qt^{l+1}z^{-1}}{1-q^k\qt^l z} \ .
\end{equation}      
Due to the functional equations 
\begin{equation} \label{double elliptic function 2nd}
   \theta(z;\qt) \, \Gamma(z;q,\qt)  = \Gamma(zq;q,\qt) \ , \qquad
   \theta(z;q) \, \Gamma(z;q,\qt)  = \Gamma(z\qt;q,\qt) \ .
\end{equation}
it is straightforward to verify this solution to the difference equation~\eqref{eq:DiffEqChiral}. As the double elliptic Gamma function has poles at $z = q^{-k} \qt^{-l}$, the contour $\cC$ of the integral~\eqref{eq:ZnpChiralInt} encircles only the poles at $z=q^{-k}$ for non-negative integers $k$. 

Let us now consider a few generalizations of this result. First of all, changing Neumann to Dirichlet boundary condition of the four-dimensional $\cN=1$ chiral multiplet along the space time boundary $T^3$ multiplies the integrand with the theta function $\theta(sy;q)$. Therefore, the integrand~$F(Q,\qt,q,y)$ of a $\Uone$-gauge theory of a chiral multiplet with Dirichlet boundary conditions reads
\begin{equation}
   F(Q,\qt,q,z) = \ee^{-\frac1{\log q} \log Q \,\cdot \, \log s} \, \Gamma(z;q,\qt) \theta(z;q)
   = \ee^{-\frac1{\log q} \log Q \,\cdot \, \log z} \, \frac{1}{\Gamma(q z^{-1};q,\qt)} \ , 
\end{equation}   
where we use the relation~\eqref{double elliptic function 2nd} together with the inversion formula for double elliptic Gamma functions
\begin{equation}
  \Gamma(z; q, \qt)\, \Gamma(\tfrac{q \qt}{z};q,\qt) = 1 \ .
\end{equation}
Second of all, considering a chiral multiplet with $\Uone$-gauge charge $\beta$ and non-zero R-charge $\mathfrak{q}$ replaces the argument $z=sy$ of the double elliptic Gamma function by $z=s^\beta yq^{\mathfrak{q}/2}$ because the $\Uone$ gauge charge scales and the R-charge shifts the location of the poles of the integrand accordingly. 

In summary, for a $\Uone$-gauge theory with several four-dimensional $\cN=1$ chiral multiplets with gauge charges $\beta_I$ and R-charges $\mathfrak{q}_I$, we arrive at the  partition function in the integral form
\begin{equation} \label{eq:ZnpIntegrand}
     Z_\text{4d} (Q,\qt,q,y_I)
     = \frac1{2\pi\ii} \int_\cC \frac{\dd s}{s} \ee^{-\frac1{\log q} \log Q \,\cdot \, \log s} 
        \frac{\prod_{\{I | \text{N bdry}\}} \Gamma(s^{\beta_I}y_Iq^{\mathfrak{q}_I/2};q,\qt)}
        {\prod_{\{J | \text{D bdry}\}} \Gamma(\frac{q^{1-\mathfrak{q}_J/2}}{s^{\beta_J}y_J};q,\qt)} \ .
\end{equation}
Here, $y_I$ are the fugacities of the classical $\Uone$ flavor symmetries of the individual chiral multiplets. The product over the contributions of the chiral multiplets is split among multiplets with Neumann and Dirichlet boundary conditions in the numerator and denominator, respectively. The contour $\cC$ is again chosen in such a way that it only picks up poles of the integrand that do not have a dependence on $\qt$ and satisfies the Jeffrey--Kirwan residue prescription \cite{Jeffrey:1995}.\footnote{The selection of Jeffrey--Kirwan residue depends on the  Fayet--Illiopolus parameter, which appears in our integral as $\frac{\text{log}Q}{\text{log}q}$.} 

From the perspective of the supersymmetric localization, the integrand~$F$ of the partition function~\eqref{eq:ZnpIntegrand} factors into the two contributions
\begin{equation}
  F_{\Uone} = \ee^{-\frac1{\log q} \log Q \,\cdot \, \log s} \ , \qquad
  F_\text{matter} = \frac{\prod_{\{I | \text{N bdry}\}} \Gamma(s^{\beta_I}y_Iq^{\mathfrak{q}_I/2};q,\qt)}
        {\prod_{\{J | \text{D bdry}\}} \Gamma(\frac{q^{1-\mathfrak{q}_J/2}}{s^{\beta_J}y_J};q,\qt)} \ ,
\end{equation}
where these respective factors arise from the four-dimensional $\Uone$ vector multiplet and the four-dimensional chiral matter multiplets. Applying the abelian/non-abelian correspondence, we deduce that the vector multiplet of a semi-simple gauge group~$G$ contributes the factor
\begin{equation}
  F_G(\qt,q,s) = \prod_{\alpha \in \Delta} \frac{1}{\Gamma(s^\alpha;q,\qt)} \ ,
\end{equation}
where the product is taken over the roots $\Delta$ of the Lie algebra $\operatorname{Lie}(G)$ and where $s^\alpha$ is the short-hand multi-index notation for the fugacities $s_1,\ldots,s_{\operatorname{rank} G}$ of the maximal torus of the Lie group~$G$ exponentiated with the root $\alpha$. Furthermore, chiral matter multiplets in the irreducipble representations $R_I$ of the gauge group $G$ with $\Uone_R$ charge $\mathfrak{q}_I$ yield the matter factor
\begin{equation}
  F_\text{matter}(\qt,q,s) = \frac{\prod_{\{R_I | \text{N bdry}\}} \prod_{\beta_I \in w(R_I)} \Gamma(s^{\beta_I}q^{\mathfrak{q}_I/2};q,\qt)}
        {\prod_{\{R_J | \text{D bdry}\}}
        \prod_{\beta_J \in w(R_J)}\Gamma(\frac{q^{1-\mathfrak{q}_J/2}}{s^{\beta_J}};q,\qt)} \ ,
\end{equation}
Here the products run over the irreducible representations $R_I$ of the chiral matter fields and their set of weights~$w(R_I)$. As for the $\Uone$ gauge group, the contributions of the matter fields in the numerator and the denominator come from matter multiplets with Neumann and Dirichlet boundary conditions, respectively. Notice that these building blocks are the same as the holomorphic blocks on $T_{\tau}^2\times_{\varepsilon_1,\varepsilon_2} D^2$ proposed in refs.~\cite{Nieri_2015, Longhi:2019hdh} up to a phase factor that become constant for anomaly free theories. Hence, the building blocks obtained from one-loop calculation in supersymmetric localization \cite{Longhi:2019hdh} are naturally related to non-perturbative vortex partition function that we calculated in Section~\ref{sec:VorParTheoryT}. This reflects a standard lore in supersymmetric localization, particularly in lower dimensions with $4$ or more supercharges, that the vortex partition function usually admits an integral expression that is related to Coulomb branch one loop calculation. 

\subsection{Anomalies and the Vortex Partition Function}
\label{subsect: anomaly and vortex string}
In comparison to two-dimensional $\cN=(2,2)$ gauge theories and three-dimensional $\cN=2$ supersymmetric gauge theories with four supercharges, four-dimensional $\cN=1$ gauge theories are subject to anomalies.\footnote{Three-dimensional $\cN=2$ supersymmetric gauge theories are subject to parity anomalies \cite{Redlich:1983kn,Redlich:1983dv}, which can be compensated by a choice of suitable Chern--Simons terms \cite{Aharony:1997bx}.} For such four-dimensional gauge theory to be consistent at the quantum level, the gauge symmetry must be non-anomalous. For the four-dimensional gauge theory $\cT$ with the spectrum in Table~\ref{tab:TheoryT} the condition for the gauge anomaly to vanish is spelt out explicitly by the constraint~$\kG=0$ in eq.~\eqref{eq:Anom}. In addition, these field theories may possess additional classical global symmetries that are anomalous at the quantum level. For instance in the gauge theory $\cT$, such an Adler--Bell--Jackiw anomaly may occur for the $\Uone_R$-symmetry, which vanishes for $\kR=0$ according to eq.~\eqref{eq:AnomMixed}. While such an anomaly does not render the quantum field theory inconsistent, the index~\eqref{eq:4d2dIndex} and the related supersymmetric localization on the space-time geometry $T_{\tau}^2\times_{\varepsilon_1,\varepsilon_2} D^2$ is not well-defined if the $\Uone_R$~symmetry is anomalous. Furthermore, as discussed in Section~\ref{sec:BPSVorMSpace} for the gauge theory $\cT$ an Adler--Bell--Jackiw anomaly breaks the classical flavor symmetry group $\Uone^{n+1+a}$ to a quantum flavor subgroup $\Uone^{n+a}$. In addition to these anomalies, the four-dimensional quantum field theory may possess global quantum symmetries that cannot be gauged consistently due to the presence of 't~Hooft anomalies \cite{tHooft:1979rat}. 

To investigate the incarnation of anomalies on the level of the non-perturbative vortex partition function, we examine gauge transformations on the space-time background $T_{\tau}^2\times_{\varepsilon_1,\varepsilon_2} D^2$ in some detail.  In particular, we study large $\Uone$ gauge transformations $\mathcal{G}_{n,m}$ on the topological retraction of the space-time $T_{\tau}^2\times_{\varepsilon_1,\varepsilon_2} D^2$ to $T_\tau^2$. Such large gauge transformations are topologically characterized by the winding numbers $n,m\in\mathbb{Z}$ and induce a shift of the holonomy~$\mu$ of the $\Uone$ connection as $\mu \mapsto \mu + n + \tau m$. Thus the fugacity $s = \ee^{2\pi\ii \mu}$ of the $\Uone$ gauge theory transforms as (c.f., eqs.~\eqref{eq:Defmunu} and \eqref{eq:Defsf})
\begin{equation} \label{eq:LargeGauge}
   \mathcal{G}_{n,m}:\ s \mapsto s \,\qt^m \ .
\end{equation}
For explicitness, let us now consider for the gauge theory $\cT$ the degree $d$ vortex contributions 
\begin{equation} \label{eq:ZDefF}
   F_{\text{vor},d}(s,\qt,q,y) =  \frac{\prod_{j=1}^{a}
\prod_{k=1}^{d\,l_j}\theta(y_{j+n+1}s^{-l_j} q^{k-1+\mathfrak{q}_j/2} ;\qt)}
    {\prod_{J=1}^{n+1} \prod_{k=1}^d\theta(y_J sq^{-k+\mathfrak{q}_\phi/2};\qt)} \ ,
\end{equation}
which according to eq.~\eqref{eq:ZnpT} determine the vortex partition functions as
\begin{equation} \label{eq:ZvorDefF}
    Z_\text{vor} = \sum_{d=0}^{+\infty} \sum_{I=1}^{n+1} \left. F(s,\qt,q,y) \right|_{s=y_I^{-1}q^{-\mathfrak{q}_\phi/2}} Q^d\ .
\end{equation}
Due to the transformation behavior~\eqref{eq:ThetaTrans} of the theta function~\eqref{eq:deftheta}, we readily compute that the degree $d$ contribution $F_{\text{vor},d}$ transforms under the transformations $\mathcal{G}_{n,m}$ as
\begin{equation}
    F_{\text{vor},d}(s\,\qt^m,\qt,q,y)= f_m(s,\qt,q,y)\, F_{\text{vor},d}(s,\qt,q,y) \ ,
\end{equation}
where 
\begin{multline}
  f_m(s,\qt,q,y) = 
  \left( \prod_{j=1}^a y_{j+n+1}^{l_j^2} \prod_{J=1}^{n+1} y_J \right)^{m d}
  (-s)^{m d\, \kG} q^{\frac12 m d \left(d\,\kG+\kR\right)} \qt^{\frac12 m(m-1) d \, \kG} \\
  \cdot \qt^{\frac12 m d  \sum_{j=1}^a l_j^2(l_j-1)} \ .
\end{multline}
Here $\kG$ and $\kR$ encode the anomaly~\eqref{eq:Anom} of the gauge symmetry and the anomaly~\eqref{eq:AnomMixed} of the R-symmetry.

The transition functions $f_m$ under large gauge transformations $\mathcal{G}_{n,m}$ determine that the degree~$d$ vortex contribution~$F_{\text{vor},d}$ is a meromorphic section of a degree $d\,\kG$ line bundle~$\cL_{\text{G},d}$ over the Jacobian $\mathcal{J}(T_\tau^2) \simeq E_{\qt,s}$, which is associated to the gauge group $\Uone$ as introduced in Section~\ref{sec:N02BPSDefects}. In particular, for a consistent anomaly-free quantum gauge theory $\cT$ with $\kG=0$, the transition function $f_m$ simplifies to
\begin{equation}
  \left. f_m(s,\qt,q,y) \right|_{\kG = 0} =
  \left( \prod_{j=1}^a y_{j+n+1}^{l_j^2} \prod_{J=1}^{n+1} y_J \right)^{m d}  
  q^{\frac12 m d \, \kR} \,
  \qt^{\frac12 m d  \sum_{j=1}^a l_j^2(l_j-1)} \ .
\end{equation}  
Recall further that the index~\eqref{eq:4d2dIndex} is only consistent for an anomaly-free $\Uone_R$ symmetry, and that it is only a well-defined as function of the flavor fugacities of the quantum flavor symmetries. The former requirement imposes $\kR=0$, whereas the latter condition yields
\begin{equation} \label{eq:FlavorAnom}
  \mathcal{Y}_\text{F}(y) =1 \quad \text{with} \quad  \mathcal{Y}_F(y) = \prod_{j=1}^a y_{j+n+1}^{l_j^2} \prod_{J=1}^{n+1} y_J \ ,
\end{equation}  
because this is the condition for the flavor fugacities to parametrize the anomaly-free flavor subgroup $\Uone^{n+a}$ of the classical flavor group $\Uone^{n+1+a}$, c.f., Section~\ref{sec:BPSVorMSpace}. Imposing these further consistency requirements yields the transition functions
\begin{equation} \label{eq:fNoAnom}
  \left. f_m(s,\qt,q,y) \right|_{\kG = 0,\kR=0,\mathcal{Y}_\text{F}(y)=1}
   = \qt^{\frac12 m d  \sum_{j=1}^a l_j^2(l_j-1)} \ .
\end{equation}
Since $\sum_{j=1}^a l_j^2(l_j-1)$ is always even, the right hand side of this transition function is a monomial $\qt^\alpha$ with an integral exponent $\alpha \in\mathbb{Z}$. As the factor $\qt^\alpha$, $\alpha\in\mathbb{Z}$ in the transition function is a global holomorphic function on the covering space $\mathbb{C}^*$ of the elliptic curve $\Ecurve \simeq \mathbb{C}^*/{\qt^\mathbb{Z}}$, this factor simply changes the local trivialization of the line bundle $\cL_{\text{G},d}$, c.f., for instance ref.~\cite{beauville2013theta}. Hence, the transition function~\eqref{eq:fNoAnom} implies that the line bundle  $\cL_{\text{G},d}$ is isomorphic to the trivial line bundle $\cO_{\Ecurve}$, and $F_{\text{vor},d}$ is a meromorphic section thereof. The trivialization of the line bundle $\cL_{G,d}$ depends on the chosen regularization scheme and can be modified by adding counter terms to the gauge theory action. Thus the factors~$q^\alpha$, $\alpha\in \bZ$, on the right-hand side of eq.~\eqref{eq:fNoAnom} do not have a scheme-independent physical meaning. Therefore, we often do not keep track of such factors in the following.

Putting our findings in the geometric context of Section~\ref{sec:N02BPSDefects}, we find that the vortex contributions~$F_{\text{vor},d}$ at degree $d$ are meromorphic sections of the line bundles~$\cL_{\text{vor},d}$ over the abelian variety
\begin{equation}
  \cM_{\text{flat}, T_{\tau}^2} = \cM_{\text{flat,G}} \times \cM_{\text{flat,R}} \times \cM_{\text{flat,F}} \ ,
\end{equation}
where $\cM_{\text{flat,G}} = \bigtimes_{k=1}^{r} E_{\qt,s_i}$ is the $r$-fold Cartesian product associated to the rank $r$ maximal torus $T$ of the gauge group $G$ \footnote{When $G$ is non-abelian, $\cM_{\text{flat,G}}$ needs to be further quotient by Weyl group action $W(G)$.}, $\cM_{\text{flat,R}} = E_{\qt,q}$ is the elliptic curve for the $\Uone_R$-symmetry, and $\cM_{\text{flat,F}}=\bigtimes_{k=1}^{r_F} \Ecurve$ is the $r_F$-fold Cartesian product for the rank $r_F$ abelian flavor symmetries. The vortex partition function $Z_\text{vor}(\qt,q,y)$ is now a consistent observable as function of the fugacity $q$ and $y$, if for any embedding
\begin{equation}
   \iota: \cM_{\text{flat,G}} \hookrightarrow  \cM_{\text{flat}, T_{\tau}^2} \ ,
\end{equation}   
the line bundles $\iota^* \cL_{\text{vor},d}$ is isomorphic to the trivial line bundle $\cO_{\cM_{\text{flat,G}}}$. This condition ensures that the gauge symmetry $G$ and the $\Uone_R$-symmetry are anomaly free and that the abelian flavor symmetries corresponding to the fugacities $y$ prevail at the quantum level. Finally, let us remark that studying the line bundles $j^*\cL_{\text{vor},d}$ obtained from embeddings $j: \cM_{\text{flat,F}} \hookrightarrow  \cM_{\text{flat}, T_{\tau}^2}$ allows us to analyze 't~Hooft anomalies of the flavor symmetries by the same technique. 

\subsection{The Twisted Chiral Ring} \label{sec:twring}
The operator product expansion of chiral quantum fields of supersymmetric gauge theories with four supercharges form a ring, which is called the chiral ring of such supersymmetric gauge theories \cite{Lerche:1989uy}. In addition to the chiral ring, two-dimensional $\cN=(2,2)$ supersymmetric gauge theories possess a twisted chiral ring that arises from the operator product expansion of twisted chiral fields \cite{Witten:1993yc,Hori:2000kt}, which arise from the twisted chiral field strength of the vector multiplets \cite{Gates:1984nk}. Upon compactifying three-dimensional $\cN=2$ supersymmetric gauge theories on a circle $S^1$ and four-dimensional $\cN=1$ gauge theories on two-dimensional a torus $T^2$, we obtain an effective two-dimensional $\cN=(2,2)$ supersymmetric field theory with an infinity number of two-dimensional fields arising from the infinite tower of Kaluza--Klein modes. From this perspective, the resulting effective two-dimensional $\cN=(2,2)$ field theories from such compactifications define again a twisted chiral ring \cite{Nekrasov2:2009uh}. 

The twisted chiral ring of two-dimensional $\cN=(2,2)$ supersymmetric gauge theories plays a very important role in many different contexts such as the Bethe/Gauge correspondence \cite{Nekrasov1:2009ui,Nekrasov2:2009uh,Nekrasov3:2009zz,Nekrasov4:2009rc}. Of particular interest to us is that the twisted chiral ring determines the quantum cohomology of the gauge theory Higgs branch. Specifically, for massive and isolated vacua in a generic flavor symmetry background, the twisted chiral ring is one-loop exact and gives the equivariant quantum cohomology ring of the Higgs branch \cite{Morrison:1994fr}.

To illustrate the concepts, we focus on a two-dimensional $\cN=(2,2)$ supersymmetric gauge theory with gauge group $\Uone$. The spectrum consists of the $\cN=(2,2)$ chiral multiplets~$\phi^I$, $I=1,\ldots,n+1$, with gauge charges one and with the canonical $\Uone^{n+1}$-flavor symmetry.\footnote{The diagonal $\Uone$-flavor subgroup of the flavor symmetry $\Uone^{n+1}$ coincides with the $\Uone$-gauge symmetry and hence is redundant. Nevertheless, it is convenient to include this symmetry in the flavor group $\Uone^{n+1}$ as well.} For this basic example of the two-dimensional $\cN=(2,2)$ supersymmetric $\Uone$-gauge theory the Higgs branch is the projective space $\mathbb{P}^n$. For a generic flavor background the one-loop exact twisted superpotential $\cW$ reads \cite{Witten:1993yc,Morrison:1994fr,Nekrasov:2009uh,Nekrasov1:2009ui}\footnote{The one-loop correction~$\cW_\text{2d,1-loop}$ to the effective twisted superpotential $\cW_\text{2d}$ depends on the energy scale $\Lambda$, i.e., $\cW_\text{2d,1-loop} =\sum_I (\sigma+\nu_I)\left(\log\frac{\sigma+\nu_I}\Lambda-1\right)$. Here we set the energy scale $\Lambda$ to one.}
\begin{equation} \label{(2,2) twisted superpotential}
 \cW_\text{2d}(\sigma) =  \sum_{I=1}^{n+1} \left(\sigma +\nu_I \right)\left(\log(\sigma +\nu_I) -1\right) + 2\pi \ii \, t \, \sigma \ ,
 \quad
 t = \frac{1}{2\pi \ii} \left( -\xi + 2 \pi \ii \vartheta \right) \ .
\end{equation}  
Here $\sigma$ is the complex scalar field of the two-dimensional twisted chiral super-field strength of the $\Uone$-vector multiplet. The field $\sigma$ and the constants $\nu_I$, $I=1,\ldots,n+1$, parametrize the Coulomb branch and the flavor background of the two-dimensional gauge theory, respectively, and $t$ is the complexified Fayet--Iliopoulos parameter. In this effective twisted superpotential the logarithmic terms arise at one loop, whereas the last term corresponds to the tree-level twisted superpotential. 

The twisted chiral ring of the described two-dimensional $\cN=(2,2)$ supersymmetric gauge theory becomes the polynomial ring of the complex scalar field $\sigma$ modulo the relation
\begin{equation} \label{eq:QCoh1}
  \partial_\sigma \cW_\text{2d} = 0 \quad \Rightarrow \quad
  \prod_{I=1}^{n+1} \left( \sigma +\nu_I \right) = Q \quad \text{with} \quad
  Q = \ee^{-2\pi \ii t} \ ,
\end{equation}
such that we arrive at the well-known equivariant quantum cohomology ring \cite{MR1492534}
\begin{equation}
  QH^\bullet_T(\mathbb{P}^n) \simeq \left.\mathbb{C}[\sigma,\nu_I][[Q]] \middle/ \left( \prod_{I=1}^{n+1}  {\left(\sigma +\nu_I \right)} - Q\right)\right. \ ,
\end{equation}
where $\sigma$ becomes the hyperplane class $c_1(\cO(1)_{\mathbb{P}^n})$ and $\nu_I$ are the characters of the canonical $(\mathbb{C}^*)^{n+1}$-action on the homogeneous coordinates of the projective space $\mathbb{P}^n$. 

This simple example already illustrates the correspondence between supersymmetric gauge theories and geometry of its Higgs branch. Namely, the flavor symmetry of the gauge theory realizes a group action on the Higgs branch target space, such that the flavor background parameters become the equivariant parameters of the quantum cohomology of the Higgs branch target space. 

Let us now consider the three-dimensional $\cN=2$ supersymmetric $\Uone$--gauge theory with $(n+1)$ chiral multiplets of charge one and flavor symmetry $\Uone^{n+1}$. For a vanishing flavor background the semi-classical Higgs branch target space is again the projective space $\mathbb{P}^n$. Compactified on a circle $S^1$ of radius $R$ the resulting effective $\cN=(2,2)$ twisted superpotential reads
\begin{equation} \label{eq:W3d}
   \cW_\text{3d}(\sigma) =  \sum_{I=1}^{n+1} \sum_{k \in\mathbb{Z}} \left(\sigma +\nu_I 
   + \frac{ \ii k}R \right)\left(\log\left(\sigma +\nu_I +\frac{\ii k}R\right) -1\right) + 2\pi \ii \, t \, \sigma \ .
\end{equation}  
Here the effective two-dimensional complex scalar field $\sigma = \sigma_\mathbb{R} + \frac{\ii}{R} \int_{S^1} A$ is the real scalar field $\sigma_\mathbb{R}$ of the three-dimensional $\cN=2$ vector multiplet complexified with the $\Uone$ Wilson line modulus along the space-time circle $S^1$. In a similar fashion the three-dimensional real Fayet--Iliopoulos~$\xi_\mathbb{R}$ and the three-dimensional real flavor backgrounds~$\nu_{\mathbb{R},I}$ get complexified to the effective two-dimensional complex Fayet--Iliopoulos parameter $t$ and the complexified flavor background parameters~$\nu_I$. The infinite sum over the integers $k$ in the twisted superpotential $\cW_\text{3d}$ occurs due to the Kaluza--Klein modes of the three-dimensional $\cN=2$ chiral multiplets along the space-time circle $S^1$. The resulting effective two-dimensional $\cN=(2,2)$ chiral Kaluza--Klein multiplets carry a twisted mass because of their discrete momenta $\frac{k}{R}$ along the compact space-time circle~$S^1$ of radius $R$, which enter in at the one-loop level of the quantum-corrected twisted superpotential~$\cW_\text{3d}$. 

The twisted superpotential~\eqref{eq:W3d} determines the twisted chiral ring relations\footnote{The three-dimensional $\cN=2$ gauge theory depends on a choice of Chern--Simons terms \cite{Aharony:1997bx,Nekrasov3:2009zz,Jockers:2019lwe,Jockers:2021omw}. The stated twisted chiral ring relations arise from a standard choice of such terms.}
\begin{equation} 
   \partial_\sigma \cW_\text{3d} = 0 \quad \Rightarrow \quad
   \sum_{I=1}^{n+1} \sum_{k \in\mathbb{Z}} \log\left( \sigma + \nu_I + \frac{\ii k}R \right) = - 2\pi \ii t \ .
\end{equation}
The infinite sum over the Kaluza--Klein modes can be regularized with the Hurwitz $\zeta$-function as\footnote{Using
$$
  \sum_{k \in \mathbb{Z}}  \log\left( a (k+b) \right) \sim_\text{reg} 
  -\frac{\dd}{\dd z} \left. \left(\sum_{k\in\mathbb{Z}} \tfrac{1}{\left(a(k+b) \right)^z}\right)\right|_{z=0}
  = -\left. \frac{\dd}{\dd z}\big(\zeta(z,b) + \zeta(z,1-b) \big)\right|_{z=0} + \ii \pi \, \zeta(0,1-b) \ ,
$$
in terms of the Hurwitz $\zeta$-functions~$\zeta(z,u) = \sum_{k=0}^{+\infty} \tfrac1{(z+u)^z}$ with $\zeta(0,u) = \tfrac12 - u$ and the Lech formula $\tfrac{\dd}{\dd z} \zeta(z,u)|_{z=0} = \log\left( \tfrac{\Gamma(u)}{\sqrt{2\pi}}\right)$ one obtains
$$
  \sum_{k \in \mathbb{Z}}  \log\left( a (k+b) \right) \sim_\text{reg}  \log\left( 1- \ee^{2 \ii \pi b} \right) \ .
$$
}
\begin{equation} \label{eq:ZetaReg}
  \sum_{k\in\mathbb{Z}} \log\left( \sigma + \nu_I + \frac{\ii k}R \right)
  \sim_\text{reg}
  \log\left( 1 - \ee^{2\pi R(\sigma + \nu_I)} \right) \ ,
\end{equation}
such that we arrive at the twisted chiral ring relation
\begin{equation} \label{eq:QK1}
   \prod_{I=1}^{n+1} \left( 1 - \ee^{2\pi R(\sigma + \nu_I)} \right) = Q \quad \text{with}
   \quad Q = \ee^{-2\pi \ii t} \ .
\end{equation}
Upon identifying the factor $\ee^{2\pi R(\sigma + \nu_I)}$ with $P\Lambda_I$, where $P \simeq \mathcal{O}(-1)_{\mathbb{P}^n}$ is the tautological line bundle over the projective space $\mathbb{P}^{n}$ and $\Lambda_I$ the flavor character, we arrive at the equivariant quantum K~theory ring \cite{Givental:2001clq,Nekrasov2:2009uh,Iritani:2013qka,Jockers:2018sfl}
\begin{equation}
  QK_T(\mathbb{P}^n) \simeq \left.\mathbb{C}[P^{\pm 1},\Lambda_I^{\pm 1}][[Q]] \middle/ \left( \prod_{I=1}^{n+1}  {\left(1 - P\Lambda_I\right)}- Q\right)\right. \ ,
\end{equation}
 
Let us now consider the generalization to a four-dimensional $\cN=1$ supersymmetric gauge theory compactified on the torus $T^2_\tau$, such that each four-dimensional chiral field yields a double Kaluza--Klein tower of effective two-dimensional $\cN=(2,2)$ chiral fields that enter in the two-dimensional effective superpotential $\cW_{\text{4d}}$. For the four-dimensional $\cN=1$ supersymmetric $\Uone$-gauge theory with the same chiral field contents as in the previous three-dimensional example, we arrive at the effective twisted chiral superpotential \cite{Nekrasov4:2009rc,Nekrasov:2009uh,Closset:2017bse}
\begin{equation} \label{eq:W4d}
  \cW_\text{4d}(\sigma) = \sum_{I=1}^{n+1} \sum_{k,l \in\mathbb{Z}} \big(\sigma +\nu_I 
   + \ii (k +\tau l) \big)\big(\log\left(\sigma +\nu_I + \ii (k +\tau l) \big) -1\right) + 2\pi \ii \, t \, \sigma \ ,
\end{equation}
where the effective two-dimensional complex scalar field $\sigma$ arises now from the Wilson line moduli of the two holonomies along the space-time torus $T_\tau^2$ with complex structure modulus $\tau$, and where the parameter $t$ is defined as $t = - \int_{T_\tau^2} \widetilde{F} + \ii \xi \operatorname{vol}(T_\tau^2)$ with the $\Uone$ dual field strength $\widetilde{F}$. Then the resulting twisted chiral superpotential relation reads
\begin{equation}
  \partial_\sigma \cW_\text{4d} = 0 \quad \Rightarrow \quad
  \sum_{I=1}^{n+1} \sum_{k,l \in\mathbb{Z}} \log\big( \sigma + \nu_I + \ii (k +\tau l) \big) = - 2\pi \ii t \ .
\end{equation}
We regularize the non-negative Kaluza--Klein modes $l \ge 0$ and the negative Kaluza--Klein modes $-m= l<0$ analogously as in the three-dimensional example with the Hurwitz $\zeta$-functions regularization prescription~\eqref{eq:ZetaReg}, i.e., 
\begin{equation}
\begin{aligned}
  \sum_{k\in\mathbb{Z}} \sum_{l = 0}^{+\infty} \log\big( \sigma + \nu_I + \ii ( k +\tau l) \big) 
  &\sim_\text{reg} \log\prod_{l = 0}^{+\infty} \left( 1 - \ee^{2\pi (\sigma + \nu_I + \ii \tau l)}  \right) \ , \\
   \sum_{k\in\mathbb{Z}} \sum_{m = 1}^{\infty} \log\big((-1) (-\sigma - \nu_I + \ii (k + \tau m) \big) 
  &\sim_\text{reg} \log\prod_{m = 1}^{+\infty} \left( 1 - \ee^{2\pi (-\sigma - \nu_I + \ii \tau m)}  \right)  \ ,
\end{aligned}  
\end{equation}
such that we get altogether with eq.~\eqref{eq:deftheta} the twisted chiral superpotential relation
\begin{equation} \label{eq:4dRelW}
   \prod_{I=1}^{n+1} \theta(s y_I;\qt) = Q \quad \text{with}
   \quad s = \ee^{2\pi \sigma}\,,\ y_I = \ee^{2\pi \nu_I}\,,\ \qt = \ee^{2\pi \\i \tau}\,,\ Q = \ee^{-2\pi \ii t} \ .
\end{equation}
where $s$ and $y_I$ are identified with the gauge and flavor symmetry fugacities, respectively.

We observe that the four-dimensional twisted chiral superpotential relation~\eqref{eq:4dRelW} is not consistent as it stands, because it is not invariant with respect to large gauge transformations~\eqref{eq:LargeGauge} along the space-time torus $T_\tau^2$. Namely, the theta functions $\theta(s y_I;\qt)$ on the left-hand side transform non-trivially, whereas the background parameter $Q$ on the right-hand side is a constant.\footnote{In principle, the relation can be made covariant upon assigning a transformation behavior to $Q$. Such a modification suggests that the background parameter $t$ becomes part of the dynamics of the gauge theory. As the physics interpretation of such a phenomenon remains subtle, we do not consider this possibility further in this work.} This inconsistency in the description does not come as a surprise, as the four-dimensional gauge theory is ill-defined due to a $\Uone$ gauge anomaly, c.f., with eq.~\eqref{eq:Anom}. By introducing additional four-dimensional chiral multiplets with negative gauge charges $-l_j$, $j=1,\ldots,a$, we arrive at the generalized effective twisted chiral superpotential relation
\begin{equation} \label{eq:4dRelWMod}
      \frac{\prod_{I=1}^{n+1} \theta(sy_I;\qt) }{\prod_{j=1}^{a} \theta(s^{-l_j}y_{n+1+j};\qt )^{l_j}}   = Q\ .
\end{equation}
As the background parameter $Q$ is constant, this relation becomes consistent if the left-hand side is constant as well, which implies that the ratio of theta functions become a holomorphic section of a trivial line bundle over the elliptic curve~$E_{\qt,s}$. As $\theta(s^k y;\qt)$ is a holomorphic section of a line bundle of degree $k^2$, the ratio of theta functions on the left-hand side~\eqref{eq:4dRelWMod} is a meromorphic section of a degree zero line bundle, if and only if the anomaly constraint~\eqref{eq:Anom} is fulfilled. Furthermore, if the ratio of theta functions is restricted to those flavor fugacities $y_I$, $I=1,\ldots,n+1+a$, associated to non-anomalous flavor symmetries~\eqref{eq:FlavorAnom}, it is straight forward to check that the left-hand side is indeed invariant under large gauge transformations~\eqref{eq:LargeGauge}, and the ratio of theta functions becomes a holomorphic section of the trivial line bundle over the elliptic curve~$E_{\qt,s}$. As a result --- upon imposing these anomaly constraints --- the variable~$Q$ on the right-hand side of eq.~\eqref{eq:4dRelWMod} describes indeed a constant gauge-theory background. 

These stringent consistency conditions arising from eq.~\eqref{eq:4dRelWMod} can be relaxed, if we regard the parameter $Q$ on the right-hand side as a section of a non-trivial line bundle over the elliptic curve $E_{\qt,s}$ as well. Recall that the complex fugacity $Q$ defined in eq.~\eqref{eq:defQ} arises from a background linear multiplet. Turning $Q$ into a section of a non-trivial line bundle over $E_{\qt,s}$ thus implies that the background linear multiplet becomes dynamically. We do not further entertain this possibility in this work. 

\subsection{Ward Identities From Surface Operators} \label{subsect: WI from surface operators}
The quantum twisted chiral rings of supersymmetric gauge theories relate distinct non-perturbative sectors among each others. More specifically, for two-dimensional $\cN=(2,2)$ supersymmetric gauge theories, the exponent of the variable $Q$ labels the degree of the instanton background, such that the $Q$-dependent quantum cohomology condition~\eqref{eq:QCoh1} relates distinct instanton sectors. Similarly, for the three-dimensional $\cN=2$ supersymmetric gauge theories, the variable $Q$ enumerates the degree of vortex sectors that are non-trivially related in the K-theoretic twisted chiral ring via the $Q$-dependent quantum K~theory constraint~\eqref{eq:QK1}. It is well-established that these quantum twisted chiral ring structures are also encoded in the non-perturbative partition functions of these gauge theories. Namely, the hemisphere partition functions of two-dimensional $\cN=(2,2)$ gauge theories fulfills a differential equation with respect to the variable $Q$ \cite{Sugishita:2013jca,Honda:2013uca,Hori:2013ika}, whereas the $S^1\times D^2$-partition function of three-dimensional $\cN=2$ gauge theories obeys a certain difference equation \cite{Dimofte:2011ju,Beem:2012mb,Dimofte:2017tpi,Jockers:2018sfl}. These equations enjoy a physical interpretation in terms of Ward identities of point-like operators and of line operators, respectively. In the context of the quantum twisted chiral ring relations of four-dimensional $\cN=1$ supersymmetric gauge theories, the $T^2_\tau\times D^2$ partition function connects the quantum twisted chiral ring to Ward identities among surface operators in terms of difference equations in a similar fashion.  

For the four-dimensional $\cN=1$ gauge theory $\cT$ given in Table~\ref{tab:TheoryT}, the derived vortex partition function~\eqref{eq:ZvorDefF} arises from the degree $d$ vortex contributions~\eqref{eq:ZDefF}, which obey the recursion relation 
\begin{equation} \label{eq:Frec}
   F_{\text{vor},d+1}(s,\qt,q,y) =   \frac{\prod_{j=1}^{a}
\prod_{k=1}^{l_j}\theta(y_{j+n+1}s^{-l_j} q^{d\,l_j+k-1+\mathfrak{q}_j/2};\qt)}
    {\prod_{J=1}^{n+1} \theta(y_J sq^{-d-1+\mathfrak{q}_\phi/2};\qt)} F_{\text{vor},d}(s,\qt,q,y) \ .
\end{equation}
  From the $q$-difference operator $q^{Q\partial_Q}$ and more generally from the $q$-difference operator function $f(q^{Q\partial_Q})$ (built from an analytic function $f$) acting on the variable $Q$ as
\begin{equation}
  q^{\alpha Q\partial_Q} \cdot Q^d = q^{\alpha d} Q^d \ , \quad f(q^{\alpha Q\partial_Q})\cdot Q^d = f(q^{\alpha d}) \cdot Q^d \quad
  \text{for any}\quad \alpha, d\in \mathbb{Z} \ ,
\end{equation}
we define for the gauge theory $\cT$ the difference operator  
\begin{equation} \label{eq:DefDiffOp}
  \Lq =\prod_{J=1}^{n+1} \theta(y_J sq^{-Q \partial_Q +\mathfrak{q}_\phi/2};\qt)
  - Q \prod_{j=1}^{a} \prod_{k=1}^{l_j}\theta(y_{j+n+1}s^{-l_j} q^{l_jQ\partial_Q+k-1+\mathfrak{q}_j/2};\qt) \ .
\end{equation}
As a consequence of the recursion relation~\eqref{eq:Frec}, this difference operator annihilates the sum of vortex degree $d$ contributions, i.e., 
\begin{equation} \label{eq:DiffEq}
   \Lq \left( \sum_{d=0}^{+\infty} F_{\text{vor},d} Q^d \right) = 0 \ .
\end{equation}

We now give a physical interpret to the difference equation~\eqref{eq:DiffEq} associated to the gauge theory~$\cT$. As discussed in the context of three-dimensional $\cN=2$ supersymmetric gauge theories in ref.~\cite{Jockers:2018sfl}, the difference equations~\eqref{eq:DiffEq} annihilating the $S^1\times D^2$-partition function enjoys the interpretation of a Ward identity among line operators, as the action of a $q$-difference operator corresponds to inserting a Wilson line along the circle $S^1$ at the origin of the disk $D^2$.  These considerations can be generalized to the $T_{\tau}^2\times_{\varepsilon_1,\varepsilon_2} D^2$ partition function of four-dimensional $\cN=1$ supersymmetric gauge theories, where the difference operators~$\Lq$ encode Ward identities among $\cN=(0,2)$~surface defects along the space-time torus $T^2_\tau$ at the tip of the disk $D^2$. As derived in Section~\ref{sec:4dlocformula} (imposing Neumann boundary conditions for all chiral fields) we arrive for the partition function at the expression
\begin{equation}
     Z_\text{4d} (Q,\qt,q,y)
     = \frac1{2\pi\ii} \int_\cC \frac{\dd s}{s} \ee^{-\frac1{\log q} \log Q \,\cdot \, \log s} F_\text{matter}(\qt,q,y,s) \ ,
\end{equation}
with the matter integrand of double elliptic Gamma functions
\begin{equation}
      F_\text{matter}(\qt,q,y,s) = \prod_{I=1}^{n+1}\Gamma(sy_Iq^{\mathfrak{q}_\phi/2};q,\qt)  
      \prod_{j=1}^{a}\Gamma(s^{-l_j}y_{n+1+j}q^{\mathfrak{q}_j/2};q,\qt)   \ .
\end{equation}
The contour $\cC$ is chosen in such a way that the contributing residues to the integral in the $s$-plane yield the vortex contributions at degree $d$, namely for
\begin{equation} \label{eq:setPoles}
   s = y_I^{-1} q^{-d - \mathfrak{q}_\phi/2} \quad \text{with} \quad I=0,\ldots, n+1, \ d \in \mathbb{N}_0 \ . 
\end{equation}
Viewing the difference operator~\eqref{eq:DiffEq} as an identity between the vortex sectors at degree~$d$ and degree~$d+1$, it amounts to comparing the poles at $y_I^{-1} q^{-d - \mathfrak{q}_\phi/2}$ with the poles at $y_I^{-1} q^{-d-1 - \mathfrak{q}_\phi/2}$, which differ by the shift of the integration variable from $s$ to $s q$. Such a shift realizes the functional relation of the integrand of the form
\begin{equation}
    F_\text{matter}(\qt,q,y,sq) =\frac {\prod_{J=1}^{n+1} \theta(y_J sq^{\mathfrak{q}_\phi/2};\qt)}{\prod_{j=1}^{a}\prod_{k=1}^{l_j}\theta(y_{j+n+1}s^{-l_j} q^{-k+\mathfrak{q}_j/2} ;\qt)}
      F_\text{matter}(\qt,q,y,s) \ . 
\end{equation}
Assuming further that the residues~\eqref{eq:setPoles} are invariant with respect to the redefinition $s \mapsto s q$ --- which is the case for a suitable choice of the contour $\cC$ and together with a properly chosen flavor fugacities --- we have the integral identity
\begin{equation}
  \int_\cC \frac{\dd s}{s} \ee^{-\frac1{\log q} \log Q \,\cdot \, \log s} F_\text{matter}(\qt,q,y,sq )
 =  \int_\cC \frac{\dd s}{s} \ee^{-\frac1{\log q} \log Q \,\cdot \, \log s} F_\text{matter}(\qt,q,y,s) \ .
\end{equation} 
It further implies the integral relation
\begin{equation} \label{eq:IntRel2}
 0 = \int_\cC \frac{\dd s}{s} \ee^{-\frac1{\log q} \log Q \,\cdot \, \log s}
   \left(Q -  \frac {\prod_{J=1}^{n+1} \theta(y_J sq^{\mathfrak{q}_\phi/2};\qt)}{\prod_{j=1}^{a}\prod_{k=1}^{l_j}\theta(y_{j+n+1}s^{-l_j} q^{-k+\mathfrak{q}_j/2} ;\qt)}\right) \, F_\text{matter}(\qt,q,y,s) \ .
\end{equation}
According to the discussion in Section~\ref{sec:N02BPSDefects}, the second contribution of a ratio of theta functions enjoys the interpretation as an insertion of a $\mathcal{N}=(0,2)$ BPS surface defect with elliptic genus $I_{\text{2d}}(\qt,q,y,s)$, such that the integral identity becomes
\begin{equation}
  0 = \int_\cC \frac{\dd s}{s} \ee^{-\frac1{\log q} \log Q \,\cdot \, \log s}
   \left(Q- I_{\text{2d}}(\qt,q,y,s) \right) F_\text{matter}(\qt,q,y,s) \ .
\end{equation}
Hence, the insertion of a $\mathcal{N}=(0,2)$ BPS surface defect with elliptic genus $I_{\text{2d}}(\qt,q,y,s)$ is equivalent to reducing the vortex degree by one. Furthermore, the integration kernel of the identity~\eqref{eq:IntRel2} can be obtained from $q$-difference operator $\Lq$ by first using
\begin{equation*}
    Q \prod_{j=1}^{a} \prod_{k=1}^{l_j}\theta(y_{j+n+1}s^{-l_j} q^{l_jQ\partial_Q+k-1+\mathfrak{q}_j/2};\qt) = \prod_{j=1}^{a} \prod_{k=1}^{l_j}\theta(y_{j+n+1}s^{-l_j} q^{l_jQ\partial_Q-k+\mathfrak{q}_j/2};\qt) Q \ ,
\end{equation*} 
and then dividing by $\prod_{j=1}^{a} \prod_{k=1}^{l_j}\theta(y_{j+n+1}s^{-l_j} q^{l_jQ\partial_Q-k+\mathfrak{q}_j/2};\qt)$ and setting $Q\partial_Q \to 0$. This shows that the $q$-difference operators $\Lq$ of four-dimensional $\cN=1$ gauge theories physically realize Ward identities among $\mathcal{N}=(0,2)$ BPS surface defects.

Let us finally remark that the $q$-difference equation~\eqref{eq:DefDiffOp} realizes the quantum twisted chiral ring relation~\eqref{eq:4dRelWMod} in the limit $q \to 1$. This reflects the fact that in derivation of the quantum twisted chiral ring relations in Section~\ref{sec:twring}, the (non-anomalous) $\Uone_R$ symmetry is not used. However, as discussed in the context of the vortex partition function on the space-time background $T_{\tau}^2\times_{\varepsilon_1,\varepsilon_2} D^2$ the symmetry associated to the fugacity~$q$ is essential for the definition of the vortex partition function. As a consequence the quantum twisted chiral ring relations~\eqref{eq:4dRelWMod} are enhanced to the $q$-difference equation~\eqref{eq:DefDiffOp} associated to vortex partition function. Conversely --- analogously as in ref.~\cite{Jockers:2018sfl} for three-dimensional $\cN=2$ supersymmetric gauge theories --- we can explicitly calculate from the $q$-difference operators \eqref{eq:DefDiffOp} the quantum twisted chiral ring relations of four-dimensional $\cN=1$ supersymmetric gauge theories.

\section{Equivariant Elliptic Cohomology and the Thom Sheaf}
\label{sec: equi ell coho}
In this section, we introduce classical equivarant elliptic cohomology. Our exposition is tailored towards the applications we have in mind. On the one hand this means  to highlight the differences between K theory and elliptic cohomology, and on the other hand, we want to provide the tools to use push-forwards and localization in the rest of this paper. We start out with a discussion of the simplest possible geometry: the point. Even in this simple example, one can understand that while K theoretic rings correspond to affine spaces, elliptic cohomology enforces a generalization, replacing affine space by projective schemes. A slightly more involved example is ${\mathbb C}$ with a $\Uone$ action.  Physically this models a single flavor, and hence serves as building block for a larger class of models. Mathematically, it can be viewed as a vector bundle over a point, providing the simplest example of a bundle. We discuss the Thom isomorphism in K theory for this example, and show how it fails in elliptic cohomology. Generalizing, the Thom isomorphism is provided by a Thom class, whose pull back along the zero section of the vector bundle yields the Euler class. 

In equivariant localization, the Euler class features prominently as its inverse relates integrals over the full range to that of the fixed point sets. We investigate integration and push-forward for equivariant elliptic cohomology, with the conclusion, that push-forwards require a twist by the Thom sheaf. In K theory, this is trivial, in elliptic cohomology it is not. We propose a concrete form of the elliptic Euler class and a prescription for the push-forward in elliptic cohomology. We will make extensive use of this when discussing elliptic quantum cohomology.

\subsection{The Simplest Case: The Point}
The simplest geometry one may imagine is a single point. K~theory describes equivalence classes of vector bundles. If the geometry is just a point, a vector bundle reduces to a vector space, and its equivalence class is given by its dimension. Furthermore, we can take tensor products of vector spaces and vector bundles, so that K~theory has a ring structure. In the equivariant case, the vector spaces are equipped with a group action, and the ring structure comes from the tensor product of representations, such that
\[
K_G(pt) = R(G) \ ,
\]
where $R(G)$ is the representation ring of the group. It can also be regarded as the ring of characters. Most relevant for us in this paper is the abelian case, where $G=T= \Uone^r$ and we get
\[
K_T(pt) = {\mathbb Z} [\chi_1^{\pm 1}, \dots, \chi_r^{\pm 1}] \ .
\]
Throughout this paper, we complexify by tensoring with ${\mathbb C}$ and write
\begin{equation}
K_T (pt) = {\mathbb C} [u_1^{\pm 1}, \dots , u_r^{\pm 1}] \ .
\end{equation}
In algebraic geometry, we are instructed to associate spaces to rings, such that the geometric space is the spectrum of the ring. For K~theory of a point, it means
\begin{equation}\label{eq:KTspec}
   \spec K_T (pt) = ({\mathbb C}^*)^r \ .
\end{equation}
We can furthermore interpret this space in terms of line bundles on $S^1$, which  are characterized by their holonomies $\Hom(\pi_1(S^1), T) \cong T $.

The starting point for our discussion of elliptic cohomology is eq.~\eqref{eq:KTspec}. In the rank one case, $\spec(K_{\Uone} (pt))$ is simply ${\mathbb C}^*$. Elliptic cohomology instructs us to `periodify' the $\spec$ of the corresponding K~theory by dividing out $q_\tau^{\mathbb Z}$. This is the multiplicative version of the more familiar additive construction, where the elliptic curve is obtained as ${\bC}/(\bZ +\tau \bZ)$, and $\qt$ and $\tau$ are related by $q_\tau=\ee^{2\pi \ii \tau}$. In the simple example, this yields
\[
Ell_{\Uone} (pt) = {\mathbb C}^* / q_\tau^{\mathbb Z} \cong \Ecurve \  .
\]
For higher rank, one gets
\[
Ell_T(pt) \cong \Ecurve^r \ .
\]
This is no longer the spectrum of a ring, so we cannot go back to any elliptic cohomology ring. Furthermore, there are no interesting global holomorphic functions on $\Ecurve$. To obtain something interesting, we instead  have to consider sections of line bundles. On the level of sections, we still have a possibility to multiply, however, if we multiply two sections of line bundles, then the result will be a section of the tensor product of the two initial line bundles.

Extrapolating the fact that \eqref{eq:KTspec} may be interpreted in terms of $S^1$ bundles, we  may now regard $\Ecurve$ as the moduli space of degree zero $\Uone$ line bundles on $\Ecurve$
\[
{\mathrm{Bun}}^0_{\Uone} (\Ecurve) \cong \Ecurve \ ,
\]
and similarly for higher rank
\begin{equation} \label{eq:ptGKV}
{\mathrm{Bun}}^0_{\Uone^r} (\Ecurve) \cong \Ecurve^r \ .
\end{equation}
This is indeed also useful for generalizations to other spaces.

We omit at this point  a separate discussion of equivariant cohomology, as over ${\mathbb C}$ we can relate K~theory and cohomology theory via the Chern character. In particular, it relates  line bundles to their first Chern class, $\operatorname{ch}(L)= \ee^{c_1(L)}$. Equivariant cohomology of a point becomes
\[
H_T^\bullet(pt)  = {\mathbb C}[u_1, \dots, u_r] \ .
\]

\subsection{General Case}
Consider now equivariant K~theory in the more general case, $K_T(X)$. Examples we have in mind are projective space, as well as total spaces of bundles over projective space. K~theory `decategorifies' the derived category of coherent sheaves, taking equivalence classes. The space of equivalence classes inherits a monoidal structure from the original category, and, including formal differences,  forms a ring, the Grothendieck ring. In the equivariant case, the K~theory classes originate from equivariant vector bundles. The K~theory ring associated to a point acts naturally on this ring, as we can tensor representations in $R(T)$ with equivariant bundles.

Algebraically, K~theory is thought of as a category of ${\mathcal O}_X$ modules. For a point, ${\mathcal O}_{X=pt}= {\mathbb C}$  and we recover (virtual)  vector spaces, the case discussed in the previous paragraph. The product structure is inherited from the relative tensor product over ${\mathcal O}_X$. The action of $K_T(pt)$ is given by  the ordinary tensor product over the field ${\mathbb C}$. This means that equivariant K~theory exhibits a ring as well as a module structure. We are hence instructed to associate a space (Spec of K-theoretical ring) to a space $X$ with a $T$ action. On the level of spaces, the action of $K_T(pt)$ on $K(X)$ translates to the (decategorification of the)  fiber product. Also the ring structure may be interpreted in this way: the relative tensor product of ${\mathcal O}_X$ modules $M_1$ and $M_2$ over ${\mathcal O}_X$ corresponds to the fiber product $\spec M_1 \times_{X} \spec M_2$ on the geometric side.

To generalize to elliptic cohomology, our starting point is the idea that equivariant cohomology theories can be thought of in terms of  `spaces', which in the case of elliptic cohomology are no longer spectra of rings. The precise formulation is that $Ell_T$ is a functor from $T$-spaces to `spaces that are acted on by $Ell_T(pt)$'
\begin{equation}
     \label{equi ellip functor 1st}
     Ell_T(*): \{\text{T-spaces} \ X\} \to \text{(super) schemes over } Ell_T(pt) \ .
 \end{equation}
$Ell_T$ is covariant in $T$ and $X$. Let us elaborate on the target category. If there was an elliptic cohomology ring, then the right hand side would be its Spec. Since there is no such ring, the target is not an affine scheme, but rather a more general projective scheme. `Super' refers to the possibility that there can be odd elements in cohomology, which however do not show up in the simple examples we consider in this paper. The functor in particular associates $E_T \cong E^r$ to the point. For a general $T$-space $X$, the projection $X\to pt$  induces a map $Ell_T(X) \to Ell(pt)$, since the functor is covariant. This allows us to see $\cEL_T(X)$  as a sheaf over $Ell_T(pt)$, which in turn is the moduli space of $T$-bundles over $E$. One difference between K~theory and elliptic cohomology is that elliptic cohomology is no longer the Grothendieck ring of a single category. However, as we pointed out, the tensor product in K~theory can be understood in terms of fiber products, and this point of view can be generalized to more general schemes.
 
The work of Grojnowski \cite{Grojnowski:2007} provides a concrete construction of this sheaf in terms of fixed point sets in $X$ of subgroups of $T$. To give a flavor of the construction, Grojnowski gives a prescription how to associate to any point $a\in E_T$ a subgroup $T(a)\subset T$. While $E_T$ does not act on $X$, $T(a)$ does, and the fiber at $a$ is determined in terms of the fixed point set $X^{T(a)}$ of $T(a)$.  More precisely, the stalk at $a$ is
\[
 Ell_T (X)_a = H_T(X^{T(a)}) \  \otimes_{{\mathcal O}_{\mathbb C}^r} ({\mathcal O}_{\mathbb C}^r)_0 \ ,
\]
where ${\mathcal O}_{{\mathbb C}^r,0}$ is the local ring of germs of holomorphic functions at $0$ on ${\mathbb C}^r$. 

For a generic point $a\in E_T$, the subgroup $T(a)$ is the full group. Of interest are the `torsion points' in $E_T$, whose defining property is that their elements become  trivial if applied a finite number of times. In additive notation, the element $0$ of an elliptic curve is an example and $T(0)$ is trivial, such that the stalk at zero is $H_T^\bullet (X) \otimes_{{\mathcal O}_{\mathbb C}^r} ({\mathcal O}_{\mathbb C}^r)_0$. 

The local patches of the sheaf $Ell_T (X)$ are built from the fixed point data $X^{T(a)}$. We refer to refs.~\cite{Ganter2014, rosu, ando} for more detailed accounts of Grojnowki's construction, including a prescription of how to glue the different patches to a full sheaf.
 
The work of Ginzburg--Kapranov--Vasserot takes a more abstract point of view on equivariant elliptic cohomology. They regard equivariant elliptic cohomology as a sheaf on the moduli space of $T$-bundles, and characterize it axiomatically. Their  perspective on the geometry given by a single point is provided by eq.~\eqref{eq:ptGKV}. Their work goes beyond the case of abelian groups, and they show that in the abelian case it agrees with Grojnowski's construction.
 
Finally, Lurie takes further steps in abstraction \cite{Lurie:2009}, and ties the subject of elliptic cohomology closer to recent developments in derived algebraic geometry.
 
In a one-line summary, following the lectures \cite{Okounkov:2022Lectures}, one can learn a lot about equivariant elliptic cohomology by regarding it as `periodified K~theory', i.e.,
\begin{equation}
    \label{equi ell func from equi K 1st}
    Ell_T(X) = \spec(K_T(X))/\qt^{\Hom(\bC^*,T)} \ .
\end{equation}
This is a rough picture, where one also has to replace `functions' by `sections of line bundles'.

\subsection{Thom Isomorphism and Euler Classes}
An important difference between equivariant K theory and elliptic cohomology is the fact that for K~theory (and also ordinary equivariant cohomology) there exists a Thom isomorphism. Very roughly, the Thom isomorphism relates the compactly supported cohomology (or K~theory) of the total space of a vector bundle to that of its base. The basic idea is to `integrate over the fiber'. 
A more precise discussion requires to introduce mathematical machinery around the concepts of Thom spaces, Thom class and Thom sheaf. We will be somewhat loose in our discussion, referring to ref.~\cite{Okounkov:2022Lectures} for a mathematical discussion, and instead  explain what the individual notions mean in the very simple example of a one dimensional $\Uone$ equivariant vector bundle over a point. We take the $\Uone$-action on ${\mathbb C}$ to have weight $1$ and use the notation ${\mathbb C}_1$ for this.

To understand the difference between K theory and elliptic cohomology, we first compute the Thom space of this bundle. Geometrically, one starts with the normal bundle to the zero section embedding of the base  into the total space of the vector bundle. One then collapses infinity to a point.
An algebraic construction for this can be found in ref.~\cite{Okounkov:2022Lectures}, where
\[
 \operatorname{Thom}(V) = \bP_X(V\oplus \bC)/\bP_X(V)\ ,
\]
for a vector bundle $V$ over a base $X$. Here, the hyperplane $\bP_X(V)$ that is taken out refers to the collapsed space at infinity. Further,  $\bP_X$ refers to a projective closure. In our example, this prescription reduces to
\[
\operatorname{Thom} ({\mathbb C}_1) = \bP(\bC_1\oplus \bC_0)/\bP(\bC_1)\ ,
\]
where the $\bC_0$ refers to a weight $0$ action and should be regarded as ${\mathcal O}(pt)$. 
Note that the space in the numerator inherits a $\Uone$ action from the original space, and it is isomorphic to $\bP^1$.  The space in the denominator is simply a point, as it is the projectivization of a one-dimensional space that we need to collapse, so $\bP(\bC_1)= \{ \infty \}$. Collapsing it does not change the topology of $\bP^1$, but makes the point at $\infty$ a distinguished base point and we obtain
\[
\operatorname{Thom} ({\mathbb C}_1)= \bP_1 / \{\infty \} \ ,
\]
where, again, the slash  refers to the collapsing.
As the space inherits a $\Uone$ action, we can now compute its equivariant K theory. It is different than the K theory of $\bP_1$, as we collapsed the point, and hence restrict the K theory to those classes that vanish at the base point. We hence obtain
\[\label{K theo Thom iso}
K_{\Uone} (\bP^1 / \{\infty\}) 
\cong (1-u^{-1}) R(\Uone)\ .
\]
This is isomorphic to $K_{\Uone} (pt)$! The isomorphism, called Thom isomorphism, is in this special case given by multiplication with $(1-u^{-1})$. The isomorphism is given by multiplication with the Thom class. For a general vector bundle, the Euler class is the pullback of the Thom class along the zero-section. We record that the K-theoretic Euler class for our example is
\[
e^K = (1-u^{-1}) \ .
\]
An explicit  calculation of elliptic cohomology for our example, as well as the more general weight $n$ case, can be found in ref.~\cite{rosu}, using the gluing prescription of ref.~\cite{Grojnowski:2007}, with the result\[
{\mathcal Ell}_{U(1)} (\operatorname{Thom}(\bC_1)) = {\mathcal O}_E (-[0]) \ .
\]
Here, ${\mathcal O}_E (-[0]) $ denotes the line bundle corresponding to the divisor at the identity point of the elliptic curve. Its characteristic section is the $\theta$ function $\theta(u, q_\tau)$. A first lesson we learn at this point is that\footnote{See refs.~\cite{Ganter2014, Okounkov:2022Lectures} for relevant discussions on this point. Another way to see theta function as elliptic Euler class is applying the localization calculation in ref.~\cite{nekrasov2015bethe}.}
\[
(1-e^{2\pi i z}) \leftrightarrow \theta (z, \tau)
\]
when passing from K-theory to elliptic cohomology.

To make the last statements more precise, we first identify the space in which $e^K = (1-u^{-1})$ lives,
\begin{equation}
    e^{K}(\bC^{*})(:= 1-u^{-1}) \in \Hom_{\cO_{\bC^*}}(\cO(-[0]_{\bC^{*}}), \cO_{\bC^{*}})\ .
\end{equation}
Connecting to our previous discussion, one would further expect that
in equivariant elliptic cohomology, the corresponding cohomology of the Thom space $\bP(\bC_1\oplus \bC_0)/\bP(\bC_1)$ is the ideal sheaf
\[
{\mathcal Ell}_{\Uone} (\bP({\mathbb C_1\oplus \bC_{0}})/\bP(\bC_{1})) = {\mathcal Ell}_{\Uone} ( \bP^1/\{\infty \})={\mathcal O}_E (-[0]_E) \ .
\]
In particular, one would want to regard  the elliptic Euler class as a global section in\footnote{The explicit realizations of the elliptic Euler class in terms of theta functions is ambiguous up to factors $\qt^\alpha$, $\alpha\in\bZ$. Analogously as in Section~\ref{subsect: anomaly and vortex string} such ambiguities come from different choices of local trivialization of the line bundle $\cO_E([0])$ and are therefore geometrically unimportant.}
$$
  \text{Hom}_{\cO_E}({\mathcal O}_E (-[0]_E), \cO_{E}) = \Gamma(E; \cO_{E}([0]_E))\ ,
$$
analogously to the K theoretic discussion.

Note that in elliptic cohomology ${\mathcal Ell }_{U(1)} (pt) = {\mathcal O}_E$, the structure sheaf on the elliptic curve. We in particular find that there is no Thom isomorphism! Instead, we need to twist by a non-trivial line bundle to go from the elliptic cohomology of the Thom space of a vector bundle to the one of the point.

\subsection{The Point of View of Localization}
Localization essentially relates cohomology at a fixed locus to that of the full space. In our paper, we focus our attention on abelian groups and isolated fixed points. Let us quickly recall how localization works in ordinary equivariant cohomology. 
For an abelian group, the equivariant cohomology of a point is
\[
H_T^\bullet(pt) = {\mathbb C}[u_1, \dots, u_r] \ ,
\]
and its spectrum is ${\mathbb C}^r$. The ring acts on the equivariant cohomology of another space $X$. Once one localizes away from torsion part of the module $H_T^{\bullet}(X)$ over the base ring $H_T^\bullet(pt)$ equivariant cohomology is determined by what happens at the fixed point locus $X^T$, i.e.
\[ 
H_T^\bullet (X) \otimes_{\mathbb{C}[u_1,\dots u_r]} \mathbb C (u_1, \dots u_r) = H_T^\bullet (X^T) \otimes_{\mathbb{C}[u_1,\dots u_r]} \mathbb C (u_1, \dots u_r) \ .
\]
Here, $\mathbb C (u_1, \dots u_r)$ is the field of fractions. By tensoring with it, we localize away from the 
torsion part of $H_T^\bullet(X)$.
The equation expresses that equivariant cohomology is expressed by equivariant cohomology of the fixed locus --- once one allows for division of the equivariance parameters $u_i$. 

For ordinary cohomology, Atiyah--Bott fixed-point theorem says that integration of equivariant differential forms on $X$ can be reduced to an integration over the fixed point locus $X^T$ of the group action
\[
\int_X \alpha = \int_{X^T} \frac{f^*\alpha}{e_T({\mathcal N})} \ ,
\]
where $f$ denotes the inclusion $f: X^T \to X$, and $f^* \alpha$ the pullback of the differential form to the fixed point set. $e_T({\mathcal N})$ denotes the Euler class of the normal bundle to the fixed point locus $X^T$. It is invertible after localization, and the identity on $H^\bullet_T$ can be expressed as
\[
1= f_* \frac{1}{e_T({\mathcal N})} f^* \ .
\]
The Euler class  enters the integration formula, since it is the pullback of the Thom class. It summarizes the contributions of the normal bundle to the integration. For any $X$, we denote by $\pi^X$ the map $X\to pt$. The push-forward in cohomology is  $\pi^{X}_*: H^\bullet_T(X)\to  H^\bullet_T (pt)$. It corresponds to the integration of a class $\alpha \in H^\bullet_T(X)$. The above integration formula can then be stated as
\[
\pi_*^{X} = \pi_*^{X^T} \frac{f^*\alpha}{e_T({\mathcal N})} \ .
\]
Consider a $T$ action with weights $a_i$. In this case the Euler class is given by
\begin{equation} \label{coho euler}
  e_T({\mathcal N}) = \prod_i(a_i)\ .
\end{equation}
Similarly, localization theorems hold  in equivariant K~theory  and equivariant elliptic cohomology where the Euler class has to be chosen accordingly.  As we discussed in the last section, in equivariant K~theory it is given by
\begin{equation} \label{K euler}
  e^K_T({\mathcal N}) = \prod_i(1-e^{ -a_i})\ ,
\end{equation}
and in equivariant elliptic cohomology:
\begin{equation} \label{elliptic euler class}
  e^{ell}_T({\mathcal N}) = \prod_i\theta(e^{-a_i};q)\ ,
\end{equation}
which is a section of the non-trivial line bundle over the underlying scheme in equivariant elliptic cohomology.

In contrast to the case of equivariant cohomology and K~theory, the push-forward in equivariant elliptic cohomology needs to be twisted by the Thom sheaf. To motivate this, we start from our  simple example ${\mathbb C}_1$, which already exhibits the main features.  The simplest push-forward would be the one to a point, so we consider an equivariant map $f: \bC_{1}\to pt$ and its  push-forward  $f^{ell}_{*}$ in elliptic cohomology. Elliptic cohomology is homotopy invariant \cite{ginzburg1995elliptic}. Thus, we expect that $\cEL_{\bC^*}(\bC_{1}) \simeq \cEL_{\bC^*}(pt)$,
as $\bC_1$ is contractible to a point. $\cEL_{\bC^*}(pt)$ is given by ${\mathcal O}_E$. On the other hand, we explained in the previous section that the push-forward is given in terms of the Euler class, which in the case of elliptic cohomology is a $\theta$-function with non-trivial transformation properties. To remedy this situation, we need to compensate for the non-trivial transformation properties, multiplying with a section of a meromorphic line bundle $\cL$. Summarizing, the push-forward $f^{ell}_*$ in equivariant elliptic cohomology becomes 
\begin{equation}
    \label{equi pushforward in ell coho 2nd}
            f^{ell}_{*}( =\frac{1}{ e^{ell} (\bC_{1})}):\ \   \cEL_{\bC^*}(\bC_1)\otimes \cL  \simeq \cO_E\otimes \cL \to \cEL_{\bC^*}(pt)\simeq \cO_E\ ,  
\end{equation}
In order for the push-forward~\eqref{equi pushforward in ell coho 2nd} to exhibit the correct transformation properties, $\frac{1}{e^{ell}(\bC_1)}$ should be valued in meromorphic sections $\Gamma_{mero}(E;\cL^{-1})$ and  $\cL = \cO_E([0]_E)$. 

We see in this simple example that the push-forward in equivariant elliptic cohomology requires a twist by a line bundle. This line bundle is called Thom sheaf and is denoted by $\Theta ({\mathbb C}_1)$. 

An immediate generalization is the case of higher dimensional vector spaces with different weights
\begin{equation} \label{eq:moreflavors}
V= \bigoplus {\mathbb C}_{a_i} \ .
\end{equation}
The required twist appearing in the push-forward \eqref{equi pushforward in ell coho 2nd} is by the tensor product of line bundles $\otimes_i \mathcal{L}_i$. Here,  $\mathcal{L}_i$ is line bundle over $\Ecurve$ with one of its global sections the Euler class $\theta(u^{-a_i};\qt)$. Rather obviously in this class of simple example, the Thom sheaf should satisfy 
\[\label{Thom sheaf of vector bundles of a point}
\Theta \left(\bigoplus {\mathbb C}_{a_i}\right) = \bigotimes \Theta ({\mathbb C}_{a_i}) \ .
\]

\bigskip

In more generality, these novel phenomena  in equivariant elliptic cohomology have been addressed  in ref.~\cite{ginzburg1995elliptic}.  
We refer to it for general constructions and properties of the Thom sheaf. Let us simply state that for  a general $T$-equivariant map
\begin{equation}
    \label{general T map 1st}
    f: X \to Y \ ,
\end{equation}
the equivariant push-forward needs to be twisted by a non trivial Thom sheaf valued in $\operatorname{Pic}(Ell_T(X))$:
\begin{equation}
    \label{THom sheaf for a general map 1st}
    f_*: \cEL_{T}(X) \otimes \Theta(-N_f) \to \cEL_{T}(Y)\ ,
    \quad N_f = f^*TY\ominus TX \in K_{T}(X) \ .
\end{equation}
$\Theta$ satisfies $\Theta(V_1 \oplus V_2) = \Theta(V_1) \otimes \Theta(V_2)$ (which is an abstraction of the observations  discussed around eq.~\eqref{eq:moreflavors}) and can be regarded as a group homomorphism
\begin{equation}
    \label{general Thom map 1st}
    \Theta: K_{T}(X) \to \operatorname{Pic}(Ell_T(X)) \ .
\end{equation}
$N_f$ is the normal bundle associated to the map $f$. For the explicit construction and description of Thom sheaf and the map $\Theta$, we refer the readers to the original paper \cite{ginzburg1995elliptic}. In this paper, we focus on $X_H$ whose T-fixed loci are isolated points. With the help of localization techniques in equivariant elliptic cohomology, Thom sheaves of type \eqref{Thom sheaf of vector bundles of a point} and the associated Euler classes play a central role in our calculations. 

To conclude, equivariant localization in elliptic cohomology requires a twist given by related Thom sheaf, and this Thom sheaf can be topologically non-trivial.

\subsection{Comparison With Physics}
Our toy example suggests a direct relation between equivariant elliptic cohomology and $\cN=(0,2)$ surface defects. Indeed, we discussed how the supersymmetric index of a free chiral multiplet with a $\Uone$ action inherited from the bulk transform as sections of line bundles. This statement implies that the physical theory exhibits an anomaly, which is global from the point of the $\cN=(0,2)$ theory, but a gauge anomaly from the perspective of the four-dimensional ${\mathcal N}=1$ theory.

From the mathematical side, we see that the appearance of the non-trivial sections originates from the fact that there is  a non-trivial Thom sheaf. This is a feature of equivariant elliptic cohomology. At the classical, geometric level, there is of course nothing wrong with this. The problem arises once one tries to pass to the yet-to-be-defined version of quantum elliptic cohomology. Whatever its final interpretation and formulation will be, it requires us to deform `products of sections' by numbers. If that is to make sense, then those products need to transform trivially. To see the restrictions explicitly, we have to consider topologically non-trivial configurations. This is the purpose of the next sections.

\section{Towards a Mathematical Formulation of the Vortex Partition Function} \label{subsect: vortex partition math}
In  this  section, we outline a mathematical interpretation of the physical results obtained in Section~\ref{sec:4d N=1 localization}. For this, it is necessary to move to a framework that goes beyond classical elliptic cohomology, as we expect quantum mechanical effects from the non-perturbative sector. 
The aim of the current section is to outline a mathematical argument for the vortex partition function
\[
Z_{\text{vor}}^{\text{ell}} = \sum_{d\geq 0} Q^d Z_{\text{vor};d} \ ,
\]
and to motivate an elliptic version of Givental's I~function.

Our strategy is to relate the vortex partition function to the moduli space of quasimaps from a curve to a target space $X_H$, which physically is the Higgs branch. In the case of quantum K~theory, the starting point of the discussion is the moduli space of quasimaps with one marked point. The K-theoretic vortex partition function is then given by
\[
Z_{\text{vor}}^{K} = \sum_{d\geq 0} Q^d ev_{d,*} ({\mathcal O}^{\text{vir}}) \ .
\]
It requires the formulation of an evaluation map that uses the marked point to map the moduli space of quasimaps to $X_H$, as well as the existence of a virtual structure sheaf. Such sheafs, or their associated virtual classes, are important in the context of integration theory on moduli spaces such as the moduli space of stable maps or the moduli space of quasimaps. These moduli spaces are in particular singular and the virtual structure sheaf can be viewed as a method to develop an integration and intersection theory on them. For K~theory, the virtual structure sheaf is constructed in ref.~\cite{Lee-foundation}.

Furthermore, the construction of the K-theoretic vortex partition function makes use of the symmetries of the problem and applies localization arguments. As the relevant quasimap moduli spaces are non-compact \cite{Okounkov:2022Lectures,Dedushenko2:2023qjq}, equivariant localization techniques for K-theoretic vortex partition functions are necessary rather than just a convenient technical tool.

We subsequently extrapolate the arguments to the elliptic case and conjecture the form of the vortex partition function, which in the case of elliptic cohomology becomes a meromorphic section of a Thom sheaf. We use the same toolbox as for K~theory. In particular, symmetry and localization arguments are still available, though different since localization requires a twist by the Thom sheaf. To conclude our argument, we conjecture that a virtual object exists also in elliptic cohomology. The precise definition or form does not affect our arguments. 

Let us however comment on the possibility to invoke concepts of derived algebraic geometry to study the virtual objects we need, as initially proposed by Kontsevich \cite{Kontsevich}. Derived concepts have proven useful whenever an actual dimension does not agree with the expected, or virtual, dimension. The construction of Lee~\cite{Lee-foundation} is revisited from that perspective in ref.~\cite{kernetal}. In particular, they obtain the virtual structure sheaf in terms of an actual structure sheaf on its derived enhancement. It is suggestive that these ideas generalize to the elliptic case, in the sense that it would be natural to consider the Thom sheaf of the structure sheaf on the derived enhancement. Interpreting this again in terms of the \emph{virtual} structure sheaf, one is led to introduce a Thom sheaf associated with the virtual structure sheaf. This is a natural proposal for the virtual object in elliptic cohomology. 

The remaining part of this section is organized as follows. To start, we review the moduli space of quasimaps and explain how it encodes the ultra-violet data of the physical system. We further elaborate on its mathematical definition and properties and hint why it is the right framework to formulate generalized difference equations, as we proposed in the previous section. We then proceed and formulate the K-theoretic vortex partition functions via localization. Following an analogous strategy, we propose the form of the elliptic vortex partition function.

\subsection{The Moduli Space of Quasimaps} \label{sec:QuasiMapModuliSpace}
\paragraph{Motivation and Definition:}
A mathematical model for Higgs branch of a supersymmetric gauge theory has as its basic ingredients the compact gauge group~$G$ and the matter representations~$V$. In many cases the Higgs branch can also be described by a geometric invariant theory~(GIT) quotient with respect to the complexified gauge group $G_\bC$, see for instance ref.~\cite{Thomas:2006}. The Higgs branch $X_H$ can then be regarded as the stable locus of the quotient stack $Y_H$
\begin{equation} \label{eq:XHvsYH}
X_H \subset Y_H := [V /G_{\bC}] \ .
\end{equation}

For our main example $\cT$ of Section~\ref{sec:4d N=1 localization}, the Higgs branch $X_H$ is given by the symplectic quotient~\eqref{Higgs branch XH}, which also enjoys the description in terms of a GIT~quotient
\begin{equation}
    X_{H} = 
     \big(\Hom(\bC^{n},\bC)\oplus \Hom(\bC,\bC^{a})\big)\sslash\bC^*
    :=
    \big(\Hom(\bC^{n},\bC)\oplus \Hom(\bC,\bC^{a})\big)^\text{s}/\bC^* \ ,
\end{equation}
where $\big(\Hom(\bC^{n},\bC)\oplus\Hom (\bC,\bC^{a})\big)^\text{s}$ is the stable locus of $V=\Hom(\bC^{n},\bC)\oplus \Hom(\bC,\bC^{a})$. The stack $Y_H$ reads
\begin{equation}
    Y_{H} = [\Hom(\bC^{n},\bC)\oplus \Hom(\bC,\bC^{a}) /\bC^*] \ .
\end{equation}
The difference between $X_H$ and $Y_H$ is that  the construction of $X_H$ requires a choice of stability condition that relates to the parameter $\zeta>0$, such that the unstable locus is excluded before passing to the quotient. For $\zeta>0$ the unstable locus is given by $\phi^1_0 = \ldots = \phi^{n+1}=0$ with 
the variables $\varphi^i$ unconstrained. From the orbits of the unstable locus $\phi^1=\ldots,\phi^{n+1}=0$ the quotient $Y_H$ contains `stacky' points that are not contained in $X_H$. As a consequence, the GIT quotient~$X_H$ is a `conventional' geometric space, whereas the algebraic quotient~$Y_H$ is a stack.

Associated to the GIT~quotient~$X_H$ there are different moduli spaces of maps
\begin{equation}\label{eq:higssmap}
    f: \bP^1 \to X_{H}\ ,
\end{equation}
that are given by different stability conditions. Two prominent examples are stable maps and quasimaps. While stable maps are preferred in string theory, quasimaps naturally appear in gauge theory as discussed below. 

\paragraph{Quasimaps:}
Here we give a brief and non-rigorous description of the theory of quasimaps \cite{Ciocan-Fontanine:2010,Ciocan-Fontanine:2011,Kim:2011}. To start, consider a genus $g$ curve $C_{g,m}$ with at most nodal singularities and $m$ marked points. A quasimap $C_{g,m}\to [M/G]$ is defined
to be a principal $G$-bundle $P\to C$ together with a section $f$ of the associated $M$ bundle $P\times_G M \to C$ whose base locus $f^{-1} (M_{\text{unstable}})$ is zero-dimensional and disjoint from nodes and marked points. To unpack this definition, consider the abelian rank one gauge group $G={\mathbb C}^*$ and $M={\mathbb C}^n$ with $M\sslash\bC^*=\bP^{n-1}$. In this case, the unstable locus is $0 \in \bC^n$ and we have $M\sslash\bC^\times \subset [\bC^n/\bC^*]$. In that case, principal ${\mathbb C}^*$ bundles are equivalent to line bundles over $C$. Furthermore, the associated line bundle is $L\times_{{\mathbb C}^\times} {\mathbb C}^n \simeq L^{\oplus n}$. A section $f$ of $L^{\oplus n}$ takes the form
\[
f= (f_1, \dots, f_n) \ , \quad f_i \in H^0(C,L) \ .
\]
Instead of imposing $f(C) \subset M\sslash\bC^*=\bP^{n-1}$, for quasimaps one allows a finite number of points $x\in C$ taking values $f(x) =0 \in [\bC^n/\bC^*]$. 

Let us emphasize that quasimaps naturally have a degree --- originating from  the Picard lattice $\operatorname{Pic}(V\sslash G_\bC)$ of $V\sslash G_\bC$ --- which is given by
\begin{equation}
   \cL \in \operatorname{Pic}(V\sslash G_\bC)\ , \qquad
   \deg f = \int_{C}c_{1}(f^*\cL) \ .
\end{equation}
We will discuss the difference equation of vortex partition function associated with this grading structure at the end of Section~\ref{sec:quasimaps}.

As indicated above, one may  specify a number of marked points $x_1, \dots, x_k \in C$. For the physics applications that we have in mind, the relevant case is $C={\mathbb P}_1$ and the target space $X_H$, the classical Higgs branch of the gauge theory~$\cT$. In this case, the grading of quasimaps is the same as the grading given in terms of the  second homology class $[f_*(\bP^1)] = \beta \in H_{2}(X_H;\bZ)$.

For a homology class $\beta \in H_{2}(X_{H};\bZ)$, the quasimap moduli space $\cQ\cM_{0,k}(X_{H};\beta)$ is given by
\begin{equation}
    f: (\bP^1,x_1,\dots, x_k \in \bP^1) \to Y_{H}, \ \ \ f(x_{1,\dots , k}) \in X_{H}, \quad  [f_*(\bP^1)] = \beta \in H_{2}(X_H;\bZ) \ ,
\end{equation}
together with the condition that:
\begin{equation}
    \text{Only finitely many points $x \in \bP^1$ such that}\ f(x) \notin X_H \ .
\end{equation}
That is to say, only a finite number of points in $\bP^1$ gets mapped to `stacky' points in $Y_H$. We refer to these points as defect points. These points must not coincide with the marked points. Using the grading, we can decompose 
\begin{equation}
    \cQ\cM_{0,k}(X_{H}) = \sqcup_{d\geq 0} \cQ\cM_{0,k}(X_{H};d)\ .
\end{equation}
As we will explain, the physically relevant case for us is $k=1$. The moduli space of quasimaps has many nice properties. 

\begin{itemize}
    \item $\cQ\cM_{0,1}(X_H)$ admits an equivariant action by the group $T_{F} \times \bC_{q}^*$. Here, $T_F$ physically corresponds to  the flavor symmetry group of the theory, which also naturally acts on $X_{H}$.  $\bC^{*}_q$ is the symmetry that rotates the source $\bP^1$ with fixed points $0,\infty$ and acts trivially on $X_H$. The $\bC^*_q$ fixed points in $\cQ\cM_{0,1}(X_H)$ are given by quasimaps $f$ that only map $0$ to the unstable locus, i.e., $Y_{H}\setminus X_{H}$.
    \item $\cQ\cM_{0,1}(X_H)$ has the following natural map provided by the marked point:
    \begin{equation*}
    ev: \cQ\cM_{0,1}(X_{H}) \to X_{H}
\end{equation*}
This map is $T_F \times \bC_{q}^*$ equivariant. We refer to its restrictions to $\cQ\cM_{0,1}(X_H;d)$ as $ev_d$.
\end{itemize}

\paragraph{Physical Origin of Quasimaps:}
Based on the gauge theory $\cT$, in  Section~\ref{sec:BPSVorMSpace} we discussed the physics of vortices form a geometric point of view. We argued that vortex solutions arose only for vanishing expectation values of the negatively charged fields, which led to the reduced Higgs branch $X_{H,\text{red}} \simeq \bP^n$ of the gauge theory $\cT$. Vortex solutions of degree $d$ were described in eq.~\eqref{eq:flag} in terms of the flag $\phi: {\mathcal O} (-d) \hookrightarrow {\mathcal O}^{\oplus (n+1)}$, where the embedding $\phi$ is concretely specified by $(n+1)$ polynomials~\eqref{eq:SheafHomPhi} of degree $d$. We further discussed in Section~\ref{sec:BPSVorMSpace} common zeros of the embedding $\phi$, and we referred to such field configurations as small vortices that appear at the boundary of the vortex moduli space. 

In the context of the moduli space of quasimaps, a small vortex correspond to base point of the quasimap, whereas the quasimap sends the marked point $\infty \in \bP^1$ properly to the Higgs branch $X_H$. As the boundary $\partial D^2$ of the spacetime disk $D^2 \subset T^2_\tau \times_{\epsilon_1,\epsilon_2} D^2$ collapses to the marked point $\infty \in \bP^1$ in the quasimap description, the image of the marked point in the Higgs branch $X_H$ describes the Higgs vacuum of the gauge theory. As a consequence, since defect points of vortices can be arbitrarily close to the boundary while remaining in the interior of the disk $D^2$, we see from a physics point of view that the marked point cannot coincide with defect points, which thus furnishes an open condition. From this observation we see that the quasimap moduli space $\cQ\cM_{0,1}(X_H)$ is non-compact. 

Finally, we note that the mathematically equivariant push-forward in eq.~\eqref{eq: math formulation of vortex partition} with respect to the ${\mathbb C}_q^*$-symmetry, has a physical counterpart in the space-time geometry $T^2_\tau \times_{\varepsilon_1,\varepsilon_2} D^2$, as BPS $\cN=(0,2)$ surface defects along the torus $T_\tau$ are localized at the center of the disk $D^2$.

\subsection{Vortex Partition Function and Quasimaps} \label{se:VorPart_QuasiMaps}
As the moduli space $\cQ\cM_{0,1}(X_H)$ is in general non-compact, the symmetries discussed above are important to formulate the vortex partition function with equivariant localization techniques in the context of equivariant cohomology, K~theory, or elliptic cohomology. Specifically, we define the push-forward\footnote{More precisely, since we apply equivariant localization to the push-forward~$ev_*$, it should be defined with a localized version $E_{\bC^*_q}(\cQ\cM_{0.1}(X_{H},d))_{loc}$ of $E_{\bC^*_q}(\cQ\cM_{0.1}(X_{H},d))$. We omit the subscript $loc$ here and in the following.}
\begin{equation}\label{Pushforward in generalized cohomology}
   ev_* = \sum_{d\geq 0} Q^d ev_{d,*} 
   \quad \text{with} \quad
   ev_{d,*}:\ E(\cQ\cM_{0,1}(X_H;d)) \to E(X_H) \ ,
\end{equation}
where $E(-)$ refers to a generalized cohomology theory. There are three important choices
\begin{itemize}
    \item $E(-) = H^\bullet_{T_{F}\times \bC^*_q}(-)$: $T_{F}\times \bC^*_q$-equivariant cohomology theory.
    \item $E(-) = K_{T_{F}\times \bC^*_q}(-)$: $T_{F}\times \bC^*_q$-equivariant K~theory.
    \item $E(-) = \cEL^*_{T_{F}\times \bC^*_q}(-)$:  $T_{F}\times \bC^*_q$-equivariant elliptic cohomology theory.
\end{itemize}
With the defined push-forward~\eqref{Pushforward in generalized cohomology}, the vortex partition function is formulated as
\begin{equation}\label{eq: math formulation of vortex partition}
  Z_{\text{vor}} =\sum_{d\geq 0} Q^k Z_{\text{vor},k} = \sum_{d\geq 0} Q^k ev_{d,*}(\mathbf{1}^\text{vir}) \ \in\ E(X_{H})[[Q]] \ .
\end{equation}
In this mathematical context $Q$ is now a formal variable. Furthermore, we assume the existence of a virtual fundamental object $\mathbf{1}^\text{vir}$, which is constructed for quantum cohomology in ref.~\cite{Li:1996} and for quantum K~theory in ref.~\cite{Lee-foundation}, but remains an assumption for quantum elliptic cohomology.

In the following, we discuss the computation of the evaluation map~\eqref{eq: math formulation of vortex partition} using equivariant localization techniques. Before we explicitly carry out this computation for our main example of the gauge theory $\cT$, we outline the general strategy. 

We compute the push-forward~$ ev_{d,*}( \mathbf{1}^{vir})$ by applying equivariant localization according to the commutative diagram
\begin{equation} \label{localization of quasi map}
\begin{tikzcd}
    \cQ\cM_{0,1}(X_H;d)^{T_F\times \bC_q^*} \arrow[r, "\iota_1"] \arrow[d, "ev_{d}^T"'] &  \cQ\cM_{0,1}(X_H;d) \arrow[d, "ev_{d}"] \\
    X_H^{T_F\times \bC_q^*} \arrow[r, "\iota_2"'] & X_H
\end{tikzcd}
\end{equation}
where
\begin{itemize}
\item $X_H^{T_F\times \bC_q^*}$ and $\cQ\cM_{0,1}(X_H;d)^{T_F\times \bC_q^*}$ are the $T_F\times \bC_q^*$ fixed loci of the Higgs branch~$X_H$ and the quasimap moduli space $\cQ\cM_{0,1}(X_H;d)$, respectively. For our main example $\cT$, the fixed points $X_H^{T_F\times \bC_q^*}$ are the points $p_1=[1:0:\ldots :0],\ldots,p_{n+1}=[0:0:\ldots :1]$ in the base $\bP^n$ at the zero section of the bundle~$\cE$, whereas the fixed quasimaps $\cQ\cM_{0,1}(X_H;d)^{T_F\times \bC_q^*}$ are $f_1=[z^d:0:\ldots :0],\ldots,f_{n+1}=[0:0:\ldots:z^d]$ with $f_I(\infty)=p_I$ and contracting to the basepoint $f_I(0)\notin X_H$, $I=1,\ldots,n=1$. Thus, the image of the marked point $\infty\in\bP$ provides a canonical identification between the fixed quasimaps $f_I$ and the fixed points $p_I$.
\item $\iota_1$ and $\iota_2$ are the embeddings of the fixed loci $\cQ\cM_{0,1}(X_H;d)^{T_F\times \bC_q^*}$ and  $X_H^{T_F\times \bC_q^*}$ into the quasimap moduli space $\cQ\cM_{0,1}(X_H;d)$ and into the Higgs branch $X_H$, respectively.
\item $ev_{d}^T$ is the restriction of the map $ev_d$ to the fixed loci. Thus, the evaluation map $ev_{d}^T$ identifies fixed points in $\cQ\cM_{0,1}(X_H;d)$ and with fixed points in $X_H$, as detailed for the example $\cT$.
\end{itemize}
Applying equivariant localization to the diagram \eqref{localization of quasi map} gives
\begin{equation}
    ev_{d,*} = \iota_{2,*}\circ ev_{d,*}^T \circ \iota_{1,*}^{-1} \ ,
\end{equation}
such that
\begin{multline}\label{quasimap loc in generalized cohomology}
    ev_{d,*}(\mathbf{1}^\text{vir}_d) = \iota_{2,*}\circ ev_{d,*}^T \circ \iota_{1,*}^{-1}(\mathbf{1}^\text{vir}_d) \\
    = \sum_{I \in \text{Fixed points}} \frac{e^{E}(\cN_{X_H,I})}{e^{E}(\cN_{\cQ\cM,I})}
    = \sum_{I \in \text{Fixed points}} \frac{1}{e^{E}(\cN^\text{vir}_{\cQ\cM,I}\ominus \cN_{X_H,I})}\ .
\end{multline}
Here $\cN_{X_H,I}$ and $\cN^{vir}_{\cQ\cM,I}$ are the (virtual) normal bundles  of the $I$-th fixed point in $X_H$ and $\cQ\cM_{0,1}(X_H;d))$. Furthermore, $e^{E}(\cN^\text{vir}_{\cQ\cM,I})$ and $e^{E}(\cN_{X_H,I})$ are the (multiplicative) Euler classes of these bundles in equivariant generalized cohomology theory $E(-)$.

\subsection{Explicit Computation for the Quasimap Moduli Space} \label{sec:quasimaps}
The next step is to determine for the example of the gauge theory~$\cT$ the sections of the normal bundles $\cN^{vir}_{\cQ\cM,I}$ and $\cN_{X_H,I}$ at the fixed loci of the quasimap moduli space~$\cQ\cM_{0,1}(X_H;d)$ and the Higgs branch $X_H$. Recall that according to eqs.~\eqref{eq:DefXH} and \eqref{eq:XHvsYH} the relevant quasimaps are
\begin{equation}\label{quasimap to XH}
    f: \bP^1 \dashrightarrow X_{H}=\text{Tot}(\oplus_{j }\cO(-l_j) \to \bP^n) \ \subset\
    Y_H = \left[\left(\bigoplus_{j=1}^a\bC_{-l_j}\oplus \bC^{n+1}_1\right)\middle/\bC^{*}\right] \ .
\end{equation}
Let the hyperplane bundle $L =\cO_{\bP^n}(1)$ of the projective space $\bP^n$ be the generator of the Picard group $\operatorname{Pic}(X_H)$, then the degree of quasimap~\eqref{quasimap to XH} is given by
\begin{equation}
    \deg f = \int_{\bP^1} f^{*}(c_1(L))
    = \int_{\bP^1} f^{*}(c_1(\cO_{\bP^n}(1)))
    = \int_{\bP^1} \cO_{\bP^1}(d) = d \ .
\end{equation}
We now work out the action of the symmetry group~$T_F \times \bC_{q}^{*}$ on the moduli stack $Y_H$ and hence on the quasimap moduli space:
\begin{itemize}
\item The characters of the symmetry group $T_F$ are $y_{I}$, $I=1,\ldots,n+1$, for the summand $\bC^{n+1}_1$ and $y_{n+1+j}$, $j=1,\ldots,a$, for the summand $\bigoplus_{j}\bC_{-l_j}$. The symmetry factor $\bC_{q}^*$ acts trivially on the vector space~$\bigoplus_{j=1}^a\bC_{-l_j}\oplus \bC^{n+1}_1$. 
\item While the symmetry $T_F$ acts trivially on the domain curve $\bP^1$ of the quasimap~\eqref{quasimap to XH}, the symmetry $\bC^*_q$ acts on $\bP^1$ as 
\begin{equation}\label{Cq action}
    \bC^{*}_q: \ [u:v] \in \bP^1 \mapsto [qu:v]\in \bP^1\ ,
\end{equation}
where we have introduced homogeneous coordinates $[u:v]$ for $\bP^1$ such that the marked point in terms of the affine coordinate $z=\frac{u}{v}$ at $z=\infty$.
\item According to eq.~\eqref{quasimap to XH} the $T_F \times \bC_{q}^*$-group action on the quasimap moduli space is now obtained according from pre-composition of the $\bC_{q}^*$-action on the domain curve $\bP^1$ and from post-composition of the $T_F$-action on the codomain~$X_H$.
\end{itemize}

On the Higgs branch~$X_H$, each summand~$\bC_{-l_j}$ gives rise to a line bundle $\cO_{\bP^{n}}(-l_j)$ together with $n+1$ line bundles $\cO_{\bP^n}(1)$ from the summand $\bC_{1}^{n+1}$. Therefore, the virtual tangent bundle of the quasimap moduli space at the point $f$ given by a quasimap~\eqref{quasimap to XH} reads
\begin{equation}\label{virtual deformation bundle}
     H^{0}(\bP^1, f^*{TX_{H}})\ominus H^{1}(\bP^1, f^*TX_{H}) \ ,
\end{equation}
where each component has an explicit description. Namely, with eq.~\eqref{quasimap to XH} each quasimap can be viewed as a map from $\bP^1$ into the quotient stack $Y_H$, which in turn corresponds to a map from $\bP^{1}$ into the vector space $\bigoplus_{j}\bC_{-l_j}\oplus \bC^{n+1}_1$ modulo $\bC^*$-equivalences.

The map~$f$ has the components $f = \left( f_j(u,v); F_I(u,v) \right)$, where $f_j$, $j=1,\ldots,a$, and $F_I$, $I=1,\ldots,n+1$, take values in $\bC_{-l_j}$ and $\bC_1$, respectively. Upon taking the GIT~quotient, the vector space components $\bC_{-l_j}$ and $\bC_{1}$ become the line bundles $\cO_{\bP^n}(-l_j)$ and $\cO_{\bP^n}(1)$ over the base $\bP^n$ of $X_H$, such that the deformation part $H^{0}(\bP^1, f^*{TX_{H}})$ and obstruction part $H^{1}(\bP^1, f^*TX_{H})$ of the virtual tangent bundle become with $f^{*}(\cO_{\bP^n}(1)) = \cO_{\bP^1}(d)$
\begin{equation}
\begin{aligned}
  H^{0}(\bP^1, f^*{TX_{H}}) &\simeq 
  \bigoplus_{I=1}^{n+1} H^0(\bP^1;\cO(d))\oplus \bigoplus_{j=1}^{a} H^0(\bP^1;\cO(-dl_j)) \simeq \bigoplus_{I=1}^{n+1} H^0(\bP^1;\cO(d))\ ,\\
  H^{1}(\bP^1, f^*TX_{H}) &\simeq 
  \bigoplus_{I=1}^{n+1} H^1(\bP^1;\cO(d))\oplus \bigoplus_{j=1}^{a} H^1(\bP^1;\cO(-dl_j)) \simeq \bigoplus_{j=1}^{a} H^1(\bP^1;\cO(-dl_j)) \ .
\end{aligned}
\end{equation}
Here we use that only positive line bundles over $\bP^1$ have global holomorphic sections, i.e., $\dim H^0(\bP^1,\cO(k))=k+1$ for $k\ge0$ and $\dim H^0(\bP^1,\cO(k))=0$ for $k<0$. Hence, only the line bundles~$\cO_{\bP^1}(d)$ contribute to the deformation part of the virtual tangent bundle. Moreover, Serre duality $H^1(\bP^1,\cO(k)) \simeq H^0(\bP^1,\cO(-k-2))^\vee$ implies $\dim H^1(\bP^1,\cO(k))=-k-1$ for $k\le -1$ and $\dim H^1(\bP^1,\cO(k))=0$ for $k\ge -1$, such that only the negative line bundles~$\cO(-dl_j)$ contribute to the obstruction part of the virtual tangent bundle. 

As a result the quasimap moduli space $\cQ\cM_{0,1}(X_H,d)$ is the same as $\cQ\cM_{0,1}(\bP^n,d)$ because they share the same deformation bundle. However, the line bundles $\cO(-l_j)$ in $X_H$ introduce obstructions that modify the virtual fundamental class of the quasimap moduli spaces associated to the Higgs branch $X_H$ compared to the quasimap moduli space  of the projective space $\bP^n$.

\paragraph{The deformation bundle $H^{0}(\bP^1, f^*{TX_{H}})$ of quasi map moduli:}
The deformation bundle $H^{0}(\bP^1, f^*{TX_{H}})$ is given by $F_I(u,v) \in \Gamma(\bP^1;\cO_{\bP^1}(d))$, $I=1,..,n+1$, modulo the $\bC^*$-action
 \begin{equation}
     s\in \bC^*: \ (F_1(u,v),\ldots,F_{n+1}(u,v)) \mapsto (s^{-1}F_1(u,v),\ldots,s^{-1}F_{n+1}(u,v)) \ .
\end{equation}
As $F_I(u,v) \in \Gamma(\bP^1;\cO_{\bP^1}(d))$, it can be expanded in terms of monomials of degree $d$ in the homogeneous coordinates $u$ and $v$
\begin{equation}\label{general form of fi(u,v)}
  F_{I}(u,v) = c_{I,d}u^d + \ldots + c_{I,r}u^{r}v^{d-r} +\ldots+ c_{I,0}v^d \ , \quad I=1,\ldots,n+1 \ ,
\end{equation}
where $u^{r}v^{d-r}$, $r =0,\ldots,d$, form a basis of $\Gamma(\bP^1;\cO_{\bP^1}(d))$. The $\bC_{q}^*$-action~\eqref{Cq action} maps
\begin{equation}
    \bC_{q}^*:\ u^{r}v^{d-r} \mapsto q^r u^{r}v^{d-r}.
\end{equation}
The $\bC_q^*$-invariant quasimaps satisfy
\[
(F_1 (qu,v), \ldots, F_n(qu,v_r)) = \lambda(q) (F_1 (u,v), \ldots,  F_n(u,v_r))\ ,
\]
where $\lambda$ may depend non-trivially on $q$. Hence the $\bC_q^*$-fixed point locus is
\[
\bigsqcup_{r=0, \ldots, d} [c_{1,r}:c_{2,r}: \,\ldots \, :c_{n+1,r}] q^r u^r v^{d-r} \ ,
\]
i.e., a disjoint union of $d+1$ copies of $\bP^n$. In addition we need to impose the condition that the marked point is mapped to $X_H$. Taking it to be $[1:0]$ singles out $r=0$. Hence, the $\bC_{q}^*$-fixed locus of the quasimap moduli space of degree $d$ is the same as the base $\bP^n$ of $X_H$. In addition, we need to consider the $T_F$ action, which the space of maps inherits from the target by postcomposition. Their fixed points are
\[ [1:0: \ldots, :0], \ldots , [0: \ldots, :1] \in \bP^n
\]
labeling $n+1$ points. On the level of characters, they are singled out by the condition $sy_{I}=1$, $I=1,\ldots,n+1$. The contributions to the normal bundle to these fixed points come from two places:
\begin{itemize}
\item $c_{I,r}, \forall I\in \{1,..,n+1\},r \in {1,..,d}$. The $T_F \times \bC^*_q$ weights of $c_{I,r}$ is $s^{-1}y_I^{-1}q^r$. Their contribution is 
\begin{equation}\label{weights of deformation bundle}
  \bigoplus_{J=1}^{n+1}
  \bigoplus_{r=1}^{d}\left.\bC_{s^{-1}y_{J}^{-1}q^r}
  \right|_{sy_I=1} \ .
\end{equation} 
\item The normal bundle at each fixed point in $\bP^n$ reads
\begin{equation}\label{weights of deformation bundle 2nd}
   \left.\cN_{\bP^{n}}\right|_{sy_I=1} 
   = \bigoplus_{J=1}^{n+1} \bC_{s^{-1}y_J^{-1}}\ ,
\end{equation}  
which is exactly the $r=0$ term from the first contribution. We will remove this part eventually.
\end{itemize}
The sum of the two contributions is
\begin{equation}\label{eq:normaldef}
 \bigoplus_{J=1}^{n+1}\bigoplus_{r=0}^{d}
 \left.\bC_{s^{-1}y_{J}^{-1}q^r}\right|_{sy_I =1} \ .
\end{equation}

\paragraph{The obstruction bundle $H^{1}(\bP^1, f^*{TX_{H}})$ of quasi map moduli:}
Now we turn to the obstruction bundle of the quasi map moduli space of degree $d$. Our above analysis showed that the $T_F\times \bC_{q}^*$ fixed locus of the quasimap moduli space is given by  $n+1$ points singled out by $sy_I =1$. Now we study the $T_F \times \bC_{q}^*$ weights of the obstruction bundle at these fixed points. Using Serre duality, we can analyze the one-from contribution with
\begin{equation}
    f_j \in  
    H^{0,1}_{\bar\partial}(\bP^1,\cO(-dl_j))
    \simeq H^{1}(\bP^1,\cO_{\bP^1}(-dl_j))
    \simeq H^{0}(\bP^1; \cO_{\bP^1}(dl_{j}-2))^\vee\ , \quad l_{j}>0 \ .
\end{equation}
The above equation does not yet involve the various gradings that we need to take into account. In particular, Serre duality on $\bP^1$ involves a pairing between $H^{0}(\bP^1; \cO_{\bP^1}(dl_{j}-2))$ and $H^{0,1}_{\bar\partial}(\bP^1,\cO(-dl_j))$ obtained by wedging with the canonical section and integrating over $\bP^1$. This symplectic pairing has a $-1$ shift, as $\bC^*_q$ acts non-trivially on derivatives. On the other hand, the weights of the flavor symmetries do not obtain shifts under the duality, and we merely need to take into account a shift by $n$. Our discussion hence parallels the one from above.
Compared to eq.~\eqref{eq:normaldef} we need to replace the degrees $d$ by $dl_j-2$. The summation over $r$ is thus different for each variable, and accordingly we refer to the different summation variables as $r_j$. The $r_j$ summation a priori starts at $0$, just as in eq.~\eqref{eq:normaldef}. The powers of $q$ get inverted, since Serre duality involves a dual of bundles, and as mentioned, shifted. Furthermore, the weight under the bulk gauge symmetry is given by $-l_j$, so that $s$ comes with this power, and the summation over the flavor representation needs to be shifted to the appropriate range. Taken together, we get for the obstruction bundle
\begin{equation}\label{weight of obstruction bundle}
    \bigoplus_{j=1}^a\bigoplus_{r_j=0}^{dl_j-2}
    \left.\bC_{s^{l_j} y^{-1}_{n+1+j}q^{-r_j-1}}\right|_{sy_I=1} \ .
\end{equation} 
Note that the powers of $q$ in this sum start at $q^{-1}$ and decrease till $q^{-dl_j+2}$. Now we can combine eqs.~\eqref{weights of deformation bundle}, \eqref{weights of deformation bundle 2nd} and \eqref{weight of obstruction bundle} to summarize the  weights of the virtual deformation bundle \eqref{virtual deformation bundle} at the fixed points $sy_I=1$: 
\begin{equation}\label{vietual normal bundle of quasimap moduli}
       \cN^{vir}_{\cQ\cM,I} = \left(\bigoplus_{J=1}^{n+1}\bigoplus_{r=1}^{d}
       \left.\bC_{s^{-1}y_{J}^{-1}q^r}\right|_{sy_I =1}\right)\oplus \left.\cN_{\bP^{n}}\right|_{sy_I=1} \ominus \left(\bigoplus_{j=1}^a\bigoplus_{r_j=0}^{dl_j-2}\left.\bC_{s^{l_j}y^{-1}_{n+1+j}q^{-r_j-1}}\right|_{sy_I=1}\right)
\end{equation}
However, recall that we want to construct the BPS vortex partition function with the zero sector normalized to one. This means we need to take out the contribution of the degree zero quasimaps moduli space, which is simply $X_{H}$. The weights of $TX_H$ at the fixed point $sy_I=1$ are
\begin{equation}
  \cN_{X_H,I}=  \left.  TX_H\right|_{sy_I=1} 
   = \cN_{\bP^{n}} \oplus \bigoplus_{j=1}^a
   \left.\bC_{s^{l_j}y^{-1}_{n+1+j}}\right|_{sy_I=1} \ .
\end{equation}
Hence, we obtain
\begin{equation}\label{normal bundle of degree d quasimap}
   \cN^{\text{vir}}_{\cQ\cM,I} \ominus \cN_{X_H,I} =  \left(\bigoplus_{J=1}^{n+1}\bigoplus_{r=1}^{d}
   \left.\bC_{s^{-1}y_{I}^{-1}q^r}\right|_{sy_I =1}\right)  
   \ominus 
   \left( \bigoplus_{j=1}^a\bigoplus^{r_j=dl_j-1}_{0}\left.\bC_{s^{l_j}y^{-1}_{n+1+j}q^{-r_j}}\right|_{sy_I=1}\right) \ .
\end{equation}
Substituting into eq.~\eqref{quasimap loc in generalized cohomology} gives the vortex partition function in an equivariant generalized cohomology theory $E(-)$.

\paragraph{Difference equation:}
Based on the properties of the quasimap moduli space in combination with the Euler class, we observe that the vortex partition functions \eqref{eq: math formulation of vortex partition} satisfy difference equations.

The virtual normal bundle with respect to the $T_F \times \bC_{q}^{*}$-fixed points of $\cQ\cM(X_{H};d)$ for each degree $d$ are given by eq.~\eqref{normal bundle of degree d quasimap}. Together with equivariant localization, they provide another interpretation of Ward identities formed by Wilson lines and surface operators in the three- and four-dimensional case, respectively. We note that:
\begin{itemize}
\item The degree $d$ quasimap \eqref{general form of fi(u,v)} can be naturally embedded into degree $d+1$ quasimap. This embedding is given by sending $u^{r}v^{d-r}$ to $u^{r}v^{1+d-r}$. As a result, we have:
\begin{equation}\label{difference of normal bundle}
\begin{aligned}
      \cN_{d+1}\ominus \cN_{d}|_{sy_{I}} &= \left(\bigoplus_{J=1}^{n+1}\bC_{s^{-1}y_{J}^{-1}q^{d+1}}\right) \ominus \left(\bigoplus_{j=1}^a\bigoplus_{a=q^{dl_{j}}}^{q^{(d+1)l_{j}-1}} \left.\bC_{s^{-l_j}y_{n+1+j}^{-1}q^{-a}}\right|_{sy_I=1}\right)\\
      &= \cE_{1,d} \ominus \cE_{2,d} \ .
\end{aligned}
\end{equation}
\item The (multiplicative) Euler class $e(\cE)$ in equivariant K~theory and equivariant elliptic cohomology satisfies 
    \begin{equation}
        e(\cE_{1,d}\ominus \cE_{2,d}) = \frac{e(\cE_{1,d})}{e({\cE_{2,d}})}
    \end{equation}
\end{itemize}
This makes the structure of vortex partition function clear, i.e., 
\begin{equation}
    \left.e(\cN_{d+1})\right|_{sy_{I}=1} = \left. e(\cN_{d}) \frac{e(\cE_{1,d})}{e(\cE_{2,d})}\right|_{sy_I=1} \ .
\end{equation}
By multiplying $\frac{e(\cE_{2,d})}{e(\cE_{1,d})}$, we can go from $(d+1)$-vortex sector (degree $d+1$ quasimaps) to $d$-vortex sector (degree $d$ quasimaps). There exists a canonical operator $\Lq$ as function of $q^{Q\partial_Q}$ such that for every $d\geq 0$ we get
\begin{equation}\label{difference operator from quasi map}
    \Lq\left(Q^{d+1}   \frac{1}{e(\cN_{d+1})} \right) = \left(Q^{d+1}   \frac{1}{e(\cN_{d+1})}\right)\frac{e(\cE_{1,d})}{e(\cE_{2,d})} =Q^{d+1}\frac{1}{e(\cN_{d})} \ .
\end{equation}
Hence, the operator $\Lq - Q$ annihilates the vortex partition function and gives Ward identities  in three- and four-dimensions. Note that the above arguments are not special to the example. We have described the structure of the quasimap moduli space in terms of bundles at the beginning of Section~\ref{sec:QuasiMapModuliSpace}. This description shows that there is a manifest action of the Picard group on the moduli space, relating quasimaps of different degrees. Likewise, the cited property of the Euler class does not depend on the example.  

\subsection{K-Theoretic Vortex Partition Function and Quasimaps} \label{sec:KTheoryVorPart}
 Here we specialize $E(-)$ to be the equivariant K~theory $K_{T_{F}\times \bC^*_q}(-)$. In this case,  $\mathbf{1}^{vir} $ is  the virtual structure sheaf $\cO^{\text{vir}}_d \in K_{T_{F}\times \bC^*_q}(\cQ\cM_{0,1}(X_H;d))$ \cite{Lee-foundation}. As a result, the K-theoretic vortex partition function is
\begin{equation}\label{K vortex partition function}
  Z^{K}_{vor} = \sum_{d\geq 0} Q^k ev_{d,*}(\cO^{\text{vir}}_d)
  \ \in \ K_{T_{F}\times \bC^*_q}(X_{H}) \ ,
\end{equation}
with 
\begin{equation}
\cO^{\text{vir}}_d \in  K_{T_{F}\times \bC^*_q}(\cQ\cM_{0,1}(X_H;d)) \ , \quad
ev_{d.* }: \, K_{T_{F}\times \bC^*_q}(\cQ\cM_{0,1}(X_H;d)) \to  K_{T_{F}\times \bC^*_q}(X_{H}) \ .
\end{equation}
Applying equivariant localization \eqref{quasimap loc in generalized cohomology} yields
\begin{multline}
    ev_{d,*}(\cO^{vir}_d) = \iota_{2,*}\circ ev_{d,*}^T \circ \iota_{1,*}^{-1}(\cO^{vir}_d) \\
    = \sum_{I \in \text{Fixed points}} \frac{e^{K}(\cN_{X_H,I})}{e^{K}(\cN_{\cQ\cM,I}^{\text{vir}})}
     = \sum_{I \in \text{Fixed points}} \frac{1}{e^{K}(\cN_{\cQ\cM,I}^{\text{vir}}\ominus \cN_{X_H,I})} \ ,
\end{multline}
such that eq.~\eqref{normal bundle of degree d quasimap} gives the K-theoretic vortex partition function
\begin{equation}
Z_{\text{vor}}^{K} = \sum_{d\geq 0} Q^d \sum_{I=1}^{n+1} \left.\frac{\prod_{j=1}^{a}\prod_{r_j=0}^{l_jd-1}(1-s^{-l_j}y _{n+1+j}q^{r_j})}{\prod_{J=1}^{n+1}\prod_{k=1}^{d}(1-sy_J q^{-k})}\right|_{sy_I =1} \ .
\end{equation}
Here we used the fact that K-theoretic Euler class $e^{K}(\cV)$ is $\wedge^*\cV^{\vee}$, so for $\cV = \oplus \cL_i$, $e^K( \oplus \cL_i) = \prod_i(1-\cL_i^{\vee})$. This is exactly the same as vortex partition function of three-dimensional $\cN=2$ supersymmetric gauge theory with the same spectrum as the four-dimensional $\cN=1$ theory $\cT$ \cite{Yoshida:2014ssa,Jockers:2018sfl}. The $q$-difference operator $\Lq$ in eq.~\eqref{difference operator from quasi map} is equivalent to the $q$-difference operator
\begin{equation}
    \hat{\mathfrak{L}}_q=
    \prod_{I=1}^{n+1} \left(1-sy_Iq^{-Q\partial_Q }\right) - Q \prod_{j=1}^{a}\prod_{r_j=0}^{l_j-1}\left(1-s^{-l_j}y_{n+1+j}q^{l_jQ\partial_Q+r_j}\right)\ .
\end{equation}

\subsection{Elliptic Vortex Partition Function and Anomalies}
\label{subsec: Elliptic vortex partition function and anomalies}
Now we lift the above K-theoretic discussion to the case of equivariant elliptic cohomology and propose a mathematical formulation of the vortex partition function of $\cT$ computed from a physical perspective in Section~\ref{sec:4d N=1 localization}. While there are essential differences between K~theory and elliptic cohomology, the main steps required in the calculation in section \ref{sec:KTheoryVorPart} are also available in equivariant elliptic cohomology. Most prominently, we can make use of localization, taking into account a non-trivial Thom sheaf in the definition of the push-forward \eqref{THom sheaf for a general map 1st}. We find that the Thom sheaf is linked to anomalies in four-dimensional $\cN=1$ supersymmetric gauge theories.
 
Let us work out the equivariant elliptic cohomology of the Higgs branch~$X_H$ first. One efficient way is to start with the  underlying scheme in K~theory and subsequently quotient by $\qt^{\bZ}$ (in the spirit that $\Ecurve = \bC^*/\qt^{\bZ}$, leading to eq.~\eqref{equi ell func from equi K 1st}).
The equivariant K~theory ring is given by
\begin{equation}
    K_{T_F\times \bC_q^*}(X_H) = \left.\cO\left[(\bC^{*})^{\text{rank}_{T_F}} \times \bC_{q}^{*}\times \bC_{s}^{*}\right]\middle/
    \prod_{I=1}^{n+1}\left(1-sy_I\right) \right. = 0 \ .
\end{equation}
To this ring one associates a scheme, which is obtained from 
\[
\spec(\cO[(\bC^{*})^{\text{rank}_{T_F}} \times \bC_{q}^{*}\times \bC_{s}^{*}])=\bC^{*,\text{rank}(T_F)+2} \ ,
\]
by cutting out a hypersurface defined by
\[
\prod_{I=1}^{n+1}\left(1-sy_I\right)=0 \ .
\]

This equation is satisfied whenever $1=sy_I$ for a single $I$. Instructed by eq.~\eqref{equi ell func from equi K 1st}, quotienting by $\qt^{\bZ}$, the ambient scheme becomes\footnote{This scheme has a physical meaning in $\cT$, $E_{\qt,T_F}$ is the moduli of flat $T_F$ background on spacetime $T^{2}_\tau$, $E_{\qt,q}, E_{\qt,s}$ are the moduli of flat Omega background and $\Uone$ gauge holonomy on spacetime $T^{2}_\tau$.}
\begin{equation}\label{extended elliptic scheme}
   \bC^{*,r_F+2}/\qt^{\bZ} = E_{\qt,T_F}\times E_{\qt,q}\times E_{\qt,s} \ .
\end{equation}
The scheme $Ell_{T_F\times \bC^*_q}$ is cut out from this by the equation 
\begin{equation}\label{subscheme embeding}
    \prod_{I=1}^{n+1}\theta(sy_I;\qt) =0 \ .
\end{equation}
Indeed, the vanishing locus consists of a union of lines, each of which is the vanishing locus of  $sy_I-1$ for a single  $I$, just like in K~theory.

We furthermore have a natural embedding:
\begin{equation}\label{extened elliptic scheme}
    \iota: Ell_{T_F \times \bC^*_q}(X_H) \hookrightarrow  E_{\qt,T_F}\times E_{\qt,q}\times E_{\qt,s} \ .
\end{equation}
It is more convenient to discuss the extended scheme $E_{\qt,T_F}\times E_{\qt,q}\times E_{\qt,s}$. The natural mathematical formulation of vortex partition function would be
 \begin{equation}
     Z_{\text{vor}}^{ ell} = \sum_{d\geq 0} Q^d Z^{ell}_{\text{vor},d} \ , \quad
     Z_{\text{vor},d}^{ell} = ev_{d,*}(\mathbf{1}^\text{vir} ) \ , 
\end{equation}
with
$$
  ev_{d,*}:\ Coh(Ell_{T_F \times \bC^*_q}(\cQ\cM_{0,1}(X_H;d)))) 
  \to \cEL_{T_F \times \bC^*_q}(X_H)\ . 
$$
As we already said, there is no known explicit virtual fundamental object~$\mathbf{1}^\text{vir}$ in elliptic cohomology and we simply assume its existence and denote it by $\mathbf{1}^\text{vir}$. A natural candidate for it is the Thom sheaf~$\Theta(\cE)$ associated to the obstruction bundle~$\cE$,\footnote{Notice that the virtual fundamental sheaf $\cO^\text{vir}$ in K~theory over the respective moduli space $\cM$ is, roughly speaking  given by the Koszul complex of the obstruction bundle $\cE$, i.e., $\wedge^*\cE^{\vee} \in K_{0}(\cM)$. The elliptic lift of this class, would be valued in the Thom sheaf~$\Theta(\cE)$ using eq.~\eqref{general Thom map 1st}, which is valued in the space of coherent sheaves of the underlying scheme $Ell_{T}(\cM)$.} which takes values in the coherent sheaves over the elliptic scheme $Coh(Ell_{T_F \times \bC^*_q}(\cQ\cM_{0,1}(X_H;d))))$, and not just in the structure sheaf $\cEL_{T_F \times \bC^*_q}(\cQ\cM_{0,1}(X_H;d))$.  In our case, we can take $\mathbf{1}^\text{vir} = \Theta(\cE)$ and the virtual object is taken into account via equivariant localization, i.e., the normal bundle is a virtual bundle, see eq.~\eqref{normal bundle of degree d quasimap}. Hence, we now proceed in this way without developing in full generality a definition for the virtual object $\mathbf{1}^\text{vir}$ in equivariant elliptic cohomology.
 
With this in mind, we can now use eq.~\eqref{localization of quasi map} for the elliptic case in order to compute the push-forward~$ ev_{d,*}( \mathbf{1}^\text{vir})$ as detailed in the following:
\begin{itemize}
\item Equivariant localization gives $ev_{d,*} = \iota_{2,*}\circ ev_{d,*}^T \circ \iota_{1,*}^{-1}$.
\item $\iota_{1,*}^{-1}$ is defined by
\begin{multline}
  \iota_{1,*}^{-1}: Coh(Ell_{T_F \times \bC^*_q}(\cQ\cM_{0,1}(X_H;d)))  \to \\
  \cEL_{T_F \times \bC^*_q}(\cQ\cM_{0,1}(X_H;d)^{T_F\times \bC_q^*} )\otimes \Theta(- \cN^{\text{vir}}_{\cQ\cM})\ .
\end{multline}
More precisely, after localization we have
\begin{equation}
  \iota_{1,*}^{-1}( s )
  = \frac{s}{e^{ell}(  \cN^{vir}_{\cQ\cM} )} \ ,
\end{equation}
where $\cN^{vir}_{\cQ\cM}$ is the virtual normal bundle~\eqref{vietual normal bundle of quasimap moduli} of the fixed points in $\cQ\cM_{0,1}(X_H;d)$.
\item $ev_d^T$ is identity, hence $ev_{d,*}^T =id$.
\item $\iota_{2,*}$ is defined as
\begin{equation}
  \iota_{2,*}: \cEL_{T_F \times \bC^*_q}(X_H^{T_F \times \bC^*_q})\otimes \Theta(-\cN_{X_H}) \to  \cEL_{T_F \times \bC^*_q}( X_H) \ .
\end{equation}
After localization we have:
\begin{equation}
    \iota_{2,*}(\eta) = \eta\cdot e^{ell}(\cN_{X_H}). 
\end{equation}
\end{itemize}
Finally, we get
 \begin{multline}
    ev_{d,*}( \mathbf{1}^{\text{vir}}) = \iota_{2,*}\circ ev_{d,*}^T \circ \iota_{1,*}^{-1}( 1\otimes \Theta(\cO^\text{vir}_d)) \\
    = \sum_{I \in \text{Fixed points}} \frac{e^{ell}(\cN_{X_H,I})}{e^{ell}(\cN_{\cQ\cM,I})} = \sum_{I\in \text{Fixed points}} \frac{1}{e^{ell}(\cN_{\cQ\cM,I}\ominus \cN_{X_H,I})}\ .
\end{multline}
Using eq.~\eqref{normal bundle of degree d quasimap}, we have
\begin{equation}
     \label{Math elliptic vortex partition function for XH}
     Z_{\text{vor}}^{ell} = \left.\sum_{I=1}^{n+1} \sum_{d\geq 0}Q^d  \frac{\prod_{j=1}^{a}\prod_{k=-l_jd+1}^{0}\theta(s^{-l_j}y_{n+1+j}q^{-k};\qt)}{\prod_{J=1}^{n+1}\prod_{k=1}^{d}\theta(sy_Jq^{-k};\qt)}\right|_{sy_I =1} \ .
 \end{equation}

This agrees with the vortex partition function of the gauge theory~$\cT$ derived in the Section~\ref{sec:VorParTheoryT}, up to the dependence on the $\Uone_R$ charges. The latter  is natural as in the mathematical formulation one a priori  only sees certain fixed choices, namely R-charges $\mathfrak{q} =0$ for bundle geometry or R-charges $\mathfrak{q} =2$ for superbundle geometry. After shifting the $\Uone_R$ charge as is needed for the index counting from the physics perspective~\eqref{eq:4d2dIndex}, based on the precise $\Uone_R$ charge assignment of theory $\cT$, we obtain the vortex partition function derived in last section\footnote{Although general R-charge assignments are natural from a physics perspective, their geometric significance is more subtle. As discussed in Section~\ref{subsect: anomaly and vortex string}, in the context of the four-dimensional chiral field theory the assignment of R-charges is constraint by the Adler--Bell--Jackiw anomaly for quantum elliptic cohomology.}
\begin{equation}
     \label{Math formulation of elliptic vortex partition function for XH}
     Z_{\text{vor}}^{ell} = \left. \sum_{I=1}^{n+1} \sum_{d\geq 0}Q^d  \frac{\prod_{j=1}^{a}\prod_{k=-l_jd+1}^{0}\theta(s^{-l_j}y_{n+1+j}q^{-k+\mathfrak{q}_j/2};\qt)}{\prod_{J=1}^{n+1}\prod_{k=1}^{d}\theta(sy_Jq^{-k+\mathfrak{q}_I/2};\qt)}\right|_{sy_Iq^{\mathfrak{q}_I/2}=1} \ .
\end{equation}
Here \eqref{Math formulation of elliptic vortex partition function for XH} is derived in terms of equivariant elliptic Euler classes, which are global meromorphic sections of Thom sheaves $\Theta(\cN_{\cQ\cM,I}\ominus \cN_{X_H,I})$ that are needed for the push-forward map in equivariant elliptic cohomology. This makes this computation agree with the  vortex partition function of $\cT$. The anomaly discussions therefore imply the following   conditions on these Thom sheaves:\footnote{A priori $\Theta(\cN_{\cQ\cM,i}\ominus \cN_{X_H,i})$ is defined as a (virtual) line bundle over the scheme $Ell_{T_F\times \bC^*_q}(X_H)$, a subscheme of the extended scheme $E_{\qt,T_F}\times E_{\qt,q}\times E_{\qt,s}$, see eq.~\eqref{extened elliptic scheme}. It can be viewed as coming from a (virtual) line bundle over the extended scheme restrict to the subscheme $Ell_{T_F\times \bC^*_q}(X_H)$.   It is common to view various objects over the underlying scheme $Ell_{T}(X)$ of a large class of GIT quotient $X$ in equivariant elliptic cohomology as pulled back from some extended schemes which contain $Ell_{T}(X)$ as a subscheme. See, for example, the elliptic stable envelope constructed in ref.~\cite{Aganagic:2016jmx}.}
\begin{itemize}
\item[] As the vortex partition function~\eqref{Math formulation of elliptic vortex partition function for XH} is required to be invariant under the transformations $s \to s \qt^{\bZ}$ by the anomaly-free condition, it implies for any embedding of algebraic varieties
\begin{equation*}
    \iota: E_{\qt,s} \hookrightarrow E_{q,s} \times E_{\qt,T_F}\times E_{\qt,q} \ ,
\end{equation*} 
that the pull back of the Thom sheaf $\iota^*\Theta(\cN_{\cQ\cM}\ominus \cN_{X_H})$ should be a trivial holomorphic line bundle.
\end{itemize}

Although the R-charge shift in eq.~\eqref{Math formulation of elliptic vortex partition function for XH} is suggested from physics, we see that it has a mathematical meaning. The component $E_{\qt,q}$ of elliptic scheme requires this modification by the R-charge in order to make the Thom sheaf trivial along $E_{\qt,s}$. The choices of such a shift is the same as the physical condition that the $\Uone_R$ symmetry does not suffer from an Adler--Bell--Jackiw anomaly.

Similarly to the equivariant cohomology and equivariant K~theory case, from \eqref{Math formulation of elliptic vortex partition function for XH} we obtain the elliptic analogue of Givental's I~function
\begin{equation}\label{elliptic I function}
I^{ell} = \sum_{d\geq 0}Q^d  \frac{\prod_{j=1}^{a}\prod_{k=-l_jd+1}^{0}\theta(s^{-l_j}y_{n+1+j}q^{-k+\mathfrak{q}_j/2};\qt)}{\prod_{I=1}^{n+1}\prod_{k=1}^{d}\theta(sy_Iq^{-k+\mathfrak{q}_I/2};\qt)}\ ,
\end{equation}
and $q$-difference operator \eqref{difference operator from quasi map} becomes
\begin{equation*}
   {  \prod_{I=1}^{n+1}\theta(s^{-1}y_Iq^{-Q\partial_Q+\mathfrak{q}_I/2})}-Q{\prod_{j=1}^{a}\prod_{k=0}^{l_j-1}\theta(s^{-l_j}y_{n+1+j} q^{l_j Q\partial_Q +k + \mathfrak{q}_j/2})} \ .
\end{equation*}
\paragraph{A subtlety of `elliptic Givental's I~function':}
In cohomology and K~theory, Givental's I~function of the of $X_H$ is a function valued in cohomology and K~theory ring of $X_H$. As such an explicit ring structure is absent in equivariant elliptic cohomology, we understand $I^{ell}$ as a meromorphic section of the (extended) Thom sheaves $\Theta(\cN_{\cQ\cM,i}\ominus \cN_{X_H,i})$ restricted to the subscheme defined by eq.~\eqref{subscheme embeding}. It would be interesting to see if this elliptic generalization of the Givental's I~function contains any enumerative information, as is the case in the context of  quantum cohomology and quantum K~theory.

\subsection{Generalizations to Other Cases}
\label{subsect: generalizations}
Although we only considered  gauge group $\Uone$ so far, the discussion can be easily generalized to higher rank abelian gauge groups $\Uone^r$ as well as non-abelian unitary groups $\operatorname{U}(r)$, provided the gauge theory has a non-trivial classical Higgs branch $X_H$. Although the quasimap moduli space associated with these Higgs branches will be more complicated, the building block of the computations will be the Higgs branch of theory $\cT$ we discussed in detail, as in quantum cohomology and quantum K theory. Hence we will not repeat the details here.

\paragraph{Four-dimensional $\cN=1$ supersymmetric $\Uone^r$ gauge theories:}
Consider four-dimensional $\cN=1$ abelian gauge theories with $\Uone^r$ gauge group and $n_f$ chirals with gauge charge $(q_{v}^i)$, $v=1,\ldots,r$, $i =1,\ldots ,r_F$. Provided that a classical Higgs branch~$X_H$ exists of the form of line bundles over a compact smooth toric variety, the discussions and calculations of the `elliptic Givental's I~function' can easily be generalized to this case. The corresponding Givental's I~function reads
\begin{equation} \label{elliptic I function for multi U(1)}
     I^{ell} = \sum_{d\geq 0}Q^d ev_{d,*}(\mathbf{1}^{\text{vir}})=  
     \sum_{d_{1}\geq 0,\ldots,d_r \geq 0}
     \prod_{v=1}^{r}Q_v^{d_v}  \frac{\prod_{i=1}^{n_f}\prod_{k=-\infty}^{0}\theta( s_i y_iq^{-k+\mathfrak{q}_i/2};\qt)}{\prod_{i=1}^{n_f}\prod_{k=-\infty}^{q_{v}^id_v}\theta( s_i y_iq^{-k+\mathfrak{q}_i/2};\qt)} \ ,
\end{equation}
where $s_i = \prod_{v=1}^{r}s_{v}^{-q_{v}^i}$ and the $y_i$, $i=1,\ldots,r_F$, are the characters of (the maximal torus of) the flavor symmetry $T_F$. The topological constraints of the Thom sheaf and its relations to the  anomaly are the same as in the $\Uone$ case, namely:
\begin{equation}\label{multiple U(1) = U(1)}
    \text{$\Uone^r$ is anomaly free} \equiv \text{Any $\Uone$ subgroup of $\Uone^r$ is anomaly free}
\end{equation}

\paragraph{Four-dimensional $\cN=1$ supersymmetric $\operatorname{U}(r)$ gauge theories:}
Consider the four-dimensional $\cN=1$ supersymmetric $\operatorname{U}(r)$ gauge theory with $n_f>r$ fundamental flavors.\footnote{The condition $n_f >r$ is needed for supersymmetry not being dynamically broken.} The classical Higgs branch $X_H$ is described by $n_f$ copies of the tautological vector bundles over the Grassmannian $\operatorname{Gr}(r,n_f)$. Although four-dimensional non-abelian gauge theories have different dynamics from four-dimensional abelian ones, their vortex partition function (or elliptic Givental's I~function of the corresponding Higgs branch~$X_H$) are closely related by abelian/non-abelian correspondence \cite{horivafamirrorsymmetry,Bertram:2003, Bertram:2004,Ciocan_Fontanine:2007,Cecotti:2013mba,Webb:2018}.  

\paragraph{$\Uone_R$ charges in four-dimensional $\operatorname{U}(r)$ supersymmetric gauge symmetry:}
In the previous discussion, we saw that the physical interpretation of Givental's I function is the vortex partition function of a supersymmetric gauge theory. The existence of non-trivial vortex configurations requires a gauge group with $\pi_1(G) \neq 0$, this is why we consider the gauge group $\operatorname{U}(r)$ gauge group instead of the simple gauge group~$\operatorname{SU}(r)$. On the other hand, the center $\Uone$ of the gauge group~$\operatorname{U}(r)$ gives additional conditions for the $\Uone_R$ symmetry to be anomaly free, i.e.,
\begin{equation}
\label{U(1)R anomaly from cental U(1)}
\sum_{I=1}^{n_f} (\mathfrak{q}_I -1) +(\mathfrak{q}'_{I}-1) = 0 \ ,
\end{equation}
while the $\operatorname{SU}(r)$ gauge group imposes the constraint
\begin{equation}
\label{U(1)R anomaly from  SU(r)}
\sum_{I=1}^{n_f} (\mathfrak{q}_I -1) +(\mathfrak{q}'_I-1) = -2r \ ,
\end{equation}
where $\mathfrak{q}_I$ and $\mathfrak{q}_I'$ are the $\Uone_R$ charge of $I$-th fundamental flavor and anti-fundamental flavor chiral multiplets, respectively. The contribution~$-2r$ of the right-hand side of the anomaly constraint~\eqref{U(1)R anomaly from  SU(r)} comes from gauginos in the $\operatorname{SU}(r)$ vector multiplets. \eqref{U(1)R anomaly from  SU(r)} and \eqref{U(1)R anomaly from cental U(1)} cannot simultaneously be satisfied. For the given spectrum we see an interesting tension in the four-dimensional theory:\footnote{One possible way to resolve the tension is introducing additional chiral multiplets in the determinantal representation of $\operatorname{U}(r)$. Here we focus on the minimal setup instead of introducing these determinantal chirals.} The central $\Uone$ is needed to support non-trivial vortex configurations, while at the same time its presence spoils the $\Uone_R$ symmetry! Both are essential for applying gauge theoretic methods to define in our setup quantum geometry on the classical Higgs branch $X_H$. This problem has already appeared in various other contexts of four-dimensions gauge theories \cite{Gaiotto1:2012xa,Gaiotto2:2014ina}. Here we apply the interpretations of these works, treating the central $\Uone$ as background and likewise the corresponding BPS vortex configurations as non-trivial background, or consider the central $\Uone$ is very weakly gauged. In other words, we want \eqref{U(1)R anomaly from  SU(r)} to be satisfied. Applying abelian/non-abelian correspondence in equivariant elliptic cohomology, we can then read off the elliptic Givental's I~function for $X_H$
\begin{multline}
    \label{non Abelian I function}
    I^{ell} =   \sum_{d_{1}\geq 0,\ldots,d_r \geq 0} Q ^{\sum_{v=1}^r d_v} \\
    \times\frac{\prod_{v=1}^{r}\prod_{I=1}^{n_f}\prod_{k=-\infty}^{0}\theta( s_v y_Iq^{-k+\mathfrak{q}_I/2};\qt)\theta( s^{-1}_v y'_iq^{-k+\mathfrak{q}'_I/2};\qt)}{\prod_{v=1}^{r}\prod_{I=1}^{n_f}\prod_{k=-\infty}^{d_v}\theta( s_v y_Iq^{-k+\mathfrak{q}_I/2};\qt)\prod_{k=-\infty}^{- d_v}\theta( s^{-1}_v y'_iq^{-k+\mathfrak{q}'_I/2};\qt)}  \\  
    \times\prod_{1\leq a\neq b\leq r}\frac{ \prod_{k=-\infty}^{d_a-d_b}\theta( \frac{s_a}{s_b}  q^{-k};\qt)}{\prod_{k=-\infty}^{0}\theta( \frac{s_a}{s_b}  q^{-k};\qt)}
\end{multline}
The physical origin  of the last part $ \prod_{1\leq a\neq b\leq r}\frac{ \prod_{k=-\infty}^{d_a-d_b}\theta( \frac{s_a}{s_b}  q^{-k};\qt)}{\prod_{k=-\infty}^{0}\theta( \frac{s_a}{s_b}  q^{-k};\qt)}$ is the gaugino contribution to the BPS vortex partition function. Here the shift $\mathfrak{q}_I/2,\mathfrak{q}'_I/2$ of the argument $q$ in eq.~\eqref{non Abelian I function} are preassigned $\Uone_R$ charges for the fundamental and anti-fundamental chirals which cannot be fully determined by localization in equivariant elliptic cohomology, analogously as in the $\Uone$ case discussed before. The mathematical elliptic I function \eqref{non Abelian I function} obtained using equivariant elliptic cohomology also contains anomaly information in physics. Under large gauge transformation $s_v \to s_v q^{n_v}$, $v=1,..,r$, the I~function~\eqref{non Abelian I function} produces additional factors involving arguments $q,y_i,y_i'$, but not on $s_v$. Independence of $s_v$ reflects the fact that the related four-dimensional $\cN=1$ supersymmetric theory is free of gauge anomaly. Moreover, the gaugino contribution $\prod_{1\leq a\neq b\leq r}\frac{ \prod_{k=-\infty}^{d_a-d_b}\theta( \frac{s_a}{s_b}  q^{k};\qt)}{\prod_{k=-\infty}^{0}\theta( \frac{s_a}{s_b}  q^{k};\qt)}$ itself does not produce factor of $s_v$ under large gauge transformations. This is in agreement with the fact that gauginos transform in real representation of gauge groups and therefore do not contribute to four-dimensional gauge anomalies. Requiring that the involvement of the factors $q,y_i,y_i'$ is trivial implies:
\begin{itemize}
\item $\prod_{i=1}^{n_f}y_iy'_i =1$, the same as the condition of flavor symmetries be free of $ABJ$ anomaly
\item Due to the presence of the factor $\prod_{1\leq a\neq b\leq r}\frac{ \prod_{k=-\infty}^{d_a-d_b}\theta( \frac{s_a}{s_b}  q^{k};\qt)}{\prod_{k=-\infty}^{0}\theta( \frac{s_a}{s_b}  q^{k};\qt)}$, the I~function~\eqref{non Abelian I function} does not generate $q$-factors under large gauge transformation. The $\Uone_R$ charges $\mathfrak{q}_I, \mathfrak{q}'_I$ need to satisfy eq.~
 \eqref{U(1)R anomaly from  SU(r)}  when restricted to the $\operatorname{SU}(r)$ gauge transformation, i.e., $\sum n_v =0$. This is in agreement with the fact that the gauginos modify the $\Uone_R$ anomaly-free condition from eq.~\eqref{U(1)R anomaly from cental U(1)} to eq.~\eqref{U(1)R anomaly from  SU(r)}. 
\end{itemize}
 
The relation between Thom sheaf and anomalies in gauge theory is thus similar to the $\Uone$ case:
\begin{itemize}
\item The gauge anomaly of $\operatorname{U}(r)$ is the same as the anomaly of the Cartan torus $\Uone^r$.\footnote{The Weyl group $\operatorname{W}_{\operatorname{U}(r)}$ of $\operatorname{U}(r)$ is not relevant for the perturbative chiral anomalies discussed here. Mathematically, the underlying elliptic schemes are different but closely related, namely $Ell_{\operatorname{U}(r)}(pt) = Ell_{\operatorname{U}(1)^r}(pt)/\operatorname{W}_{\operatorname{U}(r)}$. The discussion here is compatible with this further Weyl group quotient.} As gauginos are in the adjoint representation and do not contribute to anomalies (non-abelian gauginos only contribute to the Adler--Bell--Jackiw anomaly of the $\Uone_R$ symmetry, but this can also be captured by the Thom sheaf). As the I~function~\eqref{non Abelian I function} is invariant under $\operatorname{SU}(r)$ gauge transformations $s_v \to s_v q^{n_v}, \sum n_v =0$ gives the $\Uone_R$ anomaly free condition \eqref{U(1)R anomaly from  SU(r)}. 
\item Gauge anomaly of $\Uone^r$ can be further reduced to gauge anomaly of a single $\Uone$ by eq.~\eqref{multiple U(1) = U(1)}.
\end{itemize}
To conclude, we see a close relation between anomalies in four-dimensional $\cN=1$ gauge theories and equivariant elliptic cohomology:
\begin{itemize}
    \item[] The anomaly condition in four-dimensional $\cN=1$ gauge theory imposes topological and holomorphic conditions on the class of Thom sheaves from quasimaps. These conditions are non-trivial because the underlying scheme of equivariant elliptic cohomology is projective with non-trivial Picard group in general.   
\end{itemize}

Note that this relation can also be generalized to two-dimensional $\cN=(2,2)$ supersymmetric gauge theories and three-dimensional $\cN=2$ supersymmetric gauge theory in the following sense: Requiring that such gauge theories are free of continuous chiral anomalies also imposes topological and holomorphic constraints on the Thom sheaves obtained from the quasimaps. But in quantum cohomology and in quantum K~theory the Thom sheaves are automatically trivial, and hence these conditions are always empty. From the physics perspective this reflects the fact that these supersymmetric gauge theories in two and three dimensions are always free of continuous perturbative chiral anomalies.

\section{Conclusions}
\label{Conclusion, quantum version of equivariant elliptic cohomology.}
In this paper, we studied a gauge theoretic approach towards a quantum deformation of equivariant elliptic cohomology, using four-dimensional minimal supersymmetric gauge theory. On the physics side, we studied the (normalized) vortex partition function of four-dimensional gauge theory on $T^2_{\tau} \times_{\varepsilon_1,\varepsilon_2}\bC$  by counting BPS states and analyzed how it encodes various anomaly information of the four-dimensional minimal supersymmetric gauge theory.  
Gauge theories with four supercharges can be defined in dimensions up to four, and provide a natural physical setting for the study of different cohomology theories and their deformations:
\begin{equation}
    \label{sequence of gauge theory in diemensions 1st}
    \text{$2d$ $\cN=(2,2)$} \ \rightarrow \ 
    \text{$3d$ $\cN=2$} \ \rightarrow \ \text{$4d$ $\cN=1$}
\end{equation}
There is a repeating pattern, in the sense that (B-type) boundary conditions in the two-dimensional theories are given in terms of K~theory, while the quantum deformation of the K-theoretic ring is a feature of the three-dimensional theory. Note that this is natural from a physics perspective, as the multiplicative structure is related to `merging' line operators -- an operation that is not available for boundary conditions. Going up in dimensions, the boundary conditions in three dimensions are related to elliptic cohomology, whose quantum deformation, the subject of this paper, is a property of the four-dimensional theory. 

The gauge theory in four dimensions is significantly different from two dimensions and three dimensions due to the appearance of various anomalies.\footnote{In this paper we focus on perturbative chiral anomalies} Via compactifying this theory on $T_\tau^2$ and studying the exact twisted chiral ring relation, we found a non-trivial $\varepsilon_1,\varepsilon_2$ background dependence that promoted the twisted chiral ring relations to difference equations. These difference equations, unlike the twisted chiral ring relations, will depend on the $\Uone_R$ charge of chiral multiplets via the deformation parameter $\varepsilon_1,\varepsilon_2$. The difference equations are interpreted as Ward identities formed by surface operators given by $\cN=(0,2)$ chiral and Fermi multiplets.

From the math side, we formulated the vortex partition function in four dimensions via applying localization techniques to the quasimap moduli space in equivariant elliptic cohomology. To provide a mathematical formulation of the vortex partition function, we furthermore postulated the existence of a virtual object in elliptic cohomology. As a result, we found that the anomaly information is tied to the Thom sheaf, an object that is essential in equivariant elliptic cohomology. Like perturbative chiral anomalies of four-dimensional gauge theories with four supercharges, the Thom sheaf is in general a non-trivial holomorphic line bundle in equivariant elliptic cohomology and is trivial in equivariant cohomology and K~theory. We found that the condition for unitary groups to be anomaly free in four dimensions becomes the condition for the Thom sheaf to be trivial over the underlying elliptic (sub-)scheme given by the flat group holonomies along the spacetime torus~$T^2_{\tau}$. Moreover, we argued that the Adler--Bell--Jackiw and 't Hooft anomalies can be viewed in a similar way. This relationship between the non-triviality of Thom sheaf and chiral perturbative anomalies provides a novel explanation for these anomalies to be absent in two-dimensional $\cN=(2,2)$ and three-dimensional $\cN=2$ supersymmetric gauge theories, as the Thom sheaf in equivariant cohomology and in equivariant K~theory is always trivial by construction.

Another point is that the quasimap moduli space of the Higgs branch $X_H$ contains more information than the Higgs branch itself, which is essential for our analysis at the quantum level. Namely, equivariant elliptic cohomology is described in terms of a scheme embedded in an ambient space, acted upon by the symmetries used for equivariant localization. Compared to the Higgs branch $X_H$, the associated quasimap moduli spaces features the additional factor~$E_{\qt,q}$ attributed to the $\Uone_R$-symmetry, c.f., eq.~\eqref{extended elliptic scheme}, which enters into the Thom sheaf of the push-forward~\eqref{localization of quasi map}. Therefore, the Thom sheaf of the quasimap moduli space of the Higgs branch $X_H$ --- and not the Thom sheaf of the Higgs branch~$X_H$ itself --- encodes the Adler--Bell--Jackiw anomaly of the $\Uone_R$ symmetry.

\bigskip

There are many directions to explore, and we list a few of them here:
\begin{itemize}
\item $G$-Equivariant elliptic cohomology can in general  be defined for any general algebraic variety with a $G$-equivariant action. Our four-dimensional $\cN=1$ gauge theoretic approach allows us to formulate a quantum deformation of  the Higgs branch $X_H$ for anomaly free gauge theories.\footnote{The mathematical calculation in Section~\ref{subsect: vortex partition math} applies to more general algebraic varieties, although we restrict in this work to the geometries coming from four-dimensional anomaly-free gauge theories.}  Here, we regard the variable $Q$ as a fixed parameter, specified by the value of the background FI parameter $\zeta$ (and its bosonic partner in the linear multiplet). These observations suggest further investigation in two different directions.
    
Firstly,  it is suggestive to view  the Novikov variable~$Q$ no longer as a constant, but as a section of a non-trivial line bundle that admits non-trivial transformations under large gauge transformations along the spacetime torus~$T^2_\tau$. In gauge theory, this would mean that the background fields (specified by the variable~$Q$) will transform under gauge transformation.\footnote{A consequence of making $Q$ transform non-trivially is that every term graded by $Q^d$ in the function~\eqref{Math formulation of elliptic vortex partition function for XH} will be a meromorphic section of a different line bundle.} For such an approach one needs a suitable generalization for the relevant background fields.

Secondly, while our findings offered an interpretation of the anomaly in terms of the Thom sheaf, one may ask how `anomaly-free' geometries are mathematically distinguished from geometries `with anomalies'.
\item BPS $\cN=(0,2)$ surface defects in four-dimensional $\cN=1$ gauge theory would naturally give rise to representatives of would-be equivariant elliptic cohomology classes.\footnote{To the best of our knowledge, we are not aware of a satisfying and concrete definition of elliptic equivariant elliptic cohomology class of a general algebraic variety.} 

In this paper we built surface operators from two-dimensional $\cN=(0,2)$ Fermi and chiral multiplets and coupled them to the bulk. In this way we only realized a subclass of BPS $\cN=(0,2)$ surface operators. A complete classification of such surface defects would be valuable, both from the physics and the mathematics perspective. This direction is currently systematically investigated in ref.~\cite{Cheng:WProg01}. 
\item In this paper, we focused on indices of a certain class of surface operators. These operators are, however, equipped with more structure, including defects of lower dimensions. It would be interesting to keep track of refined information. From the perspective of surface operators, the appearance of two categorical layers in elliptic cohomology seems natural.  This may well tie in with the mathematical expectation that elliptic cohomology is not 1-categorical in nature \cite{Lurie:2009, Scherotzke:2022}.
\item
In this paper, we restricted our attention to quasimaps with one non-singular point from $\bP^1$ to the Higgs branch~$X_H$. It is natural to generalize this set up, e.g., quasimaps with relative points and associated gluing properties. In quantum K~theory, ref.~\cite{OkounkovquantumK:2015spn} discusses quasimaps with non-singular and relative points, as well as associated gluing properties, where vertex partition functions serve as one of the basic building blocks. It would be quite interesting to explore such possibilities for the elliptic case.
\item For gauge theories with $8$ supercharges, there is a similar pattern:
\begin{equation} \label{sequence of gauge theory in diemensions 2nd}
  \text{$4d$ $\cN=2$} \ \rightarrow\
  \text{$5d$ $\cN=1$} \ \rightarrow\
  \text{$6d$ $\cN=(1,0)$} \ .
\end{equation}
The perturbative chiral gauge anomaly does not appear in four and five dimensions, but it imposes strong conditions in six dimensions. From the mathematical side, four- and five-dimensional gauge theories are related to counting instantons in equivariant cohomology and K~theory. It seems natural to suspect that six-dimensional gauge theories are related to counting instantons in equivariant elliptic cohomology. Some discussions on this aspect can be found in refs.~\cite{Gukov:2018iiq,Gukov:2025nmk,Kim:2025fpz, Nekrasov:2013xda}. 
Note that the gauge anomaly cancellation in six dimensions requires a generalized Green--Schwarz mechanism, which involves (anti) self-dual tensor multiplets. It would be interesting to explore their role in the context of equivariant elliptic cohomology and their implications on the to-be-constructed Thom sheaves.\footnote{One key difference between the two series \eqref{sequence of gauge theory in diemensions 1st} and \eqref{sequence of gauge theory in diemensions 2nd} are types of moduli spaces associated to their non-perturbative sectors. For theories in the sequence~\eqref{sequence of gauge theory in diemensions 1st} --- discussed in this work --- the moduli spaces of vortices play the central role. The existence of vortex configurations requires that the gauge group $G$ contains abelian factors, i.e., $\text{rank}(\pi_{1}(G))> 0$. For theories in the sequence~\eqref{sequence of gauge theory in diemensions 2nd}, we need to consider the moduli spaces of instantons. Instanton configurations require that the gauge group contains non-abelian factors such that $\text{rank}(\pi_3(G))>0$. Due to the non-abelian nature of the required gauge groups, the associated partial differential equations are non-linear. It would be interesting to compare vortex counting for the gauge theories~\eqref{sequence of gauge theory in diemensions 1st} to instanton counting for the gauge theories~\eqref{sequence of gauge theory in diemensions 2nd} in a systematic and unified manner. Notice that vortex configurations and instanton configurations are often closely related, see, e.g., refs.~\cite{Hanany:2004ea,nakajima2011handsaw}.} 
\item Cohomological and K~theoretic Gromov--Witten invariants are symplectic invariants that are formulated in terms of the moduli space of stable maps. These invariants can be related to the non-perturbative indices calculated over the moduli space of quasimaps in terms of the wall-crossing description along the lines of ref.~\cite{zbMATH06493551}. 
From the physics point of view, for two- and three-dimensional supersymmetric gauge theories the invariants of the quasimap moduli space and the invariants of the stable map moduli space are respectively attributed to the ultraviolet and the infrared regime of these gauge theories. 
This observation triggers the question whether elliptic Gromov--Witten invariants can also be defined for the moduli space of stable maps in the context of four-dimensional supersymmetric gauge theories.
If so, it would be interesting to know if one can formulate a wall-crossing description between the quasimap moduli spaces and the stable map moduli spaces along the lines of ref.~\cite{zbMATH06493551}, in order to calculate elliptic Gromov--Witten invariants explicitly.
\end{itemize}
We hope to address some of these questions in the future.

\bigskip

\section*{Acknowledgements}
We would like to thank
Cyril Closset,
Peter Mayr,
Ilarion Melnikov, 
Yuan-Pin Lee,
Ingmar Saberi,
and Eric Sharpe 
for interesting discussions and useful correspondences.
We are in particular thankful to Yuan-Pin Lee for sharing and explaining his insights on various aspects of quantum K~theory and quantum elliptic cohomology. 
We are grateful to Ziheng Cao, Eric Sharpe, Hao Zhang for patiently coordinating with us the submission of their own work on quantum elliptic cohomology. 
The authors would like to express special thanks to the Mainz Institute for Theoretical Physics (MITP) of the Cluster of Excellence PRISMA++ (Project ID 390831469), for its hospitality and support. 
I.B. thanks the organizers of the workshop `Quantum field theory with boundaries, impurities and defects' for an inspiring workshop and the Isaac Newton Institute for hospitality and support during the workshop.
This research has received funding from the Cluster of Excellence PRISMA++ (EXC 2118/2,
Project ID 390831469) as well as the the Cluster of Excellence ORIGINS EXC-2094-390783311 funded by the German Research Foundation (DFG) within the Germany Excellence Strategy.

\pagebreak

\bibliographystyle{JHEP}
\bibliography{QEllipticTheory}

\providecommand{\href}[2]{#2}\begingroup\raggedright\begin{thebibliography}{100}

\bibitem{Seiberg:1994bz}
N.~Seiberg, \emph{{Exact results on the space of vacua of four-dimensional SUSY
  gauge theories}}, \href{https://doi.org/10.1103/PhysRevD.49.6857}{\emph{Phys.
  Rev. D} {\bfseries 49} (1994) 6857}
  [\href{https://arxiv.org/abs/hep-th/9402044}{{\ttfamily hep-th/9402044}}].

\bibitem{Intriligator:1994jr}
K.A.~Intriligator, R.G.~Leigh and N.~Seiberg, \emph{{Exact superpotentials in
  four-dimensions}},
  \href{https://doi.org/10.1103/PhysRevD.50.1092}{\emph{Phys. Rev. D}
  {\bfseries 50} (1994) 1092}
  [\href{https://arxiv.org/abs/hep-th/9403198}{{\ttfamily hep-th/9403198}}].

\bibitem{Seiberg:1994rs}
N.~Seiberg and E.~Witten, \emph{{Electric - magnetic duality, monopole
  condensation, and confinement in N=2 supersymmetric Yang-Mills theory}},
  \href{https://doi.org/10.1016/0550-3213(94)90124-4}{\emph{Nucl. Phys. B}
  {\bfseries 426} (1994) 19}
  [\href{https://arxiv.org/abs/hep-th/9407087}{{\ttfamily hep-th/9407087}}].

\bibitem{Kaplunovsky:1994fg}
V.~Kaplunovsky and J.~Louis, \emph{{Field dependent gauge couplings in locally
  supersymmetric effective quantum field theories}},
  \href{https://doi.org/10.1016/0550-3213(94)00150-2}{\emph{Nucl. Phys. B}
  {\bfseries 422} (1994) 57}
  [\href{https://arxiv.org/abs/hep-th/9402005}{{\ttfamily hep-th/9402005}}].

\bibitem{Lerche:1989uy}
W.~Lerche, C.~Vafa and N.P.~Warner, \emph{{Chiral Rings in N=2 Superconformal
  Theories}}, \href{https://doi.org/10.1016/0550-3213(89)90474-4}{\emph{Nucl.
  Phys. B} {\bfseries 324} (1989) 427}.

\bibitem{Morrison:1994fr}
D.R.~Morrison and M.R.~Plesser, \emph{{Summing the instantons: Quantum
  cohomology and mirror symmetry in toric varieties}},
  \href{https://doi.org/10.1016/0550-3213(95)00061-V}{\emph{Nucl. Phys. B}
  {\bfseries 440} (1995) 279}
  [\href{https://arxiv.org/abs/hep-th/9412236}{{\ttfamily hep-th/9412236}}].

\bibitem{Closset:2015rna}
C.~Closset, S.~Cremonesi and D.S.~Park, \emph{{The equivariant A-twist and
  gauged linear sigma models on the two-sphere}},
  \href{https://doi.org/10.1007/JHEP06(2015)076}{\emph{JHEP} {\bfseries 06}
  (2015) 076} [\href{https://arxiv.org/abs/1504.06308}{{\ttfamily
  1504.06308}}].

\bibitem{Gerhardus:2018zwb}
A.~Gerhardus, H.~Jockers and U.~Ninad, \emph{{The Geometry of Gauged Linear
  Sigma Model Correlation Functions}},
  \href{https://doi.org/10.1016/j.nuclphysb.2018.06.008}{\emph{Nucl. Phys. B}
  {\bfseries 933} (2018) 65}
  [\href{https://arxiv.org/abs/1803.10253}{{\ttfamily 1803.10253}}].

\bibitem{Kapustin:2009kz}
A.~Kapustin, B.~Willett and I.~Yaakov, \emph{{Exact Results for Wilson Loops in
  Superconformal Chern-Simons Theories with Matter}},
  \href{https://doi.org/10.1007/JHEP03(2010)089}{\emph{JHEP} {\bfseries 03}
  (2010) 089} [\href{https://arxiv.org/abs/0909.4559}{{\ttfamily 0909.4559}}].

\bibitem{Beem:2012mb}
C.~Beem, T.~Dimofte and S.~Pasquetti, \emph{{Holomorphic Blocks in Three
  Dimensions}}, \href{https://doi.org/10.1007/JHEP12(2014)177}{\emph{JHEP}
  {\bfseries 12} (2014) 177} [\href{https://arxiv.org/abs/1211.1986}{{\ttfamily
  1211.1986}}].

\bibitem{Yoshida:2014ssa}
Y.~Yoshida and K.~Sugiyama, \emph{{Localization of three-dimensional
  $\mathcal{N}=2$ supersymmetric theories on $S^1 \times D^2$}},
  \href{https://doi.org/10.1093/ptep/ptaa136}{\emph{PTEP} {\bfseries 2020}
  (2020) 113B02} [\href{https://arxiv.org/abs/1409.6713}{{\ttfamily
  1409.6713}}].

\bibitem{Koroteev:2017nab}
P.~Koroteev, P.P.~Pushkar, A.V.~Smirnov and A.M.~Zeitlin, \emph{{Quantum
  K-theory of quiver varieties and many-body systems}},
  \href{https://doi.org/10.1007/s00029-021-00698-3}{\emph{Selecta Math.}
  {\bfseries 27} (2021) 87} [\href{https://arxiv.org/abs/1705.10419}{{\ttfamily
  1705.10419}}].

\bibitem{Jockers:2018sfl}
H.~Jockers and P.~Mayr, \emph{{A 3d Gauge Theory/Quantum K-Theory
  Correspondence}},
  \href{https://doi.org/10.4310/ATMP.2020.v24.n2.a4}{\emph{Adv. Theor. Math.
  Phys.} {\bfseries 24} (2020) 327}
  [\href{https://arxiv.org/abs/1808.02040}{{\ttfamily 1808.02040}}].

\bibitem{Jockers:2019lwe}
H.~Jockers, P.~Mayr, U.~Ninad and A.~Tabler, \emph{{Wilson loop algebras and
  quantum K-theory for Grassmannians}},
  \href{https://doi.org/10.1007/JHEP10(2020)036}{\emph{JHEP} {\bfseries 10}
  (2020) 036} [\href{https://arxiv.org/abs/1911.13286}{{\ttfamily
  1911.13286}}].

\bibitem{Jockers:2021omw}
H.~Jockers, P.~Mayr, U.~Ninad and A.~Tabler, \emph{{BPS indices, modularity and
  perturbations in quantum K-theory}},
  \href{https://doi.org/10.1007/JHEP02(2022)044}{\emph{JHEP} {\bfseries 02}
  (2022) 044} [\href{https://arxiv.org/abs/2106.07670}{{\ttfamily
  2106.07670}}].

\bibitem{Ueda:2019qhg}
K.~Ueda and Y.~Yoshida, \emph{{3d $ \mathcal{N} $ = 2 Chern-Simons-matter
  theory, Bethe ansatz, and quantum $K$-theory of Grassmannians}},
  \href{https://doi.org/10.1007/JHEP08(2020)157}{\emph{JHEP} {\bfseries 08}
  (2020) 157} [\href{https://arxiv.org/abs/1912.03792}{{\ttfamily
  1912.03792}}].

\bibitem{Gu:2020zpg}
W.~Gu, L.~Mihalcea, E.~Sharpe and H.~Zou, \emph{{Quantum K theory of symplectic
  Grassmannians}},
  \href{https://doi.org/10.1016/j.geomphys.2022.104548}{\emph{J. Geom. Phys.}
  {\bfseries 177} (2022) 104548}
  [\href{https://arxiv.org/abs/2008.04909}{{\ttfamily 2008.04909}}].

\bibitem{Gu:2022yvj}
W.~Gu, L.C.~Mihalcea, E.~Sharpe and H.~Zou, \emph{{Quantum K theory of
  Grassmannians, Wilson line operators, and Schur bundles}},
  \href{https://doi.org/10.1017/fms.2025.10088}{\emph{SIGMA} {\bfseries 13}
  (2025) e140} [\href{https://arxiv.org/abs/2208.01091}{{\ttfamily
  2208.01091}}].

\bibitem{Gu:2023tcv}
W.~Gu, L.~Mihalcea, E.~Sharpe, W.~Xu, H.~Zhang and H.~Zou, \emph{{Quantum K
  theory rings of partial flag manifolds}},
  \href{https://doi.org/10.1016/j.geomphys.2024.105127}{\emph{J. Geom. Phys.}
  {\bfseries 198} (2024) 105127}
  [\href{https://arxiv.org/abs/2306.11094}{{\ttfamily 2306.11094}}].

\bibitem{Closset:2023bdr}
C.~Closset and O.~Khlaif, \emph{{Grothendieck lines in 3d $ \mathcal{N} $ = 2
  SQCD and the quantum K-theory of the Grassmannian}},
  \href{https://doi.org/10.1007/JHEP12(2023)082}{\emph{JHEP} {\bfseries 12}
  (2023) 082} [\href{https://arxiv.org/abs/2309.06980}{{\ttfamily
  2309.06980}}].

\bibitem{Nekrasov4:2009rc}
N.A.~Nekrasov and S.L.~Shatashvili, \emph{{Quantization of Integrable Systems
  and Four Dimensional Gauge Theories}},  in \emph{{16th International Congress
  on Mathematical Physics}}, pp.~265--289, 2010,
  \href{https://doi.org/10.1142/9789814304634_0015}{DOI}
  [\href{https://arxiv.org/abs/0908.4052}{{\ttfamily 0908.4052}}].

\bibitem{Nekrasov:2009uh}
N.A.~Nekrasov and S.L.~Shatashvili, \emph{{Supersymmetric vacua and Bethe
  ansatz}},
  \href{https://doi.org/10.1016/j.nuclphysbps.2009.07.047}{\emph{Nucl. Phys. B
  Proc. Suppl.} {\bfseries 192-193} (2009) 91}
  [\href{https://arxiv.org/abs/0901.4744}{{\ttfamily 0901.4744}}].

\bibitem{Witten:1988xj}
E.~Witten, \emph{{Topological Sigma Models}},
  \href{https://doi.org/10.1007/BF01466725}{\emph{Commun. Math. Phys.}
  {\bfseries 118} (1988) 411}.

\bibitem{Witten:1990hr}
E.~Witten, \emph{{Two-dimensional gravity and intersection theory on moduli
  space}}, \href{https://doi.org/10.4310/SDG.1990.v1.n1.a5}{\emph{Surveys Diff.
  Geom.} {\bfseries 1} (1991) 243}.

\bibitem{MR1492534}
W.~Fulton and R.~Pandharipande, \emph{Notes on stable maps and quantum
  cohomology},  in \emph{Algebraic geometry---{S}anta {C}ruz 1995}, vol.~62,
  Part 2 of \emph{Proc. Sympos. Pure Math.}, pp.~45--96, Amer. Math. Soc.,
  Providence, RI (1997)
  [\href{https://arxiv.org/abs/alg-geom/9608011}{{\ttfamily
  alg-geom/9608011}}].

\bibitem{MR1442525}
D.R.~Morrison, \emph{Mathematical aspects of mirror symmetry},  in
  \emph{Complex algebraic geometry ({P}ark {C}ity, {UT}, 1993)}, vol.~3 of
  \emph{IAS/Park City Math. Ser.}, pp.~265--327, Amer. Math. Soc., Providence,
  RI (1997) [\href{https://arxiv.org/abs/alg-geom/9609021}{{\ttfamily
  alg-geom/9609021}}].

\bibitem{Festuccia:2011ws}
G.~Festuccia and N.~Seiberg, \emph{{Rigid Supersymmetric Theories in Curved
  Superspace}}, \href{https://doi.org/10.1007/JHEP06(2011)114}{\emph{JHEP}
  {\bfseries 06} (2011) 114} [\href{https://arxiv.org/abs/1105.0689}{{\ttfamily
  1105.0689}}].

\bibitem{Dumitrescu:2012ha}
T.T.~Dumitrescu, G.~Festuccia and N.~Seiberg, \emph{{Exploring Curved
  Superspace}}, \href{https://doi.org/10.1007/JHEP08(2012)141}{\emph{JHEP}
  {\bfseries 08} (2012) 141} [\href{https://arxiv.org/abs/1205.1115}{{\ttfamily
  1205.1115}}].

\bibitem{Closset:2013vra}
C.~Closset, T.T.~Dumitrescu, G.~Festuccia and Z.~Komargodski, \emph{{The
  Geometry of Supersymmetric Partition Functions}},
  \href{https://doi.org/10.1007/JHEP01(2014)124}{\emph{JHEP} {\bfseries 01}
  (2014) 124} [\href{https://arxiv.org/abs/1309.5876}{{\ttfamily 1309.5876}}].

\bibitem{Pestun:2007rz}
V.~Pestun, \emph{{Localization of gauge theory on a four-sphere and
  supersymmetric Wilson loops}},
  \href{https://doi.org/10.1007/s00220-012-1485-0}{\emph{Commun. Math. Phys.}
  {\bfseries 313} (2012) 71} [\href{https://arxiv.org/abs/0712.2824}{{\ttfamily
  0712.2824}}].

\bibitem{Gadde:2013wq}
A.~Gadde, S.~Gukov and P.~Putrov, \emph{{Walls, Lines, and Spectral Dualities
  in 3d Gauge Theories}},
  \href{https://doi.org/10.1007/JHEP05(2014)047}{\emph{JHEP} {\bfseries 05}
  (2014) 047} [\href{https://arxiv.org/abs/1302.0015}{{\ttfamily 1302.0015}}].

\bibitem{Dimofte:2017tpi}
T.~Dimofte, D.~Gaiotto and N.M.~Paquette, \emph{{Dual boundary conditions in 3d
  SCFT{\textquoteright}s}},
  \href{https://doi.org/10.1007/JHEP05(2018)060}{\emph{JHEP} {\bfseries 05}
  (2018) 060} [\href{https://arxiv.org/abs/1712.07654}{{\ttfamily
  1712.07654}}].

\bibitem{Hori:2013ika}
K.~Hori and M.~Romo, \emph{{Exact Results In Two-Dimensional (2,2)
  Supersymmetric Gauge Theories With Boundary}},
  \href{https://arxiv.org/abs/1308.2438}{{\ttfamily 1308.2438}}.

\bibitem{Honda:2013uca}
D.~Honda and T.~Okuda, \emph{{Exact results for boundaries and domain walls in
  2d supersymmetric theories}},
  \href{https://doi.org/10.1007/JHEP09(2015)140}{\emph{JHEP} {\bfseries 09}
  (2015) 140} [\href{https://arxiv.org/abs/1308.2217}{{\ttfamily 1308.2217}}].

\bibitem{Sugishita:2013jca}
S.~Sugishita and S.~Terashima, \emph{{Exact Results in Supersymmetric Field
  Theories on Manifolds with Boundaries}},
  \href{https://doi.org/10.1007/JHEP11(2013)021}{\emph{JHEP} {\bfseries 11}
  (2013) 021} [\href{https://arxiv.org/abs/1308.1973}{{\ttfamily 1308.1973}}].

\bibitem{Kinney:2005ej}
J.~Kinney, J.M.~Maldacena, S.~Minwalla and S.~Raju, \emph{{An Index for 4
  dimensional super conformal theories}},
  \href{https://doi.org/10.1007/s00220-007-0258-7}{\emph{Commun. Math. Phys.}
  {\bfseries 275} (2007) 209}
  [\href{https://arxiv.org/abs/hep-th/0510251}{{\ttfamily hep-th/0510251}}].

\bibitem{Romelsberger:2005eg}
C.~R{\"o}melsberger, \emph{{Counting chiral primaries in N = 1, d=4
  superconformal field theories}},
  \href{https://doi.org/10.1016/j.nuclphysb.2006.03.037}{\emph{Nucl. Phys. B}
  {\bfseries 747} (2006) 329}
  [\href{https://arxiv.org/abs/hep-th/0510060}{{\ttfamily hep-th/0510060}}].

\bibitem{Witten:1993yc}
E.~Witten, \emph{{Phases of N=2 theories in two-dimensions}},
  \href{https://doi.org/10.1016/0550-3213(93)90033-L}{\emph{Nucl. Phys. B}
  {\bfseries 403} (1993) 159}
  [\href{https://arxiv.org/abs/hep-th/9301042}{{\ttfamily hep-th/9301042}}].

\bibitem{Hopkins:2002}
M.J.~Hopkins, \emph{Algebraic topology and modular forms},  in
  \emph{Proceedings of the International Congress of Mathematicians, ICM 2002,
  Beijing, China, August 20--28, 2002. Vol. I: Plenary lectures and
  ceremonies}, pp.~291--317, Beijing: Higher Education Press; Singapore: World
  Scientific/distributor (2002)
  [\href{https://arxiv.org/abs/math/0212397}{{\ttfamily math/0212397}}].

\bibitem{Gepner:2023}
D.~Gepner and L.~Meier, \emph{On equivariant topological modular forms},
  {\emph{Compos. Math.} {\bfseries 159} (2023) 2638}
  [\href{https://arxiv.org/abs/2004.10254}{{\ttfamily 2004.10254}}].

\bibitem{stolz-teichner}
S.~Stolz and P.~Teichner, \emph{Supersymmetric field theories and generalized
  cohomology},  in \emph{Mathematical foundations of quantum field theory and
  perturbative string theory.}, pp.~279--340, Providence, RI: American
  Mathematical Society (AMS) (2011)
  [\href{https://arxiv.org/abs/1108.0189}{{\ttfamily 1108.0189}}].

\bibitem{Stelle:1978ye}
K.S.~Stelle and P.C.~West, \emph{{Minimal Auxiliary Fields for Supergravity}},
  \href{https://doi.org/10.1016/0370-2693(78)90669-X}{\emph{Phys. Lett. B}
  {\bfseries 74} (1978) 330}.

\bibitem{Sohnius:1981tp}
M.F.~Sohnius and P.C.~West, \emph{{An Alternative Minimal Off-Shell Version of
  N=1 Supergravity}},
  \href{https://doi.org/10.1016/0370-2693(81)90778-4}{\emph{Phys. Lett. B}
  {\bfseries 105} (1981) 353}.

\bibitem{MR2108214}
K.~Honda, \emph{Local properties of self-dual harmonic 2-forms on a
  4-manifold}, \href{https://doi.org/10.1515/crll.2004.2004.577.105}{\emph{J.
  Reine Angew. Math.} {\bfseries 577} (2004) 105}
  [\href{https://arxiv.org/abs/dg-ga/9705010}{{\ttfamily dg-ga/9705010}}].

\bibitem{Closset:2013sxa}
C.~Closset and I.~Shamir, \emph{{The $\mathcal{N}=1$ Chiral Multiplet on
  $T^2\times S^2$ and Supersymmetric Localization}},
  \href{https://doi.org/10.1007/JHEP03(2014)040}{\emph{JHEP} {\bfseries 03}
  (2014) 040} [\href{https://arxiv.org/abs/1311.2430}{{\ttfamily 1311.2430}}].

\bibitem{Longhi:2019hdh}
P.~Longhi, F.~Nieri and A.~Pittelli, \emph{{Localization of 4d $\mathcal{N}=1$
  theories on $\mathbb{D}^2\times \mathbb{T}^2$}},
  \href{https://doi.org/10.1007/JHEP12(2019)147}{\emph{JHEP} {\bfseries 12}
  (2019) 147} [\href{https://arxiv.org/abs/1906.02051}{{\ttfamily
  1906.02051}}].

\bibitem{Bullimore:2020jdq}
M.~Bullimore, S.~Crew and D.~Zhang, \emph{{Boundaries, Vermas, and
  Factorisation}}, \href{https://doi.org/10.1007/JHEP04(2021)263}{\emph{JHEP}
  {\bfseries 04} (2021) 263}
  [\href{https://arxiv.org/abs/2010.09741}{{\ttfamily 2010.09741}}].

\bibitem{saberietal}
R.~Eager, I.~Saberi and J.~Walcher, \emph{{Nilpotence varieties}},
  \href{https://doi.org/10.1007/s00023-020-01007-y}{\emph{Annales Henri
  Poincare} {\bfseries 22} (2021) 1319}
  [\href{https://arxiv.org/abs/1807.03766}{{\ttfamily 1807.03766}}].

\bibitem{Gukov:2006jk}
S.~Gukov and E.~Witten, \emph{{Gauge Theory, Ramification, and the Geometric
  Langlands Program}},  \href{https://arxiv.org/abs/hep-th/0612073}{{\ttfamily
  hep-th/0612073}}.

\bibitem{Gadde:2013dda}
A.~Gadde and S.~Gukov, \emph{{2d Index and Surface operators}},
  \href{https://doi.org/10.1007/JHEP03(2014)080}{\emph{JHEP} {\bfseries 03}
  (2014) 080} [\href{https://arxiv.org/abs/1305.0266}{{\ttfamily 1305.0266}}].

\bibitem{Griffiths:1994prl}
P.~Griffiths and J.~Harris, \emph{Principles of algebraic geometry.}, Wiley
  Class. Libr., New York, NY: John Wiley \& Sons Ltd., 2nd ed.~ed. (1994).

\bibitem{Witten:1986bf}
E.~Witten, \emph{{Elliptic Genera and Quantum Field Theory}},
  \href{https://doi.org/10.1007/BF01208956}{\emph{Commun. Math. Phys.}
  {\bfseries 109} (1987) 525}.

\bibitem{Witten:1987cg}
E.~Witten, \emph{{The Index of the Dirac Operator in Loop Space}},
  \href{https://doi.org/10.1007/BFb0078045}{\emph{Lect. Notes Math.} {\bfseries
  1326} (1988) 161}.

\bibitem{Benini:2013nda}
F.~Benini, R.~Eager, K.~Hori and Y.~Tachikawa, \emph{{Elliptic genera of
  two-dimensional N=2 gauge theories with rank-one gauge groups}},
  \href{https://doi.org/10.1007/s11005-013-0673-y}{\emph{Lett. Math. Phys.}
  {\bfseries 104} (2014) 465}
  [\href{https://arxiv.org/abs/1305.0533}{{\ttfamily 1305.0533}}].

\bibitem{Cheng:2026abk}
P.~Cheng and H.P.~De~Freitas, \emph{{Dai-Freed anomalies and level matching in
  heterotic asymmetric orbifolds}},
  \href{https://arxiv.org/abs/2604.19634}{{\ttfamily 2604.19634}}.

\bibitem{Hanany:2004ea}
A.~Hanany and D.~Tong, \emph{{Vortex strings and four-dimensional gauge
  dynamics}}, \href{https://doi.org/10.1088/1126-6708/2004/04/066}{\emph{JHEP}
  {\bfseries 04} (2004) 066}
  [\href{https://arxiv.org/abs/hep-th/0403158}{{\ttfamily hep-th/0403158}}].

\bibitem{Shifman:2015kla}
M.~Shifman and A.~Yung, \emph{{Critical String from Non-Abelian Vortex in Four
  Dimensions}},
  \href{https://doi.org/10.1016/j.physletb.2015.09.045}{\emph{Phys. Lett. B}
  {\bfseries 750} (2015) 416}
  [\href{https://arxiv.org/abs/1502.00683}{{\ttfamily 1502.00683}}].

\bibitem{Assel:2015nca}
B.~Assel, D.~Cassani, L.~Di~Pietro, Z.~Komargodski, J.~Lorenzen and
  D.~Martelli, \emph{{The Casimir Energy in Curved Space and its Supersymmetric
  Counterpart}}, \href{https://doi.org/10.1007/JHEP07(2015)043}{\emph{JHEP}
  {\bfseries 07} (2015) 043}
  [\href{https://arxiv.org/abs/1503.05537}{{\ttfamily 1503.05537}}].

\bibitem{Givental:2015Kvi}
A.~Givental, \emph{Permutation-equivariant quantum {K}-theory {VI}. {M}irrors},
   \href{https://arxiv.org/abs/1509.07852}{{\ttfamily 1509.07852}}.

\bibitem{Hanany:2003hp}
A.~Hanany and D.~Tong, \emph{{Vortices, instantons and branes}},
  \href{https://doi.org/10.1088/1126-6708/2003/07/037}{\emph{JHEP} {\bfseries
  07} (2003) 037} [\href{https://arxiv.org/abs/hep-th/0306150}{{\ttfamily
  hep-th/0306150}}].

\bibitem{nakajima2011handsaw}
H.~Nakajima, \emph{Handsaw quiver varieties and finite {{\(W\)}}-algebras},
  {\emph{Mosc. Math. J.} {\bfseries 12} (2012) 633}
  [\href{https://arxiv.org/abs/1107.5073}{{\ttfamily 1107.5073}}].

\bibitem{Losev:1999nt}
A.~Losev, N.~Nekrasov and S.L.~Shatashvili, \emph{{The Freckled instantons}},
  \href{https://arxiv.org/abs/hep-th/9908204}{{\ttfamily hep-th/9908204}}.

\bibitem{Losev:1999tu}
A.~Losev, N.~Nekrasov and S.L.~Shatashvili, \emph{{Freckled instantons in
  two-dimensions and four-dimensions}},
  \href{https://doi.org/10.1088/0264-9381/17/5/327}{\emph{Class. Quant. Grav.}
  {\bfseries 17} (2000) 1181}
  [\href{https://arxiv.org/abs/hep-th/9911099}{{\ttfamily hep-th/9911099}}].

\bibitem{Atiyah:1978ri}
M.F.~Atiyah, N.J.~Hitchin, V.G.~Drinfeld and Y.I.~Manin, \emph{{Construction of
  Instantons}}, \href{https://doi.org/10.1016/0375-9601(78)90141-X}{\emph{Phys.
  Lett. A} {\bfseries 65} (1978) 185}.

\bibitem{Narukawa:2004}
A.~Narukawa, \emph{The modular properties and the integral representations of
  the multiple elliptic gamma functions},
  \href{https://doi.org/10.1016/j.aim.2003.11.009}{\emph{Adv. Math.} {\bfseries
  189} (2004) 247} [\href{https://arxiv.org/abs/math/0306164}{{\ttfamily
  math/0306164}}].

\bibitem{Jeffrey:1995}
L.C.~Jeffrey and F.C.~Kirwan, \emph{Localization for nonabelian group actions},
  \href{https://doi.org/10.1016/0040-9383(94)00028-J}{\emph{Topology}
  {\bfseries 34} (1995) 291}
  [\href{https://arxiv.org/abs/alg-geom/9307001}{{\ttfamily
  alg-geom/9307001}}].

\bibitem{Nieri_2015}
F.~Nieri and S.~Pasquetti, \emph{Factorisation and holomorphic blocks in 4d},
  \href{https://doi.org/10.1007/jhep11(2015)155}{\emph{Journal of High Energy
  Physics} {\bfseries 2015} (2015) }.

\bibitem{Redlich:1983kn}
A.N.~Redlich, \emph{{Gauge Noninvariance and Parity Violation of
  Three-Dimensional Fermions}},
  \href{https://doi.org/10.1103/PhysRevLett.52.18}{\emph{Phys. Rev. Lett.}
  {\bfseries 52} (1984) 18}.

\bibitem{Redlich:1983dv}
A.N.~Redlich, \emph{{Parity Violation and Gauge Noninvariance of the Effective
  Gauge Field Action in Three-Dimensions}},
  \href{https://doi.org/10.1103/PhysRevD.29.2366}{\emph{Phys. Rev. D}
  {\bfseries 29} (1984) 2366}.

\bibitem{Aharony:1997bx}
O.~Aharony, A.~Hanany, K.A.~Intriligator, N.~Seiberg and M.J.~Strassler,
  \emph{{Aspects of N=2 supersymmetric gauge theories in three-dimensions}},
  \href{https://doi.org/10.1016/S0550-3213(97)00323-4}{\emph{Nucl. Phys. B}
  {\bfseries 499} (1997) 67}
  [\href{https://arxiv.org/abs/hep-th/9703110}{{\ttfamily hep-th/9703110}}].

\bibitem{tHooft:1979rat}
G.~'t~Hooft, \emph{{Naturalness, chiral symmetry, and spontaneous chiral
  symmetry breaking}},
  \href{https://doi.org/10.1007/978-1-4684-7571-5_9}{\emph{NATO Sci. Ser. B}
  {\bfseries 59} (1980) 135}.

\bibitem{beauville2013theta}
A.~Beauville, \emph{Theta functions, old and new},  in \emph{Open problems and
  surveys of contemporary mathematics}, pp.~99--132, Somerville, MA:
  International Press; Beijing: Higher Education Press (2013).

\bibitem{Hori:2000kt}
K.~Hori and C.~Vafa, \emph{{Mirror symmetry}},
  \href{https://arxiv.org/abs/hep-th/0002222}{{\ttfamily hep-th/0002222}}.

\bibitem{Gates:1984nk}
S.J.~Gates, Jr., C.M.~Hull and M.~Ro\v{c}ek, \emph{{Twisted Multiplets and New
  Supersymmetric Nonlinear Sigma Models}},
  \href{https://doi.org/10.1016/0550-3213(84)90592-3}{\emph{Nucl. Phys. B}
  {\bfseries 248} (1984) 157}.

\bibitem{Nekrasov2:2009uh}
N.A.~Nekrasov and S.L.~Shatashvili, \emph{{Supersymmetric vacua and Bethe
  ansatz}},
  \href{https://doi.org/10.1016/j.nuclphysbps.2009.07.047}{\emph{Nucl. Phys. B
  Proc. Suppl.} {\bfseries 192-193} (2009) 91}
  [\href{https://arxiv.org/abs/0901.4744}{{\ttfamily 0901.4744}}].

\bibitem{Nekrasov1:2009ui}
N.A.~Nekrasov and S.L.~Shatashvili, \emph{{Quantum integrability and
  supersymmetric vacua}},
  \href{https://doi.org/10.1143/PTPS.177.105}{\emph{Prog. Theor. Phys. Suppl.}
  {\bfseries 177} (2009) 105}
  [\href{https://arxiv.org/abs/0901.4748}{{\ttfamily 0901.4748}}].

\bibitem{Nekrasov3:2009zz}
N.~Nekrasov and S.~Shatashvili, \emph{{Bethe Ansatz and supersymmetric vacua}},
  \href{https://doi.org/10.1063/1.3149487}{\emph{AIP Conf. Proc.} {\bfseries
  1134} (2009) 154}.

\bibitem{Givental:2001clq}
A.~Givental and Y.-P.~Lee, \emph{{Quantum K-theory on flag manifolds,
  finite-difference Toda lattices and quantum groups}},
  \href{https://doi.org/10.1007/s00222-002-0250-y}{\emph{Invent. Math.}
  {\bfseries 151} (2003) 193}
  [\href{https://arxiv.org/abs/math/0108105}{{\ttfamily math/0108105}}].

\bibitem{Iritani:2013qka}
H.~Iritani, T.~Milanov and V.~Tonita, \emph{{Reconstruction and Convergence in
  Quantum K-Theory via Difference Equations}},
  \href{https://doi.org/10.1093/imrn/rnu026}{\emph{Int. Math. Res. Not.}
  {\bfseries 2015} (2015) 2887}
  [\href{https://arxiv.org/abs/1309.3750}{{\ttfamily 1309.3750}}].

\bibitem{Closset:2017bse}
C.~Closset, H.~Kim and B.~Willett, \emph{{$ \mathcal{N} $ = 1 supersymmetric
  indices and the four-dimensional A-model}},
  \href{https://doi.org/10.1007/JHEP08(2017)090}{\emph{JHEP} {\bfseries 08}
  (2017) 090} [\href{https://arxiv.org/abs/1707.05774}{{\ttfamily
  1707.05774}}].

\bibitem{Dimofte:2011ju}
T.~Dimofte, D.~Gaiotto and S.~Gukov, \emph{{Gauge Theories Labelled by
  Three-Manifolds}},
  \href{https://doi.org/10.1007/s00220-013-1863-2}{\emph{Commun. Math. Phys.}
  {\bfseries 325} (2014) 367}
  [\href{https://arxiv.org/abs/1108.4389}{{\ttfamily 1108.4389}}].

\bibitem{Grojnowski:2007}
I.~Grojnowski, \emph{Delocalised equivariant elliptic cohomology},  in
  \emph{Elliptic cohomology. Geometry, applications, and higher chromatic
  analogues. Selected papers of the workshop, Cambridge, UK, December 9--20,
  2002}, pp.~114--121, Cambridge: Cambridge University Press (2007).

\bibitem{Ganter2014}
N.~Ganter, \emph{The elliptic {W}eyl character formula},
  \href{https://doi.org/10.1112/s0010437x1300777x}{\emph{Compositio
  Mathematica} {\bfseries 150} (2014) 1196–1234}
  [\href{https://arxiv.org/abs/1206.0528}{{\ttfamily 1206.0528}}].

\bibitem{rosu}
I.~Rosu, \emph{Equivariant elliptic cohomology and rigidity},
  \href{https://doi.org/10.1353/ajm.2001.0027}{\emph{Am. J. Math.} {\bfseries
  123} (2001) 647} [\href{https://arxiv.org/abs/math/9912089}{{\ttfamily
  math/9912089}}].

\bibitem{ando}
M.~Ando, \emph{Power operations in elliptic cohomology and representations of
  loop groups}, {\emph{Trans. Am. Math. Soc.} {\bfseries 352} (2000) 5619}.

\bibitem{Lurie:2009}
J.~Lurie, \emph{A survey of elliptic cohomology},  in \emph{Algebraic topology.
  The Abel symposium 2007. Proceedings of the fourth Abel symposium, Oslo,
  Norway, August 5--10, 2007}, pp.~219--277, Berlin: Springer (2009).

\bibitem{Okounkov:2022Lectures}
A.~Okounkov, ``33 lectures on quasimaps and elliptic stable envelopes.''
  Lecture notes available online:
  \href{https://www.math.columbia.edu/~okounkov/33lectures.pdf}{https://www.math.columbia.edu/\textasciitilde{}okounkov/33lectures.pdf},
  2022.

\bibitem{nekrasov2015bethe}
N.A.~Nekrasov and S.L.~Shatashvili, \emph{{Bethe/Gauge correspondence on curved
  spaces}}, \href{https://doi.org/10.1007/JHEP01(2015)100}{\emph{JHEP}
  {\bfseries 01} (2015) 100} [\href{https://arxiv.org/abs/1405.6046}{{\ttfamily
  1405.6046}}].

\bibitem{ginzburg1995elliptic}
V.~Ginzburg, M.~Kapranov and E.~Vasserot, \emph{Elliptic algebras and
  equivariant elliptic cohomology},
  \href{https://arxiv.org/abs/q-alg/9505012}{{\ttfamily q-alg/9505012}}.

\bibitem{Lee-foundation}
Y.-P.~Lee, \emph{Quantum {{\(K\)}}-theory. {I}: {Foundations}},
  \href{https://doi.org/10.1215/S0012-7094-04-12131-1}{\emph{Duke Math. J.}
  {\bfseries 121} (2004) 389}
  [\href{https://arxiv.org/abs/math/0105014}{{\ttfamily math/0105014}}].

\bibitem{Dedushenko2:2023qjq}
M.~Dedushenko and N.~Nekrasov, \emph{{Interfaces and Quantum Algebras, II:
  Cigar Partition Function}},
  \href{https://arxiv.org/abs/2306.16434}{{\ttfamily 2306.16434}}.

\bibitem{Kontsevich}
M.~Kontsevich, \emph{{Enumeration of rational curves via Torus actions}},
  \href{https://arxiv.org/abs/hep-th/9405035}{{\ttfamily hep-th/9405035}}.

\bibitem{kernetal}
D.~Kern, {\'E}.~Mann, C.~Manolache and R.~Picciotto, \emph{Derived moduli of
  sections and push-forwards}, {\emph{Sel. Math., New Ser.} {\bfseries 31}
  (2025) 46} [\href{https://arxiv.org/abs/2210.11386}{{\ttfamily 2210.11386}}].

\bibitem{Thomas:2006}
R.P.~Thomas, \emph{Notes on {GIT} and symplectic reduction for bundles and
  varieties},  in \emph{Essays in geometry in memory of S. S. Chern},
  pp.~221--273, Somerville, MA: International Press (2006)
  [\href{https://arxiv.org/abs/math/0512411}{{\ttfamily math/0512411}}].

\bibitem{Ciocan-Fontanine:2010}
I.~Ciocan-Fontanine and B.~Kim, \emph{Moduli stacks of stable toric quasimaps},
  \href{https://doi.org/10.1016/j.aim.2010.05.023}{\emph{Advances in
  Mathematics} {\bfseries 225} (2010) 3022–3051}
  [\href{https://arxiv.org/abs/0908.4446}{{\ttfamily 0908.4446}}].

\bibitem{Ciocan-Fontanine:2011}
I.~Ciocan-Fontanine, B.~Kim and D.~Maulik, \emph{Stable quasimaps to {GIT}
  quotients}, \href{https://doi.org/10.1016/j.geomphys.2013.08.019}{\emph{J.
  Geom. Phys.} {\bfseries 75} (2014) 17}
  [\href{https://arxiv.org/abs/1106.3724}{{\ttfamily 1106.3724}}].

\bibitem{Kim:2011}
B.~Kim, \emph{Stable quasimaps},
  \href{https://doi.org/10.4134/CKMS.2012.27.3.571}{\emph{Commun. Korean Math.
  Soc.} {\bfseries 27} (2012) 571}
  [\href{https://arxiv.org/abs/1106.0804}{{\ttfamily 1106.0804}}].

\bibitem{Li:1996}
J.~Li and G.~Tian, \emph{Virtual moduli cycles and {Gromov}-{Witten} invariants
  of algebraic varieties},
  \href{https://doi.org/10.1090/S0894-0347-98-00250-1}{\emph{J. Am. Math. Soc.}
  {\bfseries 11} (1998) 119}
  [\href{https://arxiv.org/abs/alg-geom/9602007}{{\ttfamily
  alg-geom/9602007}}].

\bibitem{Aganagic:2016jmx}
M.~Aganagic and A.~Okounkov, \emph{{Elliptic stable envelopes}},
  \href{https://doi.org/10.1090/jams/954}{\emph{J. Am. Math. Soc.} {\bfseries
  34} (2021) 79} [\href{https://arxiv.org/abs/1604.00423}{{\ttfamily
  1604.00423}}].

\bibitem{horivafamirrorsymmetry}
K.~Hori and C.~Vafa, \emph{Mirror symmetry},
  \href{https://arxiv.org/abs/hep-th/0002222}{{\ttfamily hep-th/0002222}}.

\bibitem{Bertram:2003}
A.~Bertram, I.~Ciocan-Fontanine and B.~Kim, \emph{Two proofs of a conjecture of
  {Hori} and {Vafa}}, {\emph{Duke Math. J.} {\bfseries 126} (2005) 101}
  [\href{https://arxiv.org/abs/math/0304403}{{\ttfamily math/0304403}}].

\bibitem{Bertram:2004}
A.~Bertram, I.~Ciocan-Fontanine and B.~Kim, \emph{Gromov-{Witten} invariants
  for abelian and nonabelian quotients.}, {\emph{J. Algebr. Geom.} {\bfseries
  17} (2008) 275} [\href{https://arxiv.org/abs/math/0407254}{{\ttfamily
  math/0407254}}].

\bibitem{Ciocan_Fontanine:2007}
I.~Ciocan-Fontanine, B.~Kim and C.~Sabbah, \emph{The {A}belian/{N}onabelian
  correspondence and {F}robenius manifolds},
  \href{https://doi.org/10.1007/s00222-007-0082-x}{\emph{Inventiones
  mathematicae} {\bfseries 171} (2007) 301–343}
  [\href{https://arxiv.org/abs/math/0610265}{{\ttfamily math/0610265}}].

\bibitem{Cecotti:2013mba}
S.~Cecotti, D.~Gaiotto and C.~Vafa, \emph{{$tt^*$ geometry in 3 and 4
  dimensions}}, \href{https://doi.org/10.1007/JHEP05(2014)055}{\emph{JHEP}
  {\bfseries 05} (2014) 055} [\href{https://arxiv.org/abs/1312.1008}{{\ttfamily
  1312.1008}}].

\bibitem{Webb:2018}
R.~Webb, \emph{The abelian-nonabelian correspondence for {{\(I\)}}-functions},
  {\emph{Int. Math. Res. Not.} {\bfseries 2023} (2023) 2592}
  [\href{https://arxiv.org/abs/1804.07786}{{\ttfamily 1804.07786}}].

\bibitem{Gaiotto1:2012xa}
D.~Gaiotto, L.~Rastelli and S.S.~Razamat, \emph{{Bootstrapping the
  superconformal index with surface defects}},
  \href{https://doi.org/10.1007/JHEP01(2013)022}{\emph{JHEP} {\bfseries 01}
  (2013) 022} [\href{https://arxiv.org/abs/1207.3577}{{\ttfamily 1207.3577}}].

\bibitem{Gaiotto2:2014ina}
D.~Gaiotto and H.-C.~Kim, \emph{{Surface defects and instanton partition
  functions}}, \href{https://doi.org/10.1007/JHEP10(2016)012}{\emph{JHEP}
  {\bfseries 10} (2016) 012} [\href{https://arxiv.org/abs/1412.2781}{{\ttfamily
  1412.2781}}].

\bibitem{Cheng:WProg01}
P.~Cheng, C.~Closset, R.I.~Garcia and A.~Keyes, ``{Work in progress}.''

\bibitem{Scherotzke:2022}
S.~Scherotzke and N.~Sibilla, \emph{Equivariant elliptic cohomology, toric
  varieties, and derived equivalences},
  \href{https://doi.org/10.1093/imrn/rnae251}{\emph{Int. Math. Res. Not.}
  {\bfseries 2024} (2024) 14600}
  [\href{https://arxiv.org/abs/2210.10862}{{\ttfamily 2210.10862}}].

\bibitem{OkounkovquantumK:2015spn}
A.~Okounkov, \emph{Lectures on {{\(K\)}}-theoretic computations in enumerative
  geometry},  in \emph{Geometry of moduli spaces and representation theory.
  Lecture notes from the 2015 IAS/Park City Mathematics Institute (PCMI)
  Graduate Summer School, Park City, UT, USA, June 28 -- July 18, 2015},
  pp.~251--380, Providence, RI: American Mathematical Society (AMS); Princeton,
  NJ: Institute for Advanced Study (IAS) (2017)
  [\href{https://arxiv.org/abs/1512.07363}{{\ttfamily 1512.07363}}].

\bibitem{Gukov:2018iiq}
S.~Gukov, D.~Pei, P.~Putrov and C.~Vafa, \emph{{4-manifolds and topological
  modular forms}}, \href{https://doi.org/10.1007/JHEP05(2021)084}{\emph{JHEP}
  {\bfseries 05} (2021) 084}
  [\href{https://arxiv.org/abs/1811.07884}{{\ttfamily 1811.07884}}].

\bibitem{Gukov:2025nmk}
S.~Gukov, V.~Krushkal, L.~Meier and D.~Pei, \emph{{A new approach to
  (3+1)-dimensional TQFTs via topological modular forms}},
  \href{https://arxiv.org/abs/2509.12402}{{\ttfamily 2509.12402}}.

\bibitem{Kim:2025fpz}
H.~Kim, J.~Manschot, G.W.~Moore, R.~Tao and X.~Zhang, \emph{{Path Integral
  Derivations Of K-Theoretic Donaldson Invariants}},
  \href{https://arxiv.org/abs/2509.23042}{{\ttfamily 2509.23042}}.

\bibitem{Nekrasov:2013xda}
N.~Nekrasov, V.~Pestun and S.~Shatashvili, \emph{{Quantum geometry and quiver
  gauge theories}},
  \href{https://doi.org/10.1007/s00220-017-3071-y}{\emph{Commun. Math. Phys.}
  {\bfseries 357} (2018) 519}
  [\href{https://arxiv.org/abs/1312.6689}{{\ttfamily 1312.6689}}].

\bibitem{zbMATH06493551}
I.~Ciocan-Fontanine and B.~Kim, \emph{Wall-crossing in genus zero quasimap
  theory and mirror maps},
  \href{https://doi.org/10.14231/AG-2014-019}{\emph{Algebr. Geom.} {\bfseries
  1} (2014) 400} [\href{https://arxiv.org/abs/1304.7056}{{\ttfamily
  1304.7056}}].

\end{thebibliography}\endgroup
\end{document}